\documentclass[aps,prx,reprint,runinaddress,superscriptaddress,amsmath,amssymb,floatfix,longbibliography]{revtex4-1}

\usepackage{microtype}
\usepackage{wasysym}
\usepackage{times}
\usepackage[varg]{txfonts}
\usepackage{mathrsfs}
\usepackage{bm}

\usepackage{color}
\usepackage{graphicx}
\usepackage{hyperref}
\usepackage{booktabs}
\usepackage{natbib}

\def \la {\langle}
\def \ra {\rangle}
\renewcommand{\vec}[1]{\boldsymbol{#1}}
\newcommand{\vhat}[1]{\vec{\hat{#1}}}
\newcommand{\mat}[1]{#1}
\newcommand{\Jo}{J_{\rm O}}
\newcommand{\matone}{\vec{1}}
\newcommand{\lambdabulk}{\lambda_0}
\newcommand{\subref}[2]{\ref{#1}\hyperref[#1]{#2}}
\newcommand{\cc}[1]{{#1}^*}
\newcommand{\abo}[2]{{#1}$_2${#2}$_2$O$_7$}

\usepackage[nolist,nohyperlinks]{acronym}  
\newacro{SI}[SI]{{spin ice}}
\newacro{DSI}[DSI]{{dipolar spin ice}}
\newacro{NN}[NN]{{nearest neighbor}}

\definecolor{cred}{RGB}{188,55,84}
\hypersetup{colorlinks=true,linkcolor=cred,citecolor=cred,urlcolor=cred}

\begin{document}

\title{Spin Ice Thin Films: Large-\emph{N} Theory and Monte Carlo Simulations}

\author{\'{E}tienne Lantagne-Hurtubise}
\affiliation{Perimeter Institute for Theoretical Physics, Waterloo, Ontario, N2L 2Y5, Canada}
\affiliation{Department of Physics and Astronomy, University of British Columbia, Vancouver,  BC, V6T 1Z1, Canada}
\author{Jeffrey G. Rau}
\affiliation{Department of Physics and Astronomy, University of Waterloo, Ontario, N2L 3G1, Canada}
\author{Michel J.P. Gingras}
\affiliation{Perimeter Institute for Theoretical Physics, Waterloo, Ontario, N2L 2Y5, Canada}
\affiliation{Department of Physics and Astronomy, University of Waterloo, Ontario, N2L 3G1, Canada}
\affiliation{Canadian Institute for Advanced Research, MaRS Centre,
West Tower, 661 University Avenue,
Suite 505, Toronto, ON
M5G 1M1, Canada}

\date{\today}

\begin{abstract}
    We explore the physics of highly frustrated magnets in confined geometries, focusing on the Coulomb phase of pyrochlore spin ices. As a specific example, we  investigate thin films of nearest-neighbor spin ice, using a combination of analytic large-$N$ techniques and Monte Carlo simulations. In the simplest film geometry, with surfaces perpendicular to the $[001]$ crystallographic direction, we observe pinch points in the spin-spin correlations characteristic of a two-dimensional Coulomb phase. We then consider the consequences of crystal symmetry breaking on the surfaces of the film through the inclusion of orphan bonds. We find that when these bonds are ferromagnetic, the Coulomb phase is destroyed by the presence of fluctuating surface magnetic charges, leading to a classical $Z_2$ spin liquid. Building on this understanding, we discuss other film geometries with surfaces  perpendicular to the $[110]$ or the $[111]$ direction. We generically predict the appearance of surface magnetic charges and discuss their implications for the physics of such films, including the possibility of an unusual $Z_3$ classical spin liquid. Finally, we comment on open questions and promising avenues for future research.
\end{abstract}

\maketitle

\section{Introduction}

The emergence of gauge structures in strongly correlated systems has proven to be an essential thread in the fabric of modern condensed matter physics~\cite{Kogut,Nagaosa_book,Wen_book,Lee_RMP}. In the prototypical example of a gauge theory -- electromagnetism -- boundary conditions can play a key role in the physics~\cite{jackson2007classical}. Indeed, realizations of gauge theories in systems with \emph{confined} geometries can lead to rich and varied phenomena as, for example, in the Casimir effect~\cite{milton2001casimir,bordag2001new}. In the same spirit, questions pertaining to surface effects in \emph{emergent} gauge theories of
strongly correlated systems, have only recently begun to be addressed~\cite{senthil2016}. A paradigmatic example where such an emergent $U(1)$ gauge theory arises is in spin ice materials~\cite{Harris1997,Ramirez1999}, a class of highly frustrated three-dimensional magnets. Given the level of maturity of research on spin ice~\cite{Bramwell_book2004,Gingras2011}, with many theoretical successes and several well-understood experimental examples, it is a natural system to explore the effects of confined geometries in emergent gauge theories.

\begin{figure}[htp]
  \centering
  	\includegraphics[width=0.95\columnwidth]{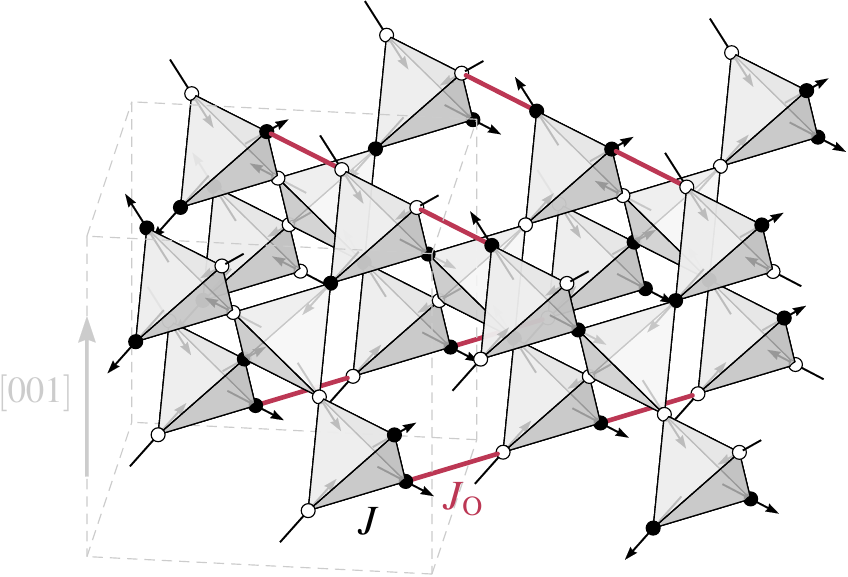}
	\caption{Spin ice film of thickness $L=1$. The magnetic ions form a pyrochlore lattice, composed of corner-sharing tetrahedra. Cleaving the surfaces perpendicular to the $[001]$ direction exposes \emph{orphan bonds} on the top and bottom surfaces, not belonging to any complete tetrahedron (shown in red). These orphan bonds have an exchange coupling $\Jo$ distinct from that of the other bonds, $J$}
	\label{fig:pyrochlore}
\end{figure}

In the prototypical spin ice materials Dy$_2$Ti$_2$O$_7$ and Ho$_2$Ti$_2$O$_7$, the magnetic moments reside on the sites of a pyrochlore lattice, which is formed of corner-sharing tetrahedra, as shown in Fig. \ref{fig:pyrochlore}. At low temperatures, these magnetic moments are forced by the crystalline electric field~\cite{RauPRB2015} to point either \emph{in} or \emph{out} of any given tetrahedron. The strongly frustrated interactions between the magnetic moments then give rise to a local ``2-in/2-out" constraint on every tetrahedron in the ground state, a close of analogue of the arrangement of protons in common water ice~\cite{Bramwell_book2004,Gingras2011}. This constraint can be rewritten in a form similar to Gauss' law in electromagnetism~\cite{Henley2005}, giving rise to an emergent \emph{Coulomb phase}~\cite{Henley2010,Castelnovo_AnnRevCMP} in these materials. This phase is characterized by algebraic spin correlations with fractionalized excitations taking the form of emergent magnetic monopoles~\cite{Castelnovo2008a,Castelnovo_AnnRevCMP}. Furthermore, quantum models of spin ice materials have been suggested as promising platforms to realize a related \emph{quantum spin liquid} phase~\cite{Gingras2014}.

Research on spin ice materials has so far mainly focused on bulk properties~\cite{Gingras2011}. In water ice, some interesting physics has been found by looking at the effects of \emph{confined} geometries. For example, by squeezing water between two sheets of graphene~\cite{algara-siller_square_2015}, one finds that the water molecules form a square lattice, reminiscent of the six-vertex model. The investigation of spin ice films in such confined geometries -- thin films, in particular -- is now developing~\cite{Bovo2014,Petrenko2014,Leusink2014}. Very recently, the first films of Dy$_2$Ti$_2$O$_7$~\cite{Bovo2014} and Ho$_2$Ti$_2$O$_7$~\cite{Leusink2014} were grown, with Ref.~[\onlinecite{Bovo2014}] reporting a vanishing residual entropy at low temperature, in strong contrast with bulk physics~\footnote{We note that the results of \citet{pomaranski2013absence} report a release of some of the spin ice entropy at very low temperatures in bulk \abo{Dy}{Ti}. The origin of this release is still a matter of debate~\cite{henelius2016,borzi2016intermediate}.}. The theoretical work to date has tackled a variety of issues; for example, Refs.~[\onlinecite{She2017},\onlinecite{Sasaki2014}] considered heterostructures involving spin ice materials,  while~\citet{Jaubert2016} investigated the dipolar spin ice model in a thin film geometry using Monte Carlo methods. However, to the best of our knowledge, there currently is no theoretical understanding of even the simplest minimal model, that of nearest-neighbor spin ice films.

\begin{figure}[t]
  \centering
  \includegraphics[width=0.95\columnwidth]{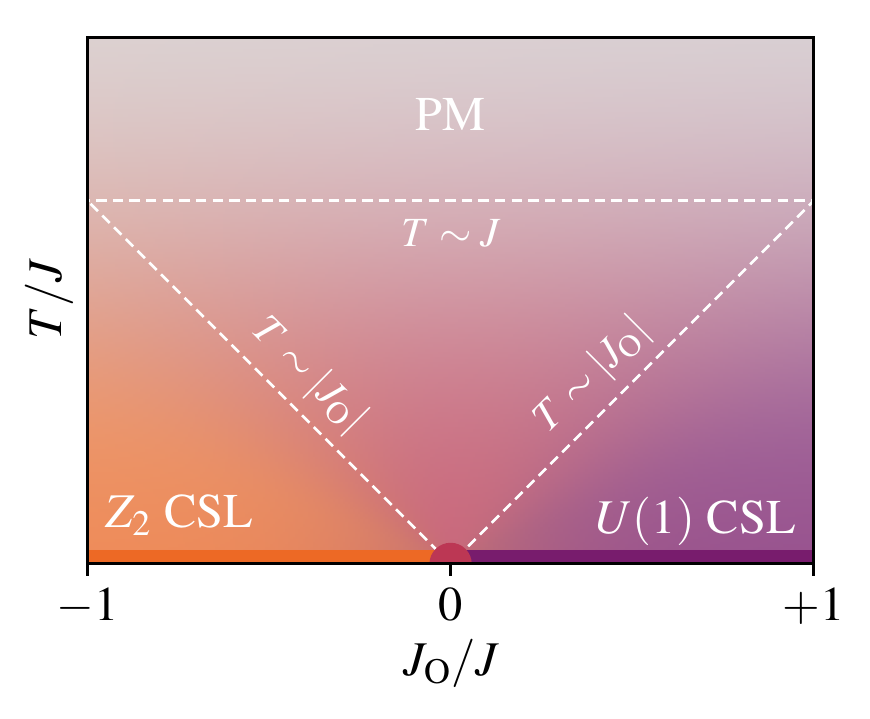}
  \caption{Phase diagram of $[001]$ thin films of spin ice as a function of temperature and orphan bond coupling $\Jo/J$ [see Eq.~(\ref{eq:Hnn_Ising_orphan})]. The state at low temperature is either a classical $U(1)$ spin liquid ($\Jo/J>0)$ or a classical $Z_2$ spin liquid ($\Jo/J<0$). Two broad crossovers at $T \sim J$ and $T \sim |\Jo|$ (dashed lines) separate these phases from the high-temperature paramagnet (PM). For thick films, there is an additional crossover (not shown) where three-dimensional bulk spin ice behavior is recovered for $T \gtrsim J/\log L$.
    \label{fig:phase-diagram}
	}
\end{figure}

In this paper, we explore the physics of nearest-neighbor spin ice films, considering the fate of the three-dimensional Coulomb phase as well as the effects of different surface terminations. We take a two-pronged approach: we use the analytical large-$N$ method, which has been successful in applications to bulk spin ice~\cite{Isakov2004} and films of ferromagnets~\cite{Dantchev2014}, and validate its predictions for nearest-neighbor spin ice films using Monte Carlo simulations. We focus our investigation on the simplest highly symmetric film geometry, with surfaces perpendicular to the crystallographic $[001]$ direction. We find that: (i) The characteristic \emph{pinch points} found in the spin-spin correlations~\cite{Henley2005,Henley2010} of bulk spin ice remain intact for momenta parallel to the surfaces, a signature of a \emph{two-dimensional} Coulomb phase (a classical $U(1)$ spin liquid). (ii) The direct space spin-spin correlations \emph{oscillate} as a function of depth in the sample, with an amplitude that \emph{increases} with \emph{decreasing} temperature. (iii) By including \emph{orphan bonds} to capture some of the crystal symmetry breaking of the film surfaces, we find that the Coulomb phase and its associated pinch points disappear when the exchange on the orphan bonds is ferromagnetic, yielding a classical $Z_2$ spin liquid~\cite{Rehn2017}. These results are summarized in the phase diagram shown in Fig. \ref{fig:phase-diagram}. Finally, building on this understanding, we extend these results to discuss the surface states of films with cleaved surfaces perpendicular to the $[110]$ or $[111]$ direction. From general considerations, we predict the appearance of surface magnetic charges in the ground state, akin to the monopoles realized as excitations in bulk spin ice. We discuss some implications of these surface charges, offering guidance for future studies on spin ice films with such geometries.

The rest of the paper is organized as follows: in Sec.~\ref{sec:model}, we detail our model and then, in Sec.~\ref{sec:methods}, develop the large-$N$ formalism used to investigate spin ice films. In Sec.~\ref{sec:results_001}, we apply the large-$N$ method to films with surfaces perpendicular to the $[001]$ direction, and compare these results to those obtained from Monte Carlo simulations. In Sec.~\ref{sec:topological}, we discuss the topological order of the classical $U(1)$ and $Z_2$ spin liquids found in these films.  Sec.~\ref{sec:non-neutral} briefly addresses other cleaving geometries, while Sec.~\ref{sec:discussion} offers concluding remarks and comments on possible avenues for future research. In Apps.~\ref{app:bulk} and \ref{app:films}, we provide details of the large-$N$ theory for bulk spin ice and $[001]$ films. In App.~\ref{app:numerics}, we discuss the numerical solution of the large-$N$ saddle point equations. Finally, in App.~\ref{app:mc}, we provide details of the Monte Carlo algorithm used to simulate Ising ($N=1$) spin ice films.

\section{Model}
\label{sec:model}
\subsection{Nearest-neighbor spin ice model}
\label{sec:model_A}

To set the stage, we first review the essential features of the nearest-neighbor (\ac{NN}) spin ice model and then, in Sec.~\ref{sec:geometries}, we minimally extend it to the context of films.

Recall that in classical spin ice~\cite{RauPRB2015}, the magnetic moments are represented by pseudospins $ \vec{S}_i= \sigma_i \vec{\hat{z}}_i $ living on pyrochlore lattice sites labeled by $i$, where $\sigma_i = \pm 1$ is a classical Ising variable and $\hat{\vec{z}}_i$ are unit vectors along the local quantization axes (see App.~\ref{app:bulk}).
In this work, we consider the simplest spin ice model, which only takes into account \ac{NN} Ising exchange interactions. In the bulk this is the celebrated \emph{pyrochlore Ising antiferromagnet} model~\cite{Anderson1956}
\begin{align}
	H = J \sum_{\langle ij\rangle} \sigma_i \sigma_j ,
	\label{eq:Hnn_Ising}
\end{align}
where $J>0$, and the sum runs over \ac{NN} bonds of the pyrochlore lattice. This Hamiltonian has a degenerate ground state manifold where every tetrahedron respects the local ice rules, i.e. the sum of Ising spins on every tetrahedron is zero. This realizes a classical $U(1)$ spin liquid, with an extensive ground-state degeneracy that remains down to zero temperature, thus giving a nonzero residual entropy. 

The structure of the ground-state manifold can be formulated in terms of a coarse-grained effective ``magnetic" field $\vec{B}(\vec{r})$ defined on each tetrahedron as  $\vec{B}(\vec{r}) \equiv (-1)^{\vec{r}} \sum_{i \in \vec{r}} \sigma_i \vec{\hat{z}}_i $, where the sign, $(-1)^{\vec{r}}$, depends on the sublattice of the tetrahedron in the dual diamond lattice (see Ref.~[\onlinecite{Henley2010}] for a review). In terms of $\vec{B}$, the ice-rule constraint amounts to a divergence-free condition, $\vec{\nabla}\cdot \vec{B} = 0$. Excitations above the spin ice ground-state manifold appear as pointlike sources or sinks of the field $\vec{B}$, behaving effectively as deconfined magnetic monopoles~\cite{Castelnovo2008a}. At low temperatures, an analogue of classical magnetostatics thus emerges and induces
a cooperative paramagnetic state dubbed a ``Coulomb phase"~\cite{Henley2010}. The divergence-free constraint also implies dipolar spin-spin correlations which manifest themselves as sharp ``pinch points" in reciprocal space~\cite{Henley2005,Henley2010}.

Even though it appears greatly simplified compared to dipolar spin-ice (\ac{DSI}) materials~\cite{DenHertog2000} such as Dy$_2$Ti$_2$O$_7$ and Ho$_2$Ti$_2$O$_7$, the \ac{NN} model captures much of the essential physics of the Coulomb phase shared with more realistic \ac{DSI} models~\cite{Castelnovo2008a,DenHertog2000,Isakov2005,Gingras2001CJP}. Although they are the best examples, dipolar interactions are not the \emph{only} route to realizing spin ice. Rare-earth magnets where super-exchange is dominant could potentially host more faithful realizations of \ac{NN} spin-ice [Eq.~(\ref{eq:Hnn_Ising})] due to the short-range nature of the exchange physics. For example, the Pr$_2$M$_2$O$_7$ family~\cite{Zhou2008,kimura2013quantum,Sibille2016}, recently discussed as \emph{quantum} spin-ice~\cite{Gingras2014} candidates, are expected to have \ac{NN} Ising exchange that is significantly larger than the magnetostatic dipolar interactions~\cite{onoda2010,onoda2011}.

\subsection{Film geometries and boundary conditions}
\label{sec:geometries}

In order to model \ac{NN} spin ice films, one must first define the
boundary conditions, such as choosing a cleaving plane along which to cut the pyrochlore lattice,  exposing free surfaces to a putative vacuum. For simplicity, we consider a free standing film and ignore complications arising from the presence of a substrate~\cite{Bovo2014,Leusink2014}. Three highly symmetric choices are planes normal to the $[001]$, $[110]$ and $[111]$ cubic crystallographic directions. We note that $[110]$ films of Dy$_2$Ti$_2$O$_7$ have been grown by~\citet{Bovo2014}, while films of Ho$_2$Ti$_2$O$_7$ for all three geometries have been grown by~\citet{Leusink2014}. In Sec.~\ref{sec:results_001}, we investigate the $[001]$ geometry in detail using the large-$N$ formalism and Monte Carlo simulations. Apart from being the simplest film geometry, it allows a direct comparison with the investigation of \ac{DSI} films recently reported in Ref.~[\onlinecite{Jaubert2016}]. We briefly explore other surface terminations, namely $[110]$ and $[111]$, in Sec.~\ref{sec:non-neutral}.

\begin{figure}
	\centering
        \includegraphics[width=\columnwidth]{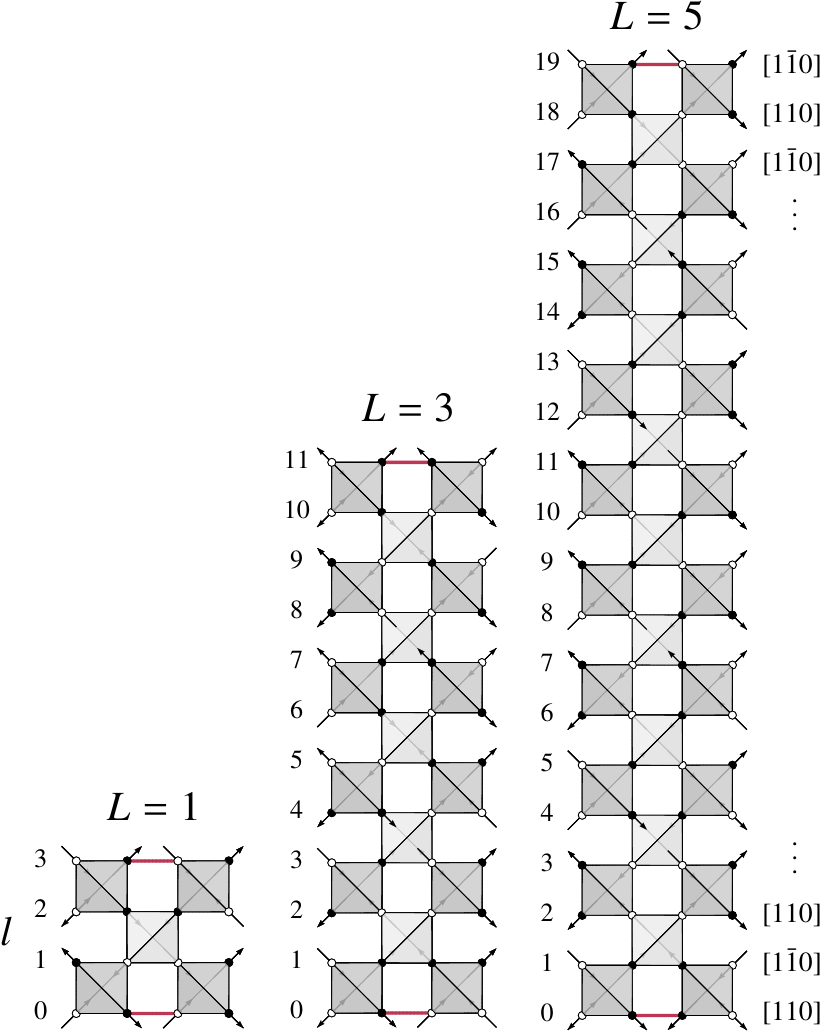}
	\caption{Structure of $[001]$ spin ice films for the three thicknesses discussed in the text ($L=1,3,5$). We show explicitly the layer index $l$ which runs from $l = 0$ to $l = 4L-1$ for a film of thickness $L$. Each layer contains two sublattices,  $\alpha = 2l,2l+1$, of the primitive cell of the film (see App.~\ref{app:films}). The one-dimensional chains that comprise each layer alternate between orientations $[110]$ and $[1\bar{1}0]$ from one layer to the next. Since the total number of layers, $4L$, is even, one surface has $[110]$ chains while the other has $[1\bar{1}0]$ chains.
	\label{fig:stacking}}
\end{figure}

Exposing surfaces perpendicular to the $[001]$ direction amounts to cutting two spins for each surface tetrahedron, as shown in Fig. \ref{fig:pyrochlore}. The resulting slab is comprised of stacked planes, or layers, on which the spins form chains oriented in the $[110]$ or $[1\bar{1}0]$ directions, alternatively, as shown in  Fig.~\ref{fig:stacking}. For simplicity, we consider thicknesses corresponding to an integer number $L$ of conventional cubic unit cells, comprising $4L$ spin layers (which we label by $l$) where the top chains are along $[1\bar{1}0]$ and the bottom chains run along $[110]$. The primitive unit cell of the film thus comprises $4L$ layers and $8L$ spins (see App.~\ref{app:films}). The associated conventional unit cells for film thicknesses $L=1,3,5$ are shown in Fig.~\ref{fig:stacking}.

As noted in Ref.~[\onlinecite{Jaubert2016}], this cleaving renders some of the surface bonds locally inequivalent to those in the bulk. Generically, one expects the bonds that join the remaining spins of a cut-off tetrahedra, which we call \emph{orphan bonds} following Ref.~[\onlinecite{Jaubert2016}], to have an Ising coupling, $\Jo$, different than the other bonds in the film. We thus consider the following minimal model
\begin{align}
	H = J \sum_{\langle ij\rangle} \sigma_i \sigma_j + (\Jo-J) \sum_{\langle ij\rangle {\in \text{O}}} \sigma_i \sigma_j,
	\label{eq:Hnn_Ising_orphan}
\end{align}
where the first sum, $\langle ij \rangle$, runs over \emph{all} \ac{NN} bonds while the second sum, $\langle ij \rangle \in {\rm O}$, runs  \emph{only} over the orphan bonds~\footnote{Our convention for the orphan-bond exchange $\Jo$ is different than that used in  Ref.~[\onlinecite{Jaubert2016}]. In the latter, an orphan bond has an exchange that differs from the bulk exchange $J$ by $\delta_O$, with $\delta_O$ positive or negative. We thus have $\Jo=J+\delta_O$ in the notation of  Ref.~[\onlinecite{Jaubert2016}].}.

\section{Methods}
\label{sec:methods}
With our model of spin ice thin films defined, we now outline the methods used to tackle these systems. We first discuss the large-$N$ method and review its application to bulk spin ice. Next, we discuss the modifications needed for an application to spin ice thin films. Finally, we discuss the Monte Carlo methods used to simulate Ising ($N=1$) spin ice films directly.

\subsection{Large-\emph{N} method in bulk spin ice}
\label{sec:methods:bulk}
The Hamiltonian for \ac{NN} spin ice, Eq.~(\ref{eq:Hnn_Ising}), can be investigated using an analytically tractable approximation scheme, the so-called large-$N$ expansion. This method allows one to obtain semi-quantitative spin-spin correlation functions at low temperature~\cite{Isakov2004}.
Consider the classical partition function, $\mathcal{Z}$, for the Hamiltonian in Eq.~(\ref{eq:Hnn_Ising}), written as
\begin{align}
\mathcal{Z} = \sum_{ \{\sigma_i\} = \pm 1 } \exp \left\{-\frac{\beta J}{2} \sum_{ij} V_{ij} \sigma_i \sigma_j \right\} ,
\label{eq:Z_largeN}	
\end{align}
where the sum over $i$ and $j$ runs over all pyrochlore lattice sites, and we have defined $V_{ij} = 1$ when $i$ and $j$ are nearest-neighbors, and $V_{ij} = 0$ otherwise.
Replacing the classical Ising spins $\sigma_i$ by continuous variables $s_i$ and enforcing the unit spin length constraint leads to the partition function
\begin{align}
\mathcal{Z} = \prod_{j} \int  ds_j\ \delta(s_j^2 - 1) \exp \left\{-\frac{\beta J}{2} \sum_{ij} V_{ij} s_i s_j \right\} .
\end{align}
In this form, the length constraints render the partition function as intractable as the original model. The large-$N$ approach circumvents this problem by extending these real variables, $s_i$, to $N$-component vectors $\vec{s}_i$ subject to the constraint
\begin{equation}
|\vec{s}_i|^2 = N .
\label{eq:selfconsistency_realspace}
\end{equation}
The interaction between the spins is also extended to be $O(N)$ symmetric, with the resulting $\mathcal{Z}_N$ partition function now taking the form
\begin{align}
\mathcal{Z}_N = \prod_{j} \int  d\vec{s}_j\ \delta(|\vec{s}_j|^2 - N) \exp \left\{-\frac{\beta J}{2} \sum_{ij} V_{ij} \vec{s}_i \cdot \vec{s}_j \right\} .
\label{eq:Z_largeN_si}	
\end{align}
Clearly, if we set $N=1$, we recover the original Ising model with $\mathcal{Z}_1 \equiv \mathcal{Z}$. The constraints at each site $i$ can be enforced using constraint fields $\mu_i$, so $\mathcal{Z}_N$ becomes
\begin{align*}
 \int \mathcal{D}\vec{s} \mathcal{D}\mu
 \exp \left\{ 
 -\frac{1}{2}\sum_j i \mu_j \left(|\vec{s}_j|^2-N\right) 
 -\frac{\beta J}{2} \sum_{ij} V_{ij} \vec{s}_i \cdot \vec{s}_j
 \right\},
\end{align*}
where $\mathcal{D}\vec{s} \equiv \prod_j d\vec{s}_j$ and $\mathcal{D}\mu \equiv \prod_j d\mu_j$.
Integrating out the $\vec{s}$ fields yields
\begin{equation}
\mathcal{Z}_N = \int \mathcal{D} \mu
\exp\left\{
-\frac{N}{2} {\rm Tr}\left[-i\mu + \log\left(
i\mat{\mu} + \beta J \mat{V}\right)
\right]
\right\},
\label{eq:integrated-out}
\end{equation}
where we have defined the diagonal matrix $\mat{\mu}$ with elements ${\mu}_{ij} \equiv \delta_{ij}\mu_i$. In this form, it is clear that as $N \rightarrow \infty$, the partition function
is dominated by the saddle points of the exponential. In the saddle-point solutions, $\mu$ is purely imaginary, so we consider the real variable $\lambda \equiv i\mu$. The saddle-point equations are then given by
\begin{equation}
(\mat{\lambda} + \beta J \mat{V})^{-1}_{ii} = 1.
\end{equation}
The correlation functions between the $\vec{s}_i$ can be readily obtained from Eqs.~(\ref{eq:Z_largeN_si},\ref{eq:integrated-out}) by taking a derivative with respect to $V_{ij}$. One obtains
\begin{equation}
\label{eq:ss-correlation}
    \langle s^a_i s^b_j\rangle = 
    \delta_{ab} (\mat{\lambda} + \beta J \mat{V})^{-1}_{ij},
\end{equation}
where $a,b = 1, \cdots N$ label the spin components. Note that by invoking this correlation function, the saddle-point condition can be interpreted as an \emph{average} length constraint on the spins $\vec{s}_i$ with
\begin{equation}
\la \vec{s}_i \cdot \vec{s}_i \ra = N.
\label{eq:avg_constraint}
\end{equation}

For bulk spin ice, the translation and rotation symmetries of the lattice enforce that the $\lambda_i$ are site independent with $\lambda_i \equiv \lambdabulk$. We can then block diagonalize the  interaction matrix $V$ using a Fourier transform which leads to
\begin{equation}	
\la \cc{s^a_\alpha({\vec{q}})} s^b_\beta({\vec{q}})\ra =\delta_{ab} \left[ \lambdabulk \matone + \beta J \mat{V}({\vec{q}}) \right]^{-1}_{\alpha \beta} ,
\label{eq:correlations_qspace}
\end{equation}
where $\alpha, \beta = 1,\dots,4$ index the four sublattices of the pyrochlore lattice, and the explicit form of the matrix $\mat{V}(\vec{q})$ is given in App.~\ref{app:bulk}. The average length constraint [Eq. (\ref{eq:avg_constraint})] can then be expressed as
\begin{align}
\frac{1}{n}\sum_{\alpha, {\vec{q}}} \la \cc{\vec{s}_\alpha({\vec{q}})}\cdot \vec{s}_\alpha({\vec{q}})\ra = N . 
\label{eq:selfconsistency_qspace}
\end{align}
where $n$ is the total number of spins in the system.

The previous derivation, leading to Eqs. (\ref{eq:correlations_qspace}) and (\ref{eq:selfconsistency_qspace}), is equivalent (in outcome) to the so-called \emph{self-consistent Gaussian approximation} or \emph{spherical approximation}~\cite{Stanley1968}. Perhaps surprisingly, this large-$N$ treatment has been found to provide semi-quantitative correlation functions when compared to Monte Carlo simulations of pyrochlore Ising ($N=1$) and Heisenberg ($N=3$) antiferromagnets~\cite{Isakov2004}. However, this method does not provide a good description of the physics for $N=2$ due to the manifestation of order-by-disorder~\cite{bramwell1994order,Moessner1998,Moessner1998a}. In principle, a $1/N$ expansion~\cite{Isakov2004} around this exactly solvable point would allow one to obtain more precise correlation functions for spins with finite $N$, but given the success of the $N \rightarrow \infty$ results, this seems unnecessary. The method has also been extended to include features found in more realistic models of spin ice; this includes further-neighbor exchange interactions~\cite{Conlon2010,Silverstein2014} and dipolar interactions~\cite{Sen2013}.

\subsection{Large-\emph{N} method for spin-ice films}
\label{sec:methods:films}

The application of the large-$N$ method to spin ice models in film geometries introduces additional complications to the methodology. In particular, the breaking of the translational symmetry in the finite direction of the film forbids a uniform constraint field, $\lambdabulk$ (as is the case in bulk spin ice). However, one can still take advantage of the translational symmetry in the plane parallel to the surfaces,  defining a constraint field \emph{on each layer} $l$. Such layer-dependent constraint fields were used previously in Ref.~\cite{Dantchev2014} to study Casimir effects in films of ferromagnets.

Proceeding as in the previous section, and integrating out the spin variables $\vec{s}_i$, we obtain the large-$N$ partition function
\begin{equation}
\mathcal{Z}_N = \int \mathcal{D} \lambda
\exp\left\{
-\frac{N}{2} {\rm Tr}\left[- \lambda + \log\left(
\lambda + \beta J \mat{V}\right)
\right]
\right\},
\label{eq:H_films}
\end{equation}
where the matrix $\lambda$ now has layer-resolved elements, $\lambda_{ij} = \lambda_l \delta_{ij}$, and the layer index $l$ is an implicit function of the lattice site $i$. Using the translational symmetry in the plane, the saddle-point solution [Eq. (\ref{eq:avg_constraint})] can be defined on each layer,
\begin{equation}
\frac{1}{n_l}\sum_{i \in l} \la{\vec{s}_i \cdot \vec{s}_i}\ra = N, 
\label{eq:selfconsistency_films_realspace}
\end{equation} 
where $n_l$ is the number of spins on layer $l$. Spin-spin correlations in reciprocal space are obtained from Eq.~(\ref{eq:ss-correlation}) after performing a Fourier transform in the plane,
	\begin{align}	
		\vec{s}_\alpha(\vec{q}_{\perp}) &= \frac{1}{\sqrt{n_c}} \sum_{\vec{r}} e^{-i (\vec{r} + \vec{r}_\alpha) \cdot { \vec{q}_\perp }} \vec{s}_\alpha(\vec{r}),
	\end{align}
where $\vec{r}$ runs over the $n_c$ primitive unit cells of the film, $\vec{q}_\perp$ are in-plane wave vectors, and $\vec{r}_\alpha$ are basis vectors locating each sublattice $\alpha$ within the unit cell. Note that we identified the pyrochlore lattice sites as $i \equiv (\mathbf{r}, \alpha)$. One ultimately finds
\begin{align}
\la \cc{s^a_\alpha({\vec{q}_\perp})} s^b_\beta({\vec{q}_\perp})\ra = \delta_{ab}  M^{-1}_{\alpha\beta}(\vec{q}_\perp) ,
\label{eq:correlations_films_qspace}
\end{align}
where $ M(\vec{q}_\perp) \equiv \lambda + \beta J V(\vec{q}_\perp)$, and $\mat{V}(\vec{q}_\perp)$ is the Fourier transform of the direct-space interaction matrix $\mat{V}$.
The numerical values of the constraint fields $\lambda_l$
~\footnote{In principle, the $\lambda_{l}$ could also depend on the sublattice index $\alpha$. However, we find that for the cases of interest, the symmetries of the film enforce uniformity of the constraint fields within each layer (independent of $\alpha$)}  
are obtained by enforcing the saddle-point conditions, given in Eq.~(\ref{eq:selfconsistency_films_realspace}), which can be expressed as

\begin{align}
\sum_{\alpha \in l} \sum_{\vec{q}_\perp} M^{-1}_{\alpha\alpha}(\vec{q}_\perp) = n_l,
\label{eq:selfconsistency_films_qspace}
\end{align}
for each layer $l$. We note that the framework provided by Eqs. (\ref{eq:correlations_films_qspace}) and (\ref{eq:selfconsistency_films_qspace}) is completely general and does not suppose a particular choice of surface geometry, which appears in the definition of the unit cell and through the structure of the matrix $\mat{M}(\vec{q}_\perp)$.
We also note that this analysis would carry through for spin ice films that include further-neighbor or dipolar interactions~\cite{Jaubert2016}, in their paramagnetic phases. One simply needs to compute the Fourier transform $\mat{V}(\vec{q}_\perp)$ of the corresponding interaction matrix in the chosen film geometry (see Refs.~[\onlinecite{Conlon2010,Silverstein2014,Sen2013}] for details in the bulk case). The inclusion of orphan bonds, as in Eq. (\ref{eq:Hnn_Ising_orphan}), is also straightforward, with the corresponding interaction matrix $V(\vec{q})$ given in App.~\ref{app:films}.

\subsection{Monte Carlo simulations}
\label{sec:methods:monte-carlo}
To confirm that the large-$N$ method correctly captures the physical behavior of the Ising ($N=1$) films at low temperatures, as it does in the bulk case~\cite{Isakov2004}, we perform a classical Monte Carlo simulation of the model of Eq.~(\ref{eq:Hnn_Ising_orphan}) for the appropriate film geometries. To avoid issues with equilibriation, we use a non-local Monte Carlo update. Specifically, we adapt the cluster algorithm of Ref.~[\onlinecite{otsuka-2014-cluster}] to the film geometry. To implement the surfaces in the $[001]$ direction we consider a periodic system of cubic cells with dimensions $L_{\perp} \times L_{\perp} \times L$. This can be modified into the appropriate film geometry by cutting the bonds that pass through a plane with normal $\vhat{z}$ and changing the remaining two bonds to carry the orphan coupling, $\Jo$.  This modifies the weights used for the surface tetrahedra in the cluster algorithm of Ref.~[\onlinecite{otsuka-2014-cluster}], but otherwise leaves the algorithm unaffected for any choice of $\Jo/J$ (see App.~\ref{app:mc} for further details). This cluster algorithm is closely related to the standard loop or worm algorithm used in spin ice simulations~\cite{Melko2001,melko2004monte}, similar to the relationship between the Swensden-Wang~\cite{swensden-1987-nonuniversal} and Wolff~\cite{wolff-1989-collective} cluster algorithms used in unfrustrated Ising models. For our purposes, one advantage of this formulation is the availability of an improved estimator~\cite{otsuka-2014-cluster} for the spin-spin correlation functions that allows to access larger system sizes at lower computational cost. Typically, accurate spin-spin correlations can be obtained with samples generated using only $10^3$ steps of the cluster algorithm when employing this improved estimator. 

For the single-layer films ($L=1$), we considered systems up to $L_{\perp}=64$, while for the thicker films ($L=3$ and $L=5$) we considered sizes up to $L_{\perp}=32$. For a given system size $n = 16 L_{\perp}^2 L$, we expect finite size effects to become important when the monopole density $\sim e^{-2J/T}$ \cite{Castelnovo2011} drops below $\sim 1/n$. Below the crossover temperature ${T}^*/J\sim 1/\log{n}$ one expects the system to be confined to the ice manifold itself and recover the $T=0$ behavior. For the lattice sizes of interest, the simulations become finite-size limited for temperatures less than ${T}^*/J \sim 0.1-0.2$. When considering the direct-space spin-spin correlators, we used smaller sizes of $L_{\perp} = 16$, but with a larger number of samples, typically of order $10^5$, to ensure small statistical errors.

\section{[001] spin ice films}
\label{sec:results_001}
To begin our exploration of spin ice films, we consider what is perhaps the simplest geometry: films with surfaces perpendicular to the $[001]$ crystallographic direction. We start with equivalent orphan and bulk bonds, $\Jo=J$, and then consider the more general and richer case, $\Jo\ne J$. In both cases, we apply the large-$N$ method described in Sec.~\ref{sec:methods:films}, and confirm its results via Monte Carlo simulations as described in Sec.~\ref{sec:methods:monte-carlo}.

\subsection{Equivalent orphan and bulk bonds}
	
\begin{figure*}[htp]
  \centering
  \includegraphics[width=0.95\textwidth]{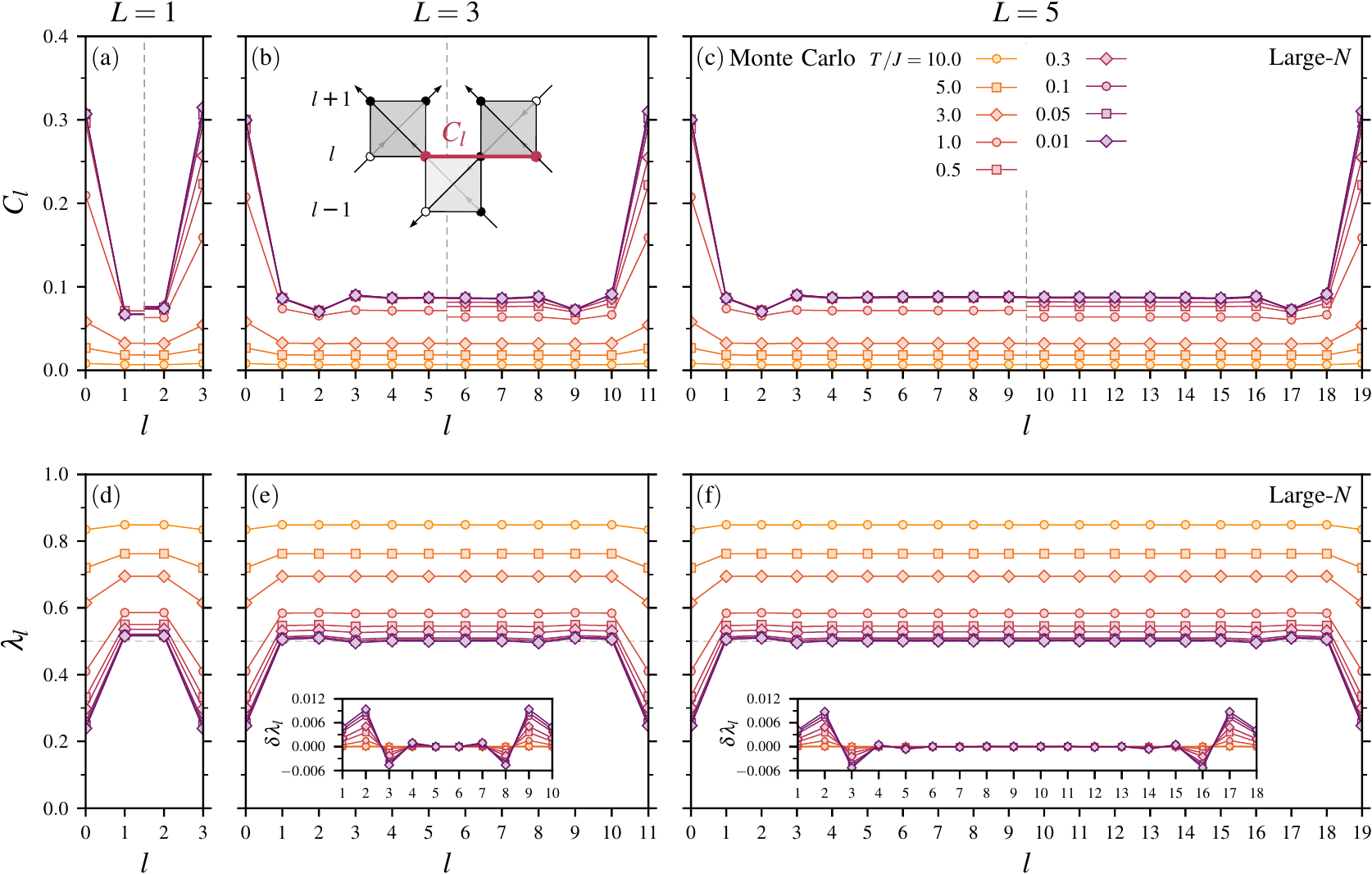}
	\caption{Layer-resolved real space correlations $C_l$ (a-c) [see the inset of (b) for a definition of $C_l$] and constraint fields $\lambda_l$ (d-f)
	of thin films of (a,d) 4 layers ($L=1$), (b,e) 12 layers ($L=3$) and (c,f) 20 layers ($L=5$), as a function of temperature $T/J$ and layer index $l$ (see Fig.~\ref{fig:stacking}).  For the real space correlations the Monte Carlo result (left) is shown for a system size of $L_{\perp}=16$, while the large-$N$ results (right) are effectively in the thermodynamic limit ($L_{\perp}=\infty$). In panels (d-f), the dashed lines show the value $\lambda_0 = 1/2$ expected for bulk spin ice at $T/J=0$, and the insets show the deviation from the central layers, $\delta \lambda_l \equiv \lambda_l - \lambda_{2L}$.
	\label{fig:thick_lambda}}
\end{figure*}

We first consider films with the orphan bond coupling equal to that of the bulk ($\Jo = J$). We start with the thinnest films ($L=1$), where we expect the most pronounced effects when compared to the bulk case. To proceed, Eq.~(\ref{eq:selfconsistency_films_qspace}) must be solved numerically in order to obtain the set of constraint fields $\lambda_l$ as a function of $T/J$, which is needed to access all other observables. This is accomplished by applying the Newton-Raphson descent algorithm (see App.~\ref{app:numerics}). Although there are four spin layers, the symmetry of the slab leads to only two distinct constraint fields, as shown in Fig. \subref{fig:thick_lambda}{(d)}. As $T/J \rightarrow 0$, the constraint fields for the middle layers approach the expected value for bulk spin ice, $\lambdabulk = 1/2$~\cite{Isakov2004}, whereas the constraint fields for the surface layers converge to a significantly lower value. As a result, the in-plane spin-spin correlations acquire a layer-resolved character, with stronger correlations at the free surfaces than in the middle of the slab. As an example of this behavior, we show in Fig. \subref{fig:thick_lambda}{(a)} the direct space correlation function between second nearest neighbors on a given layer $l$, which we denote as $C_l$ [see inset of Fig. \subref{fig:thick_lambda}{(b)}]. We compute the same correlation function in the Monte Carlo simulations [see Fig. \subref{fig:thick_lambda}{(a)}] and find reasonable agreement.

To explore the fate of the key signature of the Coulomb phase, the presence of pinch points~\cite{Henley2005,Henley2010,Castelnovo_AnnRevCMP}, we consider the spin-spin correlation function, $S({\vec {q}})$, of the films  in reciprocal space
\begin{equation}
S({\vec{q}}) = \frac{1}{n N} \sum_{ij} \la \vec{s}_i \cdot \vec{s}_j \ra e^{i \vec{q} \cdot (\vec{r}_i - \vec{r}_j )},
\end{equation} 
where $\vec{q} \equiv (\vec{q}_\perp, q_z) = 2\pi (h\vhat{x}+k\vhat{y}+l\vhat{z})$ is a three-dimensional wave vector, expressed using Miller indices $[hkl]$. In terms of the spin-spin correlation matrix $\mat{M}(\vec{q}_\perp)$, we obtain
\begin{align}
S({\vec{q}}) &= \frac{1}{8L} \sum_{\alpha\beta} e^{ i q_z ( \vec{r}_\alpha - \vec{r}_\beta) \cdot \vec{\hat{z}} }  
M_{\alpha \beta}^{-1}({\vec{q}_\perp}) .
\label{eq:SF_largeN}
\end{align}

\begin{figure*}
\centering
	\includegraphics[width=0.95\textwidth]{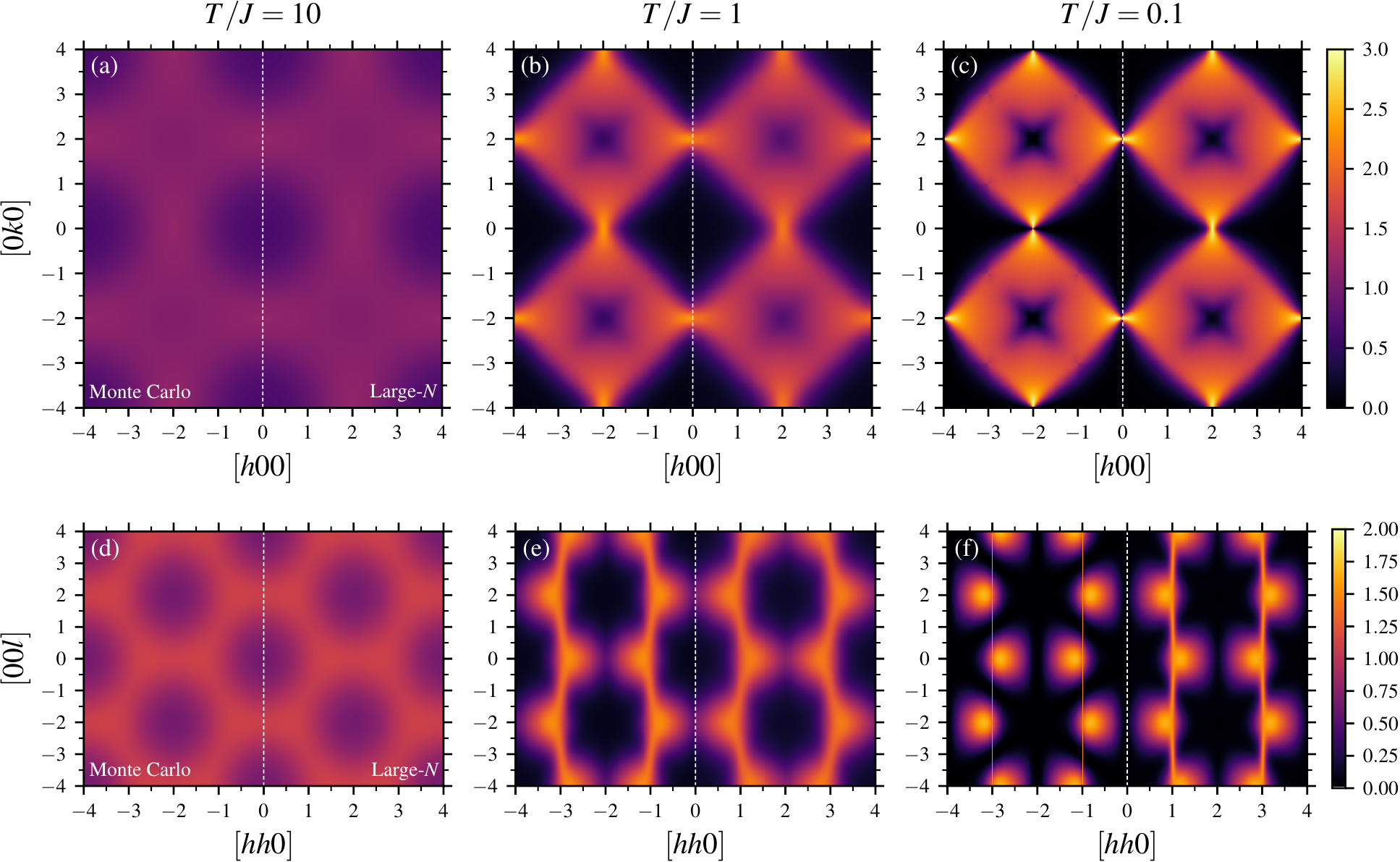}
	\caption{Spin-spin correlation functions, $S(\vec{q})$, [from large-$N$ (right) and Monte Carlo (left)] of spin ice films of thickness $L=1$, for temperatures (a,d) $T/J = 10$, (b,e) $T/J = 1$ and (c,f) $T/J = 0.1$ in the (a-c) $[hk0]$ and (d-f) $[hhl]$ planes. Monte Carlo simulations were performed with $L_{\perp}=64$. For the temperatures considered, the large-$N$ results are (effectively) in the thermodynamic limit ($L_\perp \rightarrow \infty$).
	\label{fig:L1_SF} }
	\vspace{0.5cm}
	\centering
	\includegraphics[width=0.95\textwidth]{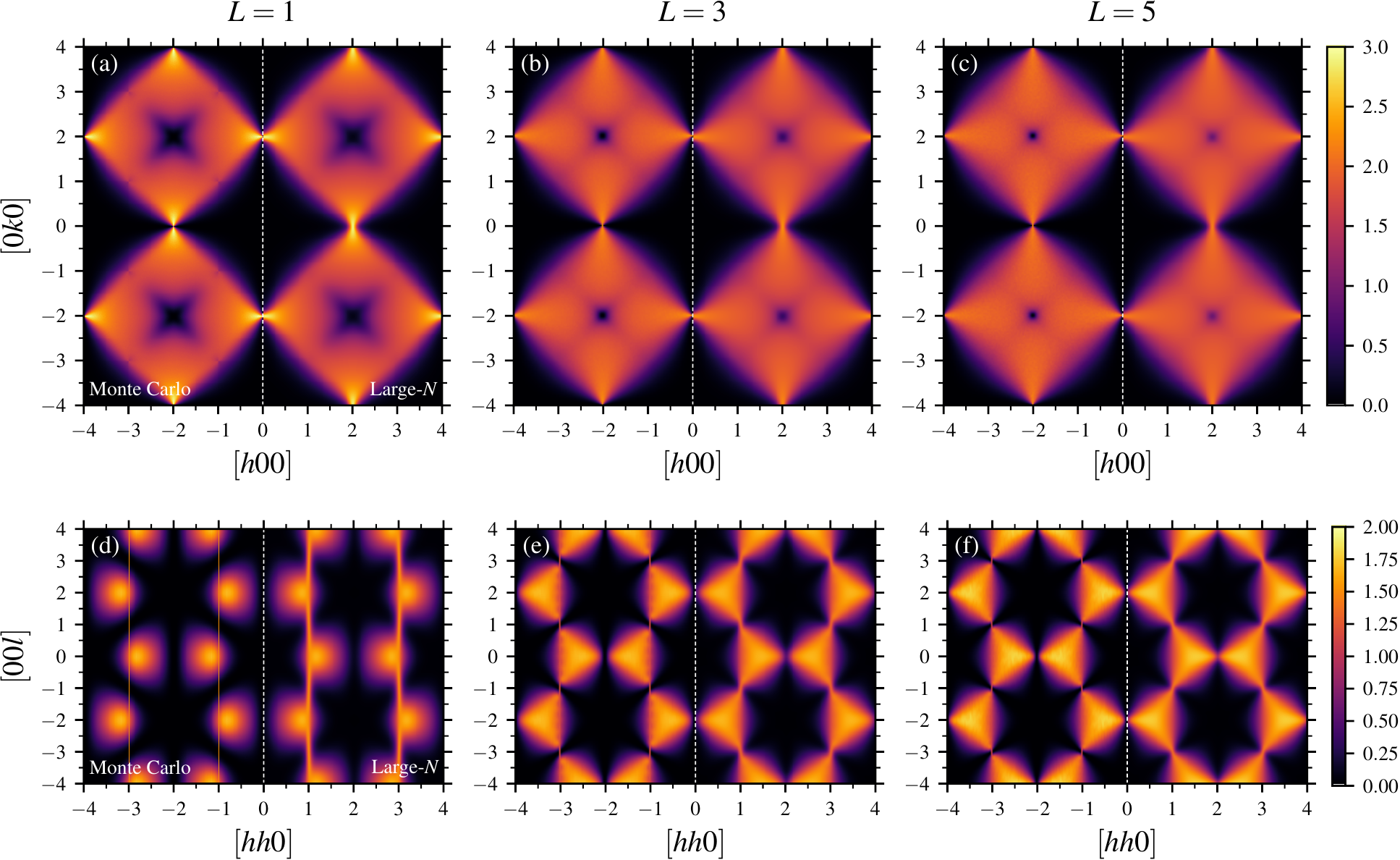}
	\caption{
	Spin-spin correlation functions, $S(\vec{q})$,  [from large-$N$ (right) and Monte Carlo (left)] of spin ice films of thickness (a,d) $L=1$, (b,e) $L = 3$ and (c,f) $L = 5$, for a temperature of $T/J = 0.1$ in the (a-c) $[hk0]$ and (d-f) $[hhl]$ planes. Monte Carlo simulations were performed with $L_{\perp}=64$ (for $L=1$) and $L_{\perp}=32$ (for $L=3,5$). For $T/J = 0.1$ the large-$N$ results are (effectively) in the thermodynamic limit ($L_\perp \rightarrow \infty$).
\label{fig:thick_SF}}
\end{figure*}

In Fig. \ref{fig:L1_SF}, we show $S(\vec{q})$ for various temperatures $T/J$ in two high-symmetry planes: $[hk0]$ and $[hhl]$. Strikingly, the characteristic pinch-points remain \emph{intact} in the $[hk0]$ scattering plane, parallel to the surfaces. However, as expected due to the finite extent in the $\vhat{z}$ direction, they are washed out in scattering planes with a non-zero $q_z$ component, normal to the film. We also observe ``scattering rods" in the $q_z$ direction, and ``secondary" pinch-points near $[110]$ and equivalent wave-vectors. 

All these features are reproduced in Monte Carlo simulations for the same geometry. The only substantive difference between the Monte Carlo and large-$N$ results lies in the temperature dependence of the build up of spin-ice correlations, as found in the bulk case~\cite{Isakov2004}. For the large-$N$ case, due to the continuous nature of the spins, the correlation functions only approach the asymptotic $T=0$ result algebraically~\cite{Henley2005}, while for discrete Ising spins this occurs exponentially \footnote{For the bulk case, there is an (ad-hoc) procedure to cure this discrepancy. Specifically, as discussed in Ref.~[\onlinecite{Sen2013}], one can simply replace the temperature dependence of the stiffness ($\lambda$) from large-$N$ by the appropriate exponential form. This is not straightforward for films since the constraint fields, $\lambda_{l}$ now have an explicit layer dependence. While one could define a inhomogeneous stiffness for each tetrahedron, it is ambiguous how to do this for each layer (since there are two layers per tetrahedron). }. This can be seen explicitly in the pinch points; as $\sim \sqrt{T/J} \rightarrow 0$, their width decays as $\sim \sqrt{T/J}$ for the large-$N$ case, as compared to $\sim e^{-J/T}$ in the Monte Carlo simulations. This is most apparent in the $[hk0]$ plane, shown in Fig.~\subref{fig:L1_SF}{(c)} for $T/J=0.1$. In the Monte Carlo data, the pinch-point width is limited by the lateral system size $L_\perp$, while in the large-$N$ results there is an appreciable width. This difference in sharpness is also apparent in the width of the scattering rods in the $[hhl]$ plane [see Fig.~\subref{fig:L1_SF}{(f)}]. 

We now examine how the properties uncovered above change as the thickness is increased towards the bulk limit. Surprisingly, the numerical solution for the constraint fields $\lambda_l$ shows \emph{oscillations} as a function of depth in the sample, as is illustrated in Fig.~\subref{fig:thick_lambda}{(e,f)} for films with $L=3$ and $L=5$. These oscillations have \emph{increasing} amplitude with decreasing temperature and a characteristic damping length scale which appears \emph{independent} of thickness, indicative of a surface effect. They are also seen in the layer-resolved direct-space correlations $C_l$ and are well reproduced in the Monte Carlo simulations, as shown in Fig. \subref{fig:thick_lambda}{(b,c)}, keeping in mind the difference in temperature dependence expected between the large-$N$ theory and the Ising model. We have verified that a large-$N$ treatment of thin films of a pyrochlore Ising \emph{ferromagnet} [i.e. Eq.~(\ref{eq:Hnn_Ising}) with $J \rightarrow -J$] with the same geometry does \emph{not} show oscillations in the constraint fields $\lambda_l$, but rather a monotonic behavior from the surface to the middle of the sample, as long as the system remains in the high-temperature paramagnetic phase. These results thus suggest that the oscillations are directly related to the geometrical frustration.

We also compute the spin-spin correlation functions $S(\vec{q})$ [given by Eq.~(\ref{eq:SF_largeN})] at $T/J = 0.1$, deep in the Coulomb phase, for thicknesses of $L=1,3$ and $5$ (see Fig.~\ref{fig:thick_SF}). In the $[hk0]$ plane, pinch-points are always present, with an increased contrast for thinner films. In the $[hhl]$ plane, the washed-out pinch-points are progressively restored with increasing thickness, as the system crosses over from a two-dimensional to three-dimensional Coulomb phase. The restoration of the ``three-dimensional" pinch-points in the $[hhl]$ plane is set by the thickness of the film $L$ which cuts off the Coulomb correlations in the $\vec{\hat{z}}$ direction. Roughly, we expect quasi-two-dimensional behaviour when the correlation length, $\xi \sim e^{J/T}$, becomes of the order of the film thickness, $L$, giving a crossover temperature $T \sim J/\log{L}$.  As with the $L=1$ films and bulk results, the Monte Carlo and large-$N$ results agree well, aside from the aforementioned temperature dependence (algebraic compared to exponential) of the build up of Coulomb correlations.

\subsection{Inequivalent orphan and bulk bonds}

\begin{figure*}[htp]
	\includegraphics[width=0.95\textwidth]{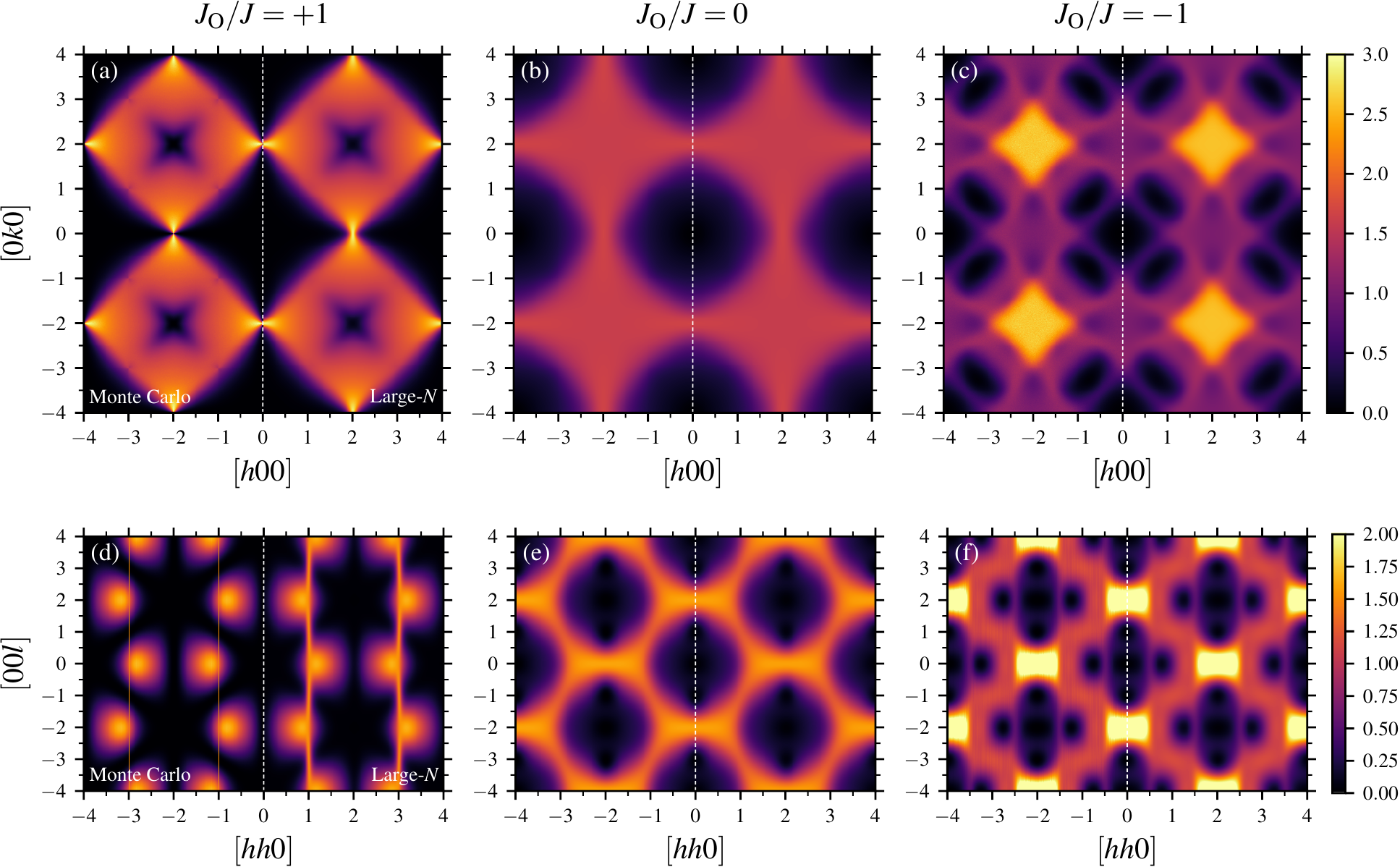}
	\caption{Spin-spin correlation functions, $S(\vec{q})$, [from large-$N$ (right) and Monte Carlo (left)] of spin ice films of thickness $L=1$ and temperature $T/J = 0.1$ in the (a-c) $[hk0]$ and (d-f) $[hhl]$ planes.  Orphan bond values of (a,d) $\Jo/J = 1$, (b,e) $\Jo/J = 0$ and (c,f) $\Jo/J = -1$ are shown. Monte Carlo simulations were performed with $L_{\perp}=64$ for all $\Jo/J$. For $T/J=0.1$ the large-$N$ results are (effectively) in the thermodynamic limit ($L_\perp \rightarrow \infty$).
	\label{fig:orphan_SF}}
\end{figure*}

\begin{figure*}
  \centering
  	\includegraphics[width=0.95\textwidth]{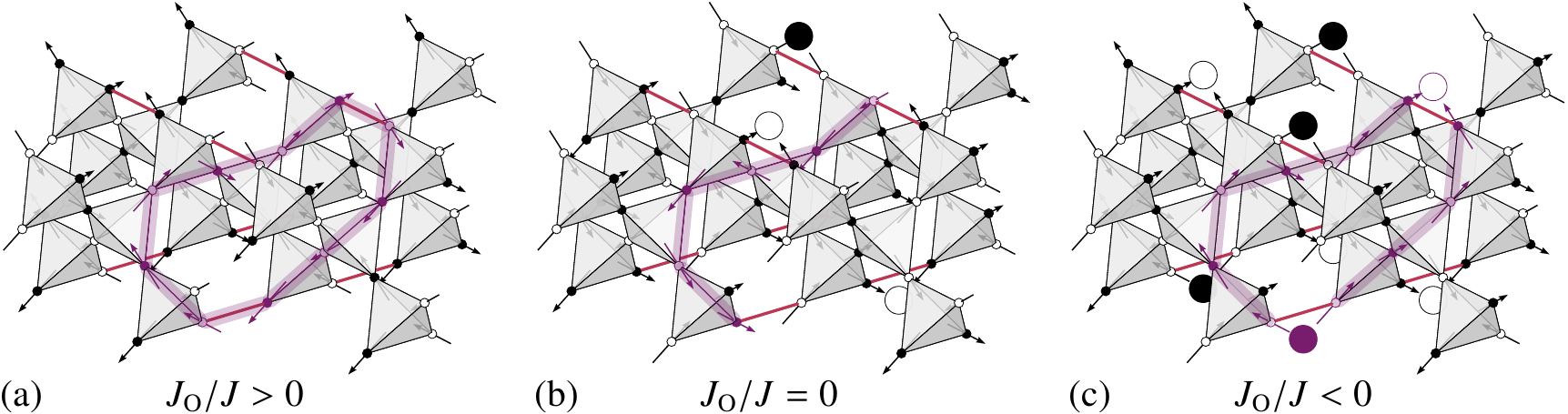}
    \caption{Representative ground states of $L=1$ $[001]$ films with $\Jo/J = +1, 0, -1$. (a) When $\Jo/J > 0$, the orphan-bond spins are anti-aligned, so that the flux lines run parallel to the surfaces. (b) When $\Jo = 0$, no specific configuration of the orphan bond spins is preferred, so that half of the orphan bonds (on average) host surface charges. (c) When $\Jo / J < 0$, the orphan-bond spins are aligned, leading to surface charges at the endpoints of flux lines. In cases (b,c), these fluctuating charges destroy the Coulomb phase and its associated pinch points. For each case we highlight a group of spins that can be flipped at zero energy cost.
    }
	\label{fig:surfaces_orphan}
\end{figure*}

We now move to the more general case and consider the influence of differing orphan and bulk bonds ($\Jo \neq J$), as defined in Eq. (\ref{eq:Hnn_Ising_orphan}), on the physics of \ac{NN} spin ice films. At low temperature ($T \ll |\Jo|, J$), we find that the system remains paramagnetic with the physics depending \emph{only} on the sign of $\Jo$ (not on its magnitude). This is expected because the ``bulk" ice rules are always compatible with minimizing the energy of the orphan bonds. We show the spin-spin correlations function, $S(\vec{q})$, for three representative values $\Jo/J = +1, 0,-1$ in Fig. \ref{fig:orphan_SF}. The case $\Jo/J = 1$ corresponds to the results of the previous section, with sharp pinch points characteristic of a two-dimensional Coulomb phase. However, these pinch points completely  disappear for $\Jo/J = 0$ and $\Jo/J = -1$, revealing only broad features in $S(\vec{q})$ in the low-temperature regime.

The preservation or destruction of the Coulomb phase (depending on the sign of $\Jo/J$) can be understood in terms of a simple picture of the ground state manifold for arbitrary film thicknesses. Note that for $\Jo/J > 0$, the flux lines of the field $\vec{B}$ (see Sec. \ref{sec:model_A}) run along the surface and are then redirected back into the bulk of the film [see \subref{fig:surfaces_orphan}{(a)}]. This choice of orphan bond coupling is thus compatible with the (bulk) constraint $\vec{\nabla} \cdot \vec{B} = 0$ and the Coulomb phase, now two-dimensional, is preserved. However, as discussed in Ref.~[\onlinecite{Jaubert2016}], when $\Jo / J < 0$, the orphan bonds host pairs of aligned spins which correspond to surface magnetic charges. These surface charges serve as endpoints to the ``flux-lines" of the effective magnetic field $\vec{B}$ \footnote{In the dumbbell picture, where each spin is represented by a pair of (effective) magnetic charges $\pm q$, these surface monopoles carry charge $Q = \pm 2q$ ~\cite{Jaubert2016}, as they are the endpoints of \emph{two} flux lines (corresponding to the two aligned orphan bond spins).} [see Fig. \subref{fig:surfaces_orphan}{(c)}]. Just as a finite density of thermally populated monopoles endows the pinch points in bulk spin ice with a finite width \cite{Henley2010,Castelnovo2011}, the finite density of these fluctuating surface charges broaden the pinch points in spin ice films. However, unlike the bulk case, these charges are present \emph{even at} $T=0$ and thus destroy the Coulomb phase. We note that the destruction of the Coulomb phase is ultimately averted in Ref.~\cite{Jaubert2016} by the ordering of the surface charges due to the long-range dipolar interactions. This static ordering inhibits thermal fluctuations of the surface charges and restores the two-dimensional Coulomb phase below the ordering temperature -- a behavior reminiscent of \emph{magnetic fragmentation}~\cite{Brooks-Bartlett2014}. In other words, the state we find for $\Jo/J < 0$ can be viewed as the ``parent" state out of which the ordering of Ref.~[\onlinecite{Jaubert2016}] arises.  

Finally, we consider the special case $\Jo/J = 0$. Since the orphan bonds provide no energetic constraint, their spin configuration is essentially random.  This leads to both the presence and absence of surface charges, as illustrated in Fig.~\subref{fig:surfaces_orphan}{(b)}. The finite density of these surface charges, while not maximal (as for $\Jo/J<0$), is sufficient to destroy the Coulomb phase. This results in the absence of pinch points in $S(\vec{q})$, as observed in Fig.~\subref{fig:orphan_SF}{(b)}.

We have verified that when $L$ is increased, the spin-spin correlation functions $S(\vec{q})$ approach the bulk result for any $\Jo/J$ (not shown). In particular, the gradual recovery of the pinch-points indicates that algebraic correlations are present up to a length scale set by the film thickness, $L$, even in the presence of surface charges. As mentioned above, this is analogous to the case of thermally activated charged defects (monopoles) in bulk spin ice which cut off the algebraic correlations at a length scale set by their average separation \cite{Castelnovo2011}. We have also investigated the depth dependence of the constraint fields $\lambda_l$ and the layer-resolved correlators $C_l$; we find that these oscillate as a function of layer index $l$ (for $L > 1$) for all values of $\Jo/J$ considered (not shown).

\section{Classical topological order in [001] films}
\label{sec:topological}

The previous section has exposed, via large-$N$ and Monte Carlo results, how boundary conditions affect the Coulomb phase present in the parent bulk system. In this section, we relate these results to the topological order that characterizes different classical spin liquids~\cite{Jaubert2013,Rehn2017}. In particular, we argue that the low-temperature state of films with $\Jo/J < 0$, while not a Coulomb phase, is nonetheless a non-trivial collective paramagnet -- a classical $Z_2$ spin liquid~\cite{Rehn2017}.

First recall that in bulk spin ice, the spins can be mapped to dimers on the dual diamond lattice~\cite{Hermele2003}. Specifically, we identify $\sigma = +1$  with the presence of a dimer on the corresponding bond of the diamond lattice (similarly, $\sigma = -1$ is identified with the absence of such a dimer). The ice rules then correspond to the requirement that two dimers touch at each diamond lattice site. To move within the ice manifold, one uses ``loop moves" that reverse all the spins on an alternating spin loop~\cite{Melko2001, melko2004monte}. In the dimer picture, this corresponds to swapping the occupied and unoccupied bonds on this loop. Any ``short" loop (not spanning the system) thus preserves three \emph{winding numbers}, corresponding to the number of dimers crossing three planes oriented in the $\vhat{x}$, $\vhat{y}$ or $\vhat{z}$ directions. These winding numbers are \emph{topological invariants} that characterize the classical $U(1)$ spin liquid (Coulomb phase), and can only be changed by ``large" loops that wind around the periodic directions of the system~\footnote{Equivalently, these winding numbers can be directly related to the three components of the total magnetization, as discussed in Ref.~[\onlinecite{Jaubert2013}]}.

How does this physics change for films? First, since our system is two-dimensional, we can define at most \emph{two} distinct winding numbers. Second, in addition to the usual ``bulk'' loop moves, we can also construct zero-cost moves that involve the surface spins. When $\Jo/J > 0$, the constraint of anti-aligned orphan bond spins only allows the construction of loops that run \textit{along} the orphan bonds, as shown in Fig.~\subref{fig:surfaces_orphan}{(a)}. The two in-plane winding numbers thus remain topological invariants and we have a two-dimensional Coulomb phase, a classical $U(1)$ spin liquid.

The $\Jo/J < 0$ case is more interesting. Here, to construct a zero-cost move, we must consider \emph{open} strings of alternating spins that end in \emph{pairs} on the orphan bonds, since preserving the surface constraint requires flipping \emph{both} orphan bond spins. One can view such a pair of strings as a usual loop, where the alternation pattern is reversed when an orphan bond is encountered [see Fig.~\subref{fig:surfaces_orphan}{(c)}]. Therefore, contributions from the two strings \emph{add up}, and these moves can change the winding numbers only by \emph{even} amounts, leading to the destruction of the aforementioned classical $U(1)$ topological order. However, not all is lost; one can still define two $Z_2$ topological invariants~\cite{Rehn2017}, corresponding to the two ``winding parities", that can only be changed by moves that wrap around the system. This leads us to identify the paramagnetic phase found for $\Jo/J<0$ films as a classical $Z_2$ spin liquid, consistent with the absence of pinch-points exposed in Sec.~\ref{sec:results_001}~\footnote{The criterion proposed in Ref.~[\onlinecite{Rehn2017}] for identifying a classical $Z_2$ spin liquid in the large-$N$ theory also holds here. For $[001]$ films the spectrum of $V(\vec{q})$ has a set of zero-energy flat bands. For $\Jo/J>0$ there is no gap between these low-lying bands and the higher bands; for $\Jo/J<0$ there is such a gap. Furthermore, the emergence of this classical $Z_2$ spin liquid is related to structure of the dual lattice~\cite{Rehn2017}; when $\Jo/J<0$ the orphan bond spins are aligned at low temperature, rendering the dual lattice non-bipartite upon identification of the two orphan bond sites}. A direct consequence of such (classical) $Z_2$ topological order is the predicted presence of fractionalized magnetic moments bound to vacancies in the film \cite{Rehn2017}. 

Finally, we turn to the $\Jo/J = 0$ case. At this point the system is neither a $Z_2$ nor a $U(1)$ spin liquid. We can see this noting that strings of alternating spins terminating on the orphan bonds can be flipped at zero energy cost [see Fig.~\subref{fig:surfaces_orphan}{(b)}]. Flipping these strings can change the winding numbers arbitrarily, even for short strings. We thus conclude that the $\Jo/J=0$ case does not support topological sectors. Given that this case sits at the critical point between the $U(1)$ and $Z_2$ spin liquids, its properties have more general implications for the regime where $|\Jo| \ll T \ll J$. In this limit, the orphan bonds are effectively at high temperature, so the system will behave more like the $\Jo/J=0$ point, rather than the classical $U(1)$ or $Z_2$ spin liquids (see the phase diagram in Fig. \ref{fig:phase-diagram}). 

\section{Magnetically charged surfaces in [110] and [111] films}
\label{sec:non-neutral}

\begin{figure}
  \centering
  \includegraphics[width=0.626\columnwidth]{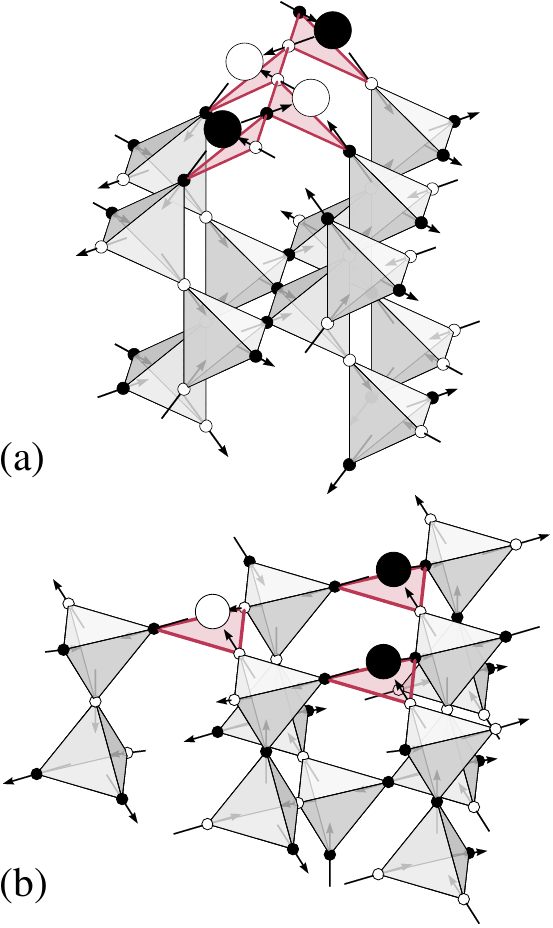}
    \caption{Representative ground states of films cleaved along the (a) $[110]$ or (b) $[111]$ directions. In both cases, the cleaving shown cuts one spin per surface tetrahedron, leaving orphan \emph{triangles} at the surface. Flux lines are therefore \emph{required} to have endpoints on each of these orphan triangles, leading to surface charges \emph{irrespective} of the sign of $\Jo/J$. }
	\label{fig:surfaces}
\end{figure}

In the previous section, we showed how specific boundary conditions (the orphan bond exchange $\Jo$) have, through the formation of fluctuating surface charges, dramatic effects on the properties of the film. With an understanding of the ``simple" case of $[001]$ films, we now proceed to briefly discuss more complicated geometries, specifically films with surfaces perpendicular to the $[110]$ and $[111]$ directions. These geometries are obtained by cutting one (or three) spins per surface tetrahedron, as shown in Fig. \ref{fig:surfaces}. The resulting slab is comprised of alternating kagome and triangular layers for $[111]$ films, but has a somewhat more complicated geometry for $[110]$ films.

These two geometries differ \emph{drastically} from the $[001]$ films which, having two spins remaining per surface tetrahedron, can still (in principle) respect the divergence-free condition $\vec{\nabla} \cdot \vec{B} = 0$ defining the Coulomb phase, by having one spin pointing in and one pointing out. Whether this is energetically favorable depends on the value of $\Jo/J$, as explained in Sec. \ref{sec:results_001}. This is, however, \emph{impossible} for $[110]$ or $[111]$ spin ice films, where the surface tetrahedra have either one or three spins remaining -- in other words, the boundary conditions imposed on the $\vec{B}$ field at the surfaces are fundamentally incompatible with the divergence-free condition of the bulk. As a result, the surface tetrahedra must host a charge of the $\vec{B}$ field, \emph{irrespective} of the value of $\Jo/J$, as illustrated in Fig. \ref{fig:surfaces}. In $[110]$ films, these effective magnetic charges sit on parallel ``zig-zag" chains running on the surfaces [see Fig.~\subref{fig:surfaces}{(a)}], whereas in $[111]$ films, they live on a triangular lattice [see Fig.~\subref{fig:surfaces}{(b)}]. As discussed in Sec.~\ref{sec:results_001}, when these charges can fluctuate, they serve as free, zero-cost end-points for the (effective) magnetic flux lines. The presence of such zero-cost end points will generically destroy the two-dimensional Coulomb correlations. We thus expect any two-dimensional pinch-points to be destroyed for $[111]$ or $[110]$ films. 

However, we do \emph{not} expect a classical $Z_2$ spin liquid in these geometries. For $\Jo/J > 0$, the surface monopoles are at the end of \emph{only one} flux line. Thus, there is no constraint on changes of the winding numbers incurred by local updates. In contrast, for $\Jo/J < 0$ the surface triangles consist of \emph{three} aligned spins, and thus three strings must terminate at each triangle. Borrowing the arguments of Sec.~\ref{sec:topological}, this would imply the presence of a classical $Z_3$ spin liquid~\cite{Motrunich2003,Rehn2016} as the winding numbers can only change in multiples of three. Whether the physics discussed above extends to zero temperature, or is preempted by some ordering phenomena requires detailed numerical study, which we leave to future work.

While this picture of surface charges applies to the Ising case ($N=1$), the $N \geq 3$ cases should be qualitatively different. Consider as an example a $[111]$ film terminating on a kagome layer; while three Ising spins cannot add to zero on orphan triangles, thus requiring surface charges, three $N=3$ (or higher) vectors \emph{can} sum to zero. This implies that such surface charge defects (i.e. a non-zero sum of spins) are not required for the $N=3$ (or higher) case. We thus expect that for $[110]$ and $[111]$ films with orphan triangles, the large-$N$ result will be qualitatively different than the Ising case, with the two-dimensional Coulomb phase not (necessarily) destroyed~\footnote{There is another possibility for $[110]$ and $[111]$ films: surface terminations where only one spin per surface tetrahedron remains. In that case, the appearance of surface charges is expected for any number $N$ of spin components.}.

\section{Discussion}
\label{sec:discussion}

We now discuss some potential extensions and implications of our work. In particular, we outline applications to continuous spin systems, possible extensions to dipolar spin ice films and the surfaces of bulk single crystal dipolar spin ices. In addition, we speculate on more theoretical aspects, such as the effects of a magnetic field and the physics of quantum spin ice films~\cite{Gingras2014}.

In view of making concrete contact with experimental realizations, we discuss some aspects of films of spin ice materials. There are several complications in connecting the ideas discussed in this work to the physics of these compounds. First, crystal symmetry breaking at the interface between spin ice and vacuum could weaken or even destroy the Ising nature of the spins and their interactions~\cite{RauPRB2015} near the surfaces. For canonical spin ices such as \abo{Dy}{Ti} and \abo{Ho}{Ti}, this may not be a serious concern. Since the crystal field ground doublets of Dy- and Ho-based pyrochlores are predominantly maximal-rank (mostly $J_z = \pm 15/2$ and $J_z = \pm 8$ respectively), strong perturbations to the crystal field would be necessary to generate significant effects on the single-ion or two-ion properties~\cite{RauPRB2015}. One might then expect that the induced transverse (quantum) exchange at the surfaces would be small for both compounds and that the splitting of the non-Kramers doublet expected in \abo{Ho}{Ti} might be negligible. For the case of the \abo{Pr}{M} family mentioned in Sec.~\ref{sec:model}, these effects are likely to be more drastic. Indeed, the presence of large random transverse fields~\cite{Wen2017} due to weak structural disorder appears to be a feature even in bulk samples. Given that the crystal field doublet in these non-Kramers compounds lacks the ``axial protection" present in \abo{Ho}{Ti}~\cite{RauPRB2015}, we expect these transverse fields to be further enhanced at any surfaces. These complications could be minimized through more clever engineering of the films. For example, one might consider heterostructures composed of a thin layer of a spin ice material sandwiched between layers of non-magnetic pyrochlore materials having the same crystal structure and similar lattice constant, such as \abo{La}{M}, \abo{Lu}{M} and \abo{Y}{Ti}.

However, one serious difficulty with all of the proposals for spin ice thin films and heterostructures is the effect of substrate-induced strain. This could be due to a lattice constant mismatch, or simply to slight chemical bonding differences at the interface. The presence of such strain will generically strengthen some of the bonds and remove the ground state degeneracy and residual entropy~\cite{Jaubert2010,Jaubert2016}. In non-Kramers compounds, this could also induce a transverse field at each site (depending on the film geometry and strain direction); as in the case of surfaces, this could be negligible for \abo{Ho}{Ti}, but significant in the \abo{Pr}{M} family. How this strain can be minimized, so that the intrinsic physics of spin ice films can be exposed, is an important but exciting challenge in the fabrication of these systems. 

In dipolar spin ice such as \abo{Dy}{Ti} or \abo{Ho}{Ti}, the long-range tail of dipolar interactions should have significant effects on the physics discussed here. For bulk spin ice, these differences are suppressed due to the structure of the spin-ice manifold -- the dipolar interactions are effectively short-range when acting on ice states~\cite{Gingras2001CJP,Isakov2005}; this is the so-called self-screening or projective equivalence. Consequently, the splitting of the ice manifold is small, and the associated ordering due to the dipolar interactions only occurs at low temperature compared to the bare scale of the dipolar interactions~\cite{Melko2001}. However, when monopoles or surface charges are present, the dipolar interaction is significant, promoting the entropic Coulomb interaction between the defects into an energetic one. Indeed, it was found in  Ref.~[\onlinecite{Jaubert2016}] that for $[001]$ surfaces with ferromagnetic orphan bonds, this attraction induces a phase transition to long-range order of the surface charges into a checkerboard pattern. For films with $[110]$ or $[111]$ surfaces, the analogous physics is likely to be even richer. For example, in the $[111]$ geometry, the charges live on a triangular lattice, with either a single monopole or anti-monopole at each site. While each surface is not required to be neutral (due the presence of the other surface), this will be favored energetically. One thus expects a one-component Coulomb gas at half-filling (the other component being treated as background) on a triangular lattice. At the nearest neighbor level, this is equivalent to an anti-ferromagnetic triangular lattice Ising model and is highly frustrated, with a macroscopically degenerate set of ground states~\cite{Wannier1950}. The effective long-range Coulomb interactions will presumably lift this degeneracy but, as in dipolar spin ice, only weakly, due to the approximate local charge neutrality. 

Some of the physics discussed here is expected to carry over from the thin film context to that of exposed surfaces of bulk crystals of spin ice materials. For example, the presence of fluctuating surface charges in certain geometries may have screening effects on the fields from monopoles in the bulk. 
 
As discussed in Sec.~\ref{sec:non-neutral}, the physics of different surface terminations depends strongly on the nature of the spins in question. While we have mainly discussed Ising spins here, it is known that the large-$N$ method also works well for $O(3)$ spins~\cite{Isakov2004,Conlon2010}. Examples of pyrochlore anti-ferromagnets with such continuous, classical spins include certain spinels~\cite{Lee2010}, as well as the recently discovered chemically disordered fluorine pyrochlores~\cite{krizan1,krizan2}. Thin films of such compounds may be an interesting playground to explore analogues of the physics discussed in this work. Other interesting avenues in this direction include the case of $O(2)$ spins, where the large-$N$ method is known to be unreliable due to the appearance of order-by-disorder~\cite{bramwell1994order,Moessner1998a,Moessner1998}. Experimentally, there are several promising candidates; for example the bulk XY pyrochlores are known to exhibit rather exotic behaviors, from the order-by-disorder physics of \abo{Er}{Ti}~\cite{Champion,Zhito,Savary} to the unusual physics of the \abo{Yb}{M} family, \abo{Yb}{Ge} in particular~\cite{Dun2014,Dun2015,Hallas} (see Ref.~[\onlinecite{hallas2017review}] for a review). Further, the limit of an anti-ferromagnetic XY model has been found to harbor several exotic phases, including a spin liquid at intermediate temperature and a ``hidden" quadrupolar order at low temperature~\cite{taillefumier2017frustrating}; the effect of a film geometry would likely lead to rich physics. 

On the more theoretical front, there are many fundamental open questions about spin ice films. For example, bulk spin ice shows a complex phase diagram in an external magnetic field, with the physics strongly dependent on the field direction~\cite{moessner2003,Jaubert2008,Ruff2005}; the effect of magnetic fields on films is likely to be similarly rich. The effects of transverse exchange also raises a host of interesting questions: in bulk spin ice this induces tunnelling between the ice states and stabilizes a $U(1)$ quantum spin liquid, \emph{quantum spin ice}~\cite{Gingras2014}. However, the quantum analogue of the two dimensional Coulomb phase is fundamentally unstable~\cite{polyakov1977quark}, meaning that a direct two-dimensional analogue of quantum spin ice does not exist~\cite{Henry2014}. The fate of quantum spin ice films is thus unclear; possibilities include magnetically ordered states, valence bond solids~\cite{Henry2014} or potentially a quantum $Z_2$ spin liquid, as found in related models on the kagome lattice~\cite{carrasquilla2015two}. How the resulting state in this quantum case depends on the film geometry and the choice of orphan bond exchange presents many directions to pursue in future studies. Given the ability to readily grow high quality rare-earth pyrochlore oxide films~\cite{Leusink2014,Bovo2014}, one might expect such theoretical investigations to motivate a range of experimental studies which will, in return, undoubtedly fuel new sets of theoretical questions.

The study of films of pyrochlore magnets is a nascent field with many open questions, and rapid development could be expected in the near future. We our hope that this work will help shed light onto these systems and provide guidance and motivation for upcoming experimental and theoretical work.

\begin{acknowledgments}
We thank Kristian Tyn-Kay Chung, Felix Flicker, Ludovic Jaubert and Peter Holdsworth for useful discussions. \'{E}. L-.H. acknowledges financial support by the FRQNT, the NSERC of Canada and the Stewart Blussom Quantum Matter Institute at the University of British Columbia. Research at the Perimeter Institute is supported by the Government of Canada through Innovation, Science and Economic Development Canada and by the Province of Ontario through the Ministry of Research, Innovation and Science.
The work at the University of Waterloo was supported by the NSERC of Canada and the Canada Research Chair program (M.J.P.G., Tier 1).
\end{acknowledgments}

\appendix
	
\section{Bulk spin ice}
\label{app:bulk}
	
\subsection{The pyrochlore lattice}

The pyrochlore lattice is a face-centered cubic lattice decorated with a tetrahedron at each site. We take the conventional cubic unit cell of side length $a = 1$; in this convention the nearest neighbor distance is $r_{\rm nn} = \sqrt{2}/4$. The primitive unit cell is an upwards-tetrahedron, with four sublattice sites located at each vertex, with the following sublattice vectors:
\begin{align}
    \vec{r}_0 &= \vec{0}, & 
    \vec{r}_1 &= \frac{\vhat{x}+\vhat{y}}{4}, &
    \vec{r}_2 &= \frac{\vhat{x}+\vhat{z}}{4}, &
    \vec{r}_3 &= \frac{\vhat{y}+\vhat{z}}{4}. 
\end{align}
The local quantization axes for spin ice are defined with respect to the primitive unit cell,
\begin{align}
    \vhat{z}_0 &= \frac{\vhat{x}+\vhat{y}+\vhat{z}}{\sqrt{3}}, &
    \vhat{z}_1 &= \frac{\vhat{z}-\vhat{x}-\vhat{y}}{\sqrt{3}}, \nonumber \\
    \vhat{z}_2 &= \frac{\vhat{y}-\vhat{z}-\vhat{x}}{\sqrt{3}}, &
    \vhat{z}_3 &= \frac{\vhat{x}-\vhat{y}-\vhat{z}}{\sqrt{3}}, 
\end{align}	
where the indexing matches that of the sublattice vectors.

\subsection{Interaction matrix}

The explicit form of the interaction matrix $\mat{V}(\vec{q})$ for \ac{NN} bulk spin ice is
\begin{align}
	\mat{V}({\vec{q}}) = \mat{A}({\vec{q}}) + 2 \matone_{4 \times 4},
\end{align}
where the term proportional to the identity matrix makes the Lagrange multiplier $\lambda$ in the large-$N$ theory $\lambda$ consistent with the stiffness parameter of the Coulomb phase. The so-called \emph{adjacency matrix} $A(\vec{q})$ is given by
\begin{align}
	\mat{A}(\vec{q}) = 2 
	\begin{pmatrix}
	    0 & c_{01} & c_{02} & c_{03} \\
	    c_{10} & 0 & c_{12} & c_{13} \\
	    c_{20} & c_{21} & 0 & c_{23} \\
	    c_{30} & c_{31} & c_{32} & 0
	\end{pmatrix},
\end{align}
where $c_{\alpha \beta} \equiv \cos[\vec{q} \cdot (\vec{r}_\alpha - \vec{r}_\beta) ]$.

\section{[001] Films}
\label{app:films}

\subsection{Unit cell}

For films with surfaces perpendicular to the $[001]$ direction, we use a primitive unit cell spanning the whole finite ($\vhat{z}$) direction, with $8L$ sublattices, where $L$ is the number of cubic (conventional) unit cells in the finite direction. The primitive lattice vectors are
\begin{align}
	\vec{a}_1 &= \vhat{x},  & \vec{a}_2 &= \frac{\vhat{x}+\vhat{y}}{2}, 
\end{align}
and the sublattice vectors $\vec{r}_\alpha$ are given by
\begin{align}
    \vec{r}_0 &=\vec{0}, &
    \vec{r}_1 &= \frac{\vhat{x}+\vhat{y}}{4}, \nonumber \\
    \vec{r}_2 &= \frac{\vhat{x}+\vhat{z}}{4}, &
    \vec{r}_3 &= \frac{\vhat{y}+\vhat{z}}{4}, \nonumber  \\
    \vec{r}_4 &= \frac{2\vhat{x}+2\vhat{z}}{4}, &
    \vec{r}_5 &= \frac{3\vhat{x}+\vhat{y}+2\vhat{z}}{4}, \nonumber  \\
    \vec{r}_6 &= \frac{3\vhat{x}+3\vhat{z}}{4}, &
    \vec{r}_7 &= \frac{2\vhat{x}+\vhat{y}+3\vhat{z}}{4}, 
\end{align}	
for the first eight spins, and $ \vec{r}_{\alpha + 8k} = \vec{r}_{\alpha} + k\vhat{z}$, $k = 1,..,L-1$ for the remaining spins. The corresponding conventional unit cell, showing the structure of stacked layers in the $\vhat{z}$ direction (each layer made of parallel chains in the $[110]$ or $[1 \bar{1} 0]$ direction, alternatively) is shown in Fig.~\ref{fig:stacking}.

\subsection{Interaction matrix}

Similar to the bulk theory, we define
\begin{equation}
	\mat{V}(\vec{q}_\perp) = \mat{A}(\vec{q}_\perp) + 2\matone_{8 L \times 8 L} .
\end{equation}
The adjacency matrix $A(\vec{q}_\perp)$ is a $8 L \times 8 L$ matrix with a tridiagonal $8 \times 8$ block structure -- that is, only the diagonal blocks and next-to-diagonal blocks are non-trivial. For spins in the same cubic unit cell, the diagonal $8 \times 8$ block reads
\begin{align}
	\mat{A}_{\rm diag}(\vec{q}_\perp) = \begin{pmatrix}
	0       & 2c_{01} & e_{02}     & e_{03}     & 0     & 0     & 0     & 0 \\
	2c_{10} & 0     & e_{12}     & e_{13}     & 0     & 0     & 0     & 0 \\
	e_{20}  & e_{21}     & 0     & 2c_{23} & e_{24}     & e_{25}     & 0     & 0 \\
	e_{30}  & e_{31}     & 2c_{32} & 0     & e_{34}     & e_{35}     & 0     & 0 \\
	0     & 0     & e_{42}     & e_{43}     & 0     & 2c_{45} & e_{46}     & e_{47} \\
	0     & 0     & e_{52}     & e_{53}     & 2c_{54} & 0     & e_{56}     & e_{57} \\
	0     & 0     & 0     & 0     & e_{64}     & e_{65}     & 0     & 2c_{67} \\
	0     & 0     & 0     & 0     & e_{74}     & e_{75}     & 2c_{76} & 0  
	\end{pmatrix} ,
\end{align}
where $e_{\alpha \beta} \equiv \exp [i \vec{q}_\perp \cdot \vec{r}_{\alpha \beta}]$, and $\vec{r}_{\alpha \beta}$ is the nearest-neighbor vector connecting sublattices $\alpha$ and $\beta$. The next-to-diagonal $ 8 \times 8$ blocks are given by:   
\begin{equation}
	\mat{A}(\vec{q}_\perp)_\text{upper} = \begin{pmatrix}
	0 & 0 &\cdots & 0 \\
	\vdots & \vdots & \ddots & 0 \\
    0 & 0 &\cdots &  0 \\
	e_{60} & e_{61} &\cdots & 0  \\
    e_{70} & e_{71} &\cdots & 0 
	\end{pmatrix} ,
\end{equation}
\begin{equation}
	\mat{A}(\vec{q}_\perp)_\text{lower} = \begin{pmatrix}
	0 & \cdots & 0 & e_{06} & e_{07} \\
	0 & \cdots & 0 & e_{16} & e_{17} \\
	\vdots & \ddots & 0 & 0 & 0 \\
	0 & \cdots & 0 & 0 & 0 
	\end{pmatrix}.
\end{equation}
One can check that the complete adjacency matrix $A(\vec{q})$ is indeed Hermitian. All other couplings are zero.

\subsection{Including orphan bonds}

When including orphan bonds, with an Hamiltonian given by Eq. (\ref{eq:Hnn_Ising_orphan}), the interaction matrix becomes
\begin{equation}
	\mat{V}(\vec{q}_\perp) = \mat{A}(\vec{q}_\perp)  + 2\matone_{8 L \times 8 L} + \left(\frac{\Jo - J}{J}\right) \mat{A}_\text{O}(\vec{q}_\perp),
\end{equation}
where the adjacency matrix corresponding to the orphan bond couplings, $\mat{A}_\text{O}(\vec{q}_\perp)$, has only the following non-zero matrix elements:
\begin{align}
    \mat{A}_\text{O}(\vec{q}_\perp)_{01} &=  \exp [-i \vec{q}_\perp \cdot ( \vec{r}_0 - \vec{r}_1) ], \nonumber\\
        \mat{A}_\text{O}(\vec{q}_\perp)_{10} &=  \exp [+i \vec{q}_\perp \cdot ( \vec{r}_0 - \vec{r}_1) ] \nonumber,\\ 
    \mat{A}_\text{O}(\vec{q}_\perp)_{8L-2, 8L-1} &=  \exp [ -i \vec{q}_\perp \cdot ( \vec{r}_6 - \vec{r}_7)], \nonumber\\
    \mat{A}_\text{O}(\vec{q}_\perp)_{8L-1, 8L-2} &=  \exp [+ i \vec{q}_\perp \cdot ( \vec{r}_6 - \vec{r}_7)].
\end{align}

\section{Numerical solution of saddle-point equations} 
\label{app:numerics}

Here we briefly describe the method used to numerically solve Eq.~(\ref{eq:selfconsistency_films_qspace}):
\begin{align}
\sum_{\alpha \in l} \sum_{\vec{q}_\perp}  M^{-1}_{\alpha \alpha}({\vec{q}}_\perp) = n_l,
\end{align}
for each layer $l$. First, let us remark that the matrix $\mat{M}(\vec{q}_\perp)$ is not block-diagonal. Therefore, the spin-spin correlation matrix $\mat{M}^{-1}(\vec{q}_\perp)$ has diagonal elements that, in general, couple all coefficients $\lambda_l$ with $l~=~1, \: ... \: , 4L$. This means that one has to solve numerically for \emph{all} the $4L$ self-consistent equations simultaneously. 

We use the Newton-Raphson descent method, which allows one to find the zeros of a real-valued function. Consider
\begin{equation}
f_l( \vec{\Lambda} ) \equiv n_l - \sum_{\alpha \in l} \sum_{\vec{q}_\perp} M^{-1}_{\alpha \alpha}(\vec{q}_\perp, \vec{\Lambda}) , 
\end{equation}
where $\vec{\Lambda}~=~(\lambda_1, \lambda_2, ..., \lambda_{4L})$. We wish to solve $ \vec{f}(\vec{\Lambda})~=~\vec{0}$, with $\vec{f} = (f_1, f_2, ..., f_{4L})$. We start with an initial configuration $\vec{ \Lambda}^{(0)} $, chosen so that the eigenvalues of $M(\vec{q}_\perp, \vec{\Lambda}^{(0)})$ are positive, iterating the configuration from step $n$ to step $n+1$ using
\begin{equation}
\vec{\Lambda}^{(n+1)} = \vec{\Lambda}^{(n)} - D^{-1} \cdot \vec{f}(\vec{\Lambda}^{(n)}) , 
\end{equation}
where $D$ is the Jacobian matrix with elements 
\begin{equation}
D_{ij} = \left[\frac{\partial f_i}{\partial \lambda_j}\right]_{\vec{\Lambda}^{(n)}} ,
\end{equation}
until we reach the condition $ \vec{f}(\vec{\Lambda}) = \vec{0}$ to satisfactory numerical accuracy. As a stability check, we verify after each iteration that the matrix $\mat{M}(\vec{q}_\perp, \vec{\Lambda}^{(n)})$ has positive eigenvalues; if not, we restart the algorithm with a different initial configuration $\vec{ \Lambda}^{(0)}$.

\section{Details of Monte Carlo algorithm}
\label{app:mc}

\begin{figure}
  \centering
  \includegraphics[width=0.72\columnwidth]{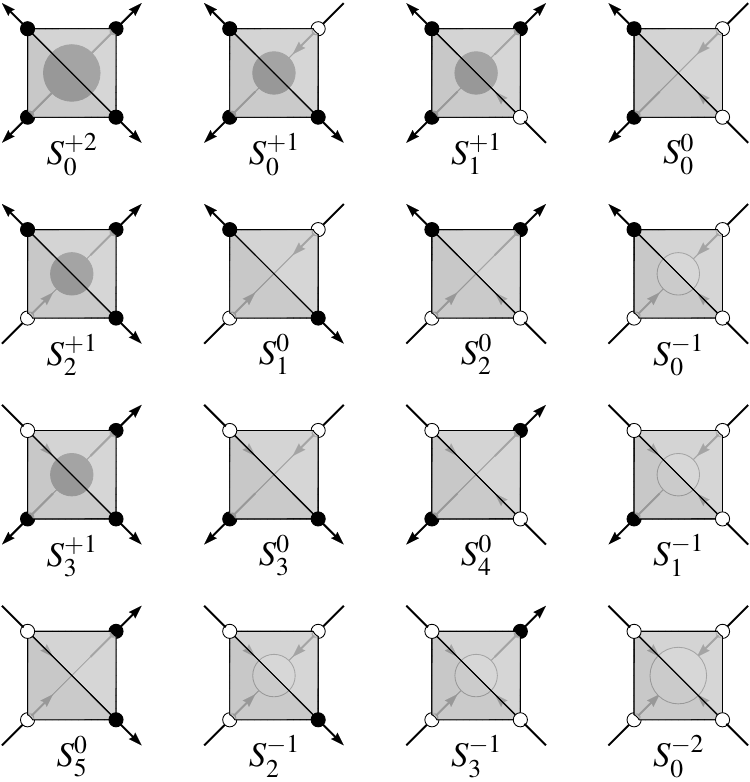}
    \caption{The sixteen states, $S^Q_\mu$, of a single tetrahedron organized by charge $Q = 0, \pm 1, \pm 2$, as used in Eq.~(\ref{eq:bulk-weights})}
	\label{fig:mc:states}
\end{figure}

Here, we provide details of the Monte Carlo methods, reviewing and extending the method first introduced in Ref.~[\onlinecite{otsuka-2014-cluster}]. This method decomposes the pyrochlore lattice into tetrahedral clusters; we label the sixteen states of each tetrahedron as $S^Q_\mu$ where $Q = 0, \pm 1, \pm 2$ is the charge and the index $\mu$ runs over the number of distinct states with the given charge. For $Q=0$ one has six states, for $Q=+1$ or $-1$ one has four states each and for $Q=+2$ or $-2$ one has a single state each (see Fig.~\ref{fig:mc:states} for an illustration). Generally, we can write the partition function of a nearest neighbor model on the pyrochlore lattice as
\begin{equation}
Z \equiv \sum_{\sigma} \prod_I \omega(S^{Q_I}_{\mu_I})
\end{equation}
where $I$ is a tetrahedron and $\omega(S^{Q_I}_{\mu_I})$ is the statistical weight of the configuration on that tetrahedron. For bulk nearest neighbour spin ice one can define the weights
\begin{align}
\label{eq:bulk-weights}
\omega(S^{0}_{\mu}) &= 1,  &
\omega(S^{\pm 1}_{\mu}) &= z,  &
\omega(S^{\pm 2}_{\mu}) &= z^4,  
\end{align}
where $z \equiv e^{-2\beta J}$, since the energy depends only on the charge $Q_I$ of a given tetrahedron,

Films in the $[001]$ direction can be implemented simply using this formalism. First, consider (bulk) nearest neighbor spin ice composed of $L_{\perp} \times L_{\perp} \times L$ conventional cubic unit cells with periodic boundary conditions. The desired film geometry can then be realized by cutting the bonds that pass thorough a fixed plane with normal $\vhat{z}$, changing the remaining bonds on those cut tetrahedra to carry the orphan coupling $\Jo$. For $\Jo/J>0$ we define the weights, $\omega^+_{\rm O}(S^Q_\mu)$, on such ``orphan" tetrahedron to be
\begin{align}
{\omega}^+_{\rm O}(S^{0}_1) &= {\omega}^+_{\rm O}(S^{0}_4) = z_{\rm O}^2,  \nonumber \\
{\omega}^+_{\rm O}(S^{0}_0) &= {\omega}^+_{\rm O}(S^{0}_2)={\omega}^+_{\rm O}(S^{0}_3)={\omega}^+_{\rm O}(S^{0}_5) = 1, \nonumber \\
{\omega}^+_{\rm O}(S^{\pm 1}_{\mu}) &= z_{\rm O}, \nonumber \\ 
{\omega}^+_{\rm O}(S^{\pm 2}_{\mu}) &= z_{\rm O}^2,
\end{align}
where $z_{\rm O} \equiv e^{-2\beta|\Jo|}$. Similarly, for $\Jo/J<0$ the weights, $\omega^-_{\rm O}(S^Q_\mu)$, can be defined as
\begin{align}
{\omega}^-_{\rm O}(S^{0}_1) &= {\omega}^-_{\rm O}(S^{0}_4) = 1,  \nonumber \\
{\omega}^-_{\rm O}(S^{0}_0) &= {\omega}^-_{\rm O}(S^{0}_2)={\omega}^-_{\rm O}(S^{0}_3)={\omega}^-_{\rm O}(S^{0}_5) = z_{\rm O}^2, \nonumber \\
{\omega}^-_{\rm O}(S^{\pm 1}_{\mu}) &= z_{\rm O}, \nonumber \\ 
{\omega}^-_{\rm O}(S^{\pm 2}_{\mu}) &= 1,
\end{align}
after a constant shift of the energy. The remaining non-orphan tetrahedra simply have the bulk weights given in Eq.~(\ref{eq:bulk-weights}) independent of $\Jo/J$.
\begin{figure}
  \centering
    \includegraphics[width=0.776\columnwidth]{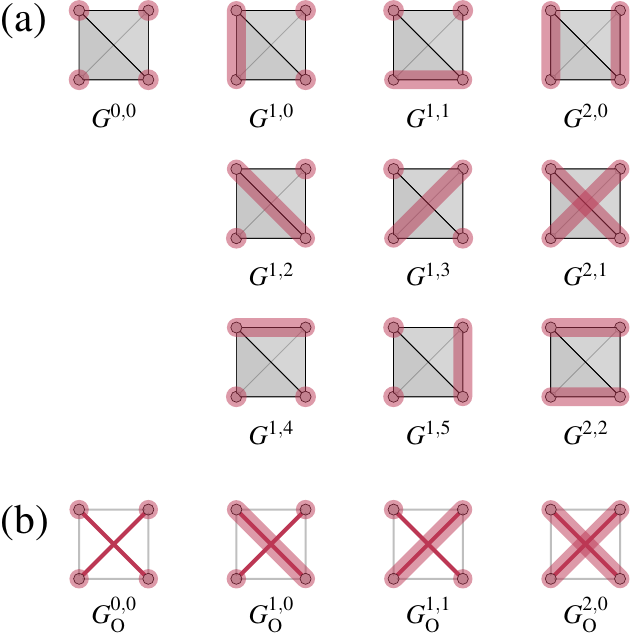}
    \caption{Tetrahedron graphs for the (a) bulk and (b) orphan tetrahedra for $[001]$ spin ice films, as used in Eq.~(\ref{eq:compat}).
	\label{fig:mc:graphs}
	}
\end{figure}

Next we define the probabilistic graph assignments that define the clusters, following the framework of Refs.~[\onlinecite{kandel1991},\onlinecite{evertz2003loop}]. To this end, we decompose a weight as
\begin{equation}
\label{eq:compat}
\omega(S^Q_\mu) \equiv \sum_G \Delta(S^Q_\mu,G) W(G),
\end{equation}
where $G$ is a \emph{graph} defined on a tetrahedron and the $\Delta(S^Q_\mu,G) = 0$ or $1$ are compatibility factors, with zero being incompatible and one being compatible. For our purposes, these graphs consist of isolated spins or possible pairings of two spins, as illustrated in Fig.~\subref{fig:mc:graphs}{(a)}. For the orphan tetrahedra, we do not include graphs that connect spins across the cut; the allowed graphs are shown in Fig.~\subref{fig:mc:graphs}{(b)}. The bulk tetrahedra and the $\Jo/J>0$ orphan tetrahedra graphs are defined to be compatible with a state if the pairs of spins joined take on opposite values while, for the $\Jo/J < 0$ graphs, the two joined spins must be \emph{equal}.

With these definitions, one can solve Eq.~(\ref{eq:compat}) to obtain the graph weights $W(G)$. As in the case of the partition function weights, we denoted the orphan tetrahedra to have graph weights as $W^{\pm}_{\rm O}(G)$. A solution for the bulk tetrahedra is given in Ref.~[\onlinecite{otsuka-2014-cluster}] as
\begin{align}
W(G^{0,0}) &= z^4, \nonumber \\
W(G^{1,a}) &= (z-z^4)/3, \nonumber \\
W(G^{2,a}) &= (3-4z+z^4)/6.
\label{eq:prob-bulk}
\end{align}
At low temperature, $T \ll J$, one has $z \rightarrow 0$ and thus only the three "icelike" graphs ($G^{2,a}$) have non-zero assignment probability. In this limit, the algorithm (for the bulk case) reduces to a variant of the usual loop algorithm~\cite{Melko2001}. For the orphan tetrahedra one finds a solution
\begin{align}
W^\pm_{\rm O}(G^{0,0}_{\rm O}) &= z_{\rm O}^2, \nonumber \\
W^\pm_{\rm O}(G^{1,a}_{\rm O}) &= z_{\rm O}(z_{\rm O}-1), \nonumber \\
W^\pm_{\rm O}(G^{2,0}_{\rm O}) &= (z_{\rm O}-1)^2.
\label{eq:prob-orphan}
\end{align}
These weights are positive and satisfy the required sum rules for any choice of $\Jo$~\cite{otsuka-2014-cluster}. At low temperature, $T \ll \Jo$, one has $z_{\rm O} \rightarrow 0$, with only the two pair graphs ($G^{2,0}_{\rm O}$) having non-zero probability. For $\Jo/J>0$ this corresponds to continuing the usual loops along the surface, while for the $\Jo/J<0$ case it corresponds to a loop where the alternation pattern is reversed when an orphan bond is encountered, as discussed in Sec.~\ref{sec:topological} [see Fig.~\subref{fig:surfaces_orphan}{(c)}]. For $\Jo/J = 0$, one has $z_{\rm O} = 1$ and thus only the free graph ($G^{0,0}_{\rm O}$) has non-zero assignment probability. This corresponds to allowing the strings of alternating spins to end at the orphan bonds. Note that these probabilities factorize; we could also define the weights on the orphan bonds separately at each surface, rather than using a combined orphan tetrahedron.

The method then proceeds as usual, as discussed in Refs.~[\onlinecite{kandel1991},\onlinecite{evertz2003loop},\onlinecite{otsuka-2014-cluster}]. A Monte Carlo step consists of first assigning graphs to each tetrahedron following the probabilities, $W(G)$, given in Eq.~(\ref{eq:prob-bulk}) and (\ref{eq:prob-orphan}). When assigned to the whole lattice, these graphs form clusters, in this case strings and loops, which are identified and then flipped or not flipped with equal probability~\cite{swensden-1987-nonuniversal}.

\bibliography{main}

\begin{thebibliography}{88}%
\makeatletter
\providecommand \@ifxundefined [1]{%
 \@ifx{#1\undefined}
}%
\providecommand \@ifnum [1]{%
 \ifnum #1\expandafter \@firstoftwo
 \else \expandafter \@secondoftwo
 \fi
}%
\providecommand \@ifx [1]{%
 \ifx #1\expandafter \@firstoftwo
 \else \expandafter \@secondoftwo
 \fi
}%
\providecommand \natexlab [1]{#1}%
\providecommand \enquote  [1]{``#1''}%
\providecommand \bibnamefont  [1]{#1}%
\providecommand \bibfnamefont [1]{#1}%
\providecommand \citenamefont [1]{#1}%
\providecommand \href@noop [0]{\@secondoftwo}%
\providecommand \href [0]{\begingroup \@sanitize@url \@href}%
\providecommand \@href[1]{\@@startlink{#1}\@@href}%
\providecommand \@@href[1]{\endgroup#1\@@endlink}%
\providecommand \@sanitize@url [0]{\catcode `\\12\catcode `\$12\catcode
  `\&12\catcode `\#12\catcode `\^12\catcode `\_12\catcode `\%12\relax}%
\providecommand \@@startlink[1]{}%
\providecommand \@@endlink[0]{}%
\providecommand \url  [0]{\begingroup\@sanitize@url \@url }%
\providecommand \@url [1]{\endgroup\@href {#1}{\urlprefix }}%
\providecommand \urlprefix  [0]{URL }%
\providecommand \Eprint [0]{\href }%
\providecommand \doibase [0]{http://dx.doi.org/}%
\providecommand \selectlanguage [0]{\@gobble}%
\providecommand \bibinfo  [0]{\@secondoftwo}%
\providecommand \bibfield  [0]{\@secondoftwo}%
\providecommand \translation [1]{[#1]}%
\providecommand \BibitemOpen [0]{}%
\providecommand \bibitemStop [0]{}%
\providecommand \bibitemNoStop [0]{.\EOS\space}%
\providecommand \EOS [0]{\spacefactor3000\relax}%
\providecommand \BibitemShut  [1]{\csname bibitem#1\endcsname}%
\let\auto@bib@innerbib\@empty
\bibitem [{\citenamefont {Kogut}(1979)}]{Kogut}%
  \BibitemOpen
  \bibfield  {author} {\bibinfo {author} {\bibfnamefont {John~B.}\ \bibnamefont
  {Kogut}},\ }\bibfield  {title} {\enquote {\bibinfo {title} {An introduction
  to lattice gauge theory and spin systems},}\ }\href {\doibase
  10.1103/RevModPhys.51.659} {\bibfield  {journal} {\bibinfo  {journal} {Rev.
  Mod. Phys.}\ }\textbf {\bibinfo {volume} {51}},\ \bibinfo {pages} {659--713}
  (\bibinfo {year} {1979})}\BibitemShut {NoStop}%
\bibitem [{\citenamefont {Nagaosa}(1999)}]{Nagaosa_book}%
  \BibitemOpen
  \bibfield  {author} {\bibinfo {author} {\bibfnamefont {N.}~\bibnamefont
  {Nagaosa}},\ }\href@noop {} {\emph {\bibinfo {title} {Quantum Field Theory in
  Strongly Correlated Electronic Systems}}}\ (\bibinfo  {publisher}
  {Springer-Verlag, Berlin, Heidelberg},\ \bibinfo {year} {1999})\BibitemShut
  {NoStop}%
\bibitem [{\citenamefont {Wen}(2004)}]{Wen_book}%
  \BibitemOpen
  \bibfield  {author} {\bibinfo {author} {\bibfnamefont {X.-G.}\ \bibnamefont
  {Wen}},\ }\href@noop {} {\emph {\bibinfo {title} {Quantum Field Theory of
  Many-body Systems}}}\ (\bibinfo  {publisher} {Oxford University Press Inc.,
  New York},\ \bibinfo {year} {2004})\BibitemShut {NoStop}%
\bibitem [{\citenamefont {Lee}\ \emph {et~al.}(2006)\citenamefont {Lee},
  \citenamefont {Nagaosa},\ and\ \citenamefont {Wen}}]{Lee_RMP}%
  \BibitemOpen
  \bibfield  {author} {\bibinfo {author} {\bibfnamefont {Patrick~A.}\
  \bibnamefont {Lee}}, \bibinfo {author} {\bibfnamefont {Naoto}\ \bibnamefont
  {Nagaosa}}, \ and\ \bibinfo {author} {\bibfnamefont {Xiao-Gang}\ \bibnamefont
  {Wen}},\ }\bibfield  {title} {\enquote {\bibinfo {title} {Doping a mott
  insulator: Physics of high-temperature superconductivity},}\ }\href {\doibase
  10.1103/RevModPhys.78.17} {\bibfield  {journal} {\bibinfo  {journal} {Rev.
  Mod. Phys.}\ }\textbf {\bibinfo {volume} {78}},\ \bibinfo {pages} {17--85}
  (\bibinfo {year} {2006})}\BibitemShut {NoStop}%
\bibitem [{\citenamefont {Jackson}(2007)}]{jackson2007classical}%
  \BibitemOpen
  \bibfield  {author} {\bibinfo {author} {\bibfnamefont {John~David}\
  \bibnamefont {Jackson}},\ }\href@noop {} {\emph {\bibinfo {title} {Classical
  electrodynamics}}}\ (\bibinfo  {publisher} {John Wiley \& Sons},\ \bibinfo
  {year} {2007})\BibitemShut {NoStop}%
\bibitem [{\citenamefont {Milton}(2001)}]{milton2001casimir}%
  \BibitemOpen
  \bibfield  {author} {\bibinfo {author} {\bibfnamefont {Kimball~A}\
  \bibnamefont {Milton}},\ }\href@noop {} {\emph {\bibinfo {title} {The Casimir
  effect: physical manifestations of zero-point energy}}}\ (\bibinfo
  {publisher} {World Scientific},\ \bibinfo {year} {2001})\BibitemShut
  {NoStop}%
\bibitem [{\citenamefont {Bordag}\ \emph {et~al.}(2001)\citenamefont {Bordag},
  \citenamefont {Mohideen},\ and\ \citenamefont
  {Mostepanenko}}]{bordag2001new}%
  \BibitemOpen
  \bibfield  {author} {\bibinfo {author} {\bibfnamefont {Michael}\ \bibnamefont
  {Bordag}}, \bibinfo {author} {\bibfnamefont {Umar}\ \bibnamefont {Mohideen}},
  \ and\ \bibinfo {author} {\bibfnamefont {Vladimir~M}\ \bibnamefont
  {Mostepanenko}},\ }\bibfield  {title} {\enquote {\bibinfo {title} {New
  developments in the casimir effect},}\ }\href {\doibase
  10.1016/S0370-1573(01)00015-1} {\bibfield  {journal} {\bibinfo  {journal}
  {Phys. Rep.}\ }\textbf {\bibinfo {volume} {353}},\ \bibinfo {pages} {1--205}
  (\bibinfo {year} {2001})}\BibitemShut {NoStop}%
\bibitem [{\citenamefont {Wang}\ and\ \citenamefont
  {Senthil}(2016)}]{senthil2016}%
  \BibitemOpen
  \bibfield  {author} {\bibinfo {author} {\bibfnamefont {Chong}\ \bibnamefont
  {Wang}}\ and\ \bibinfo {author} {\bibfnamefont {T.}~\bibnamefont {Senthil}},\
  }\bibfield  {title} {\enquote {\bibinfo {title} {Time-reversal symmetric
  {$U(1)$} quantum spin liquids},}\ }\href {\doibase 10.1103/PhysRevX.6.011034}
  {\bibfield  {journal} {\bibinfo  {journal} {Phys. Rev. X}\ }\textbf {\bibinfo
  {volume} {6}},\ \bibinfo {pages} {011034} (\bibinfo {year}
  {2016})}\BibitemShut {NoStop}%
\bibitem [{\citenamefont {{Harris}}\ \emph {et~al.}(1997)\citenamefont
  {{Harris}}, \citenamefont {{Bramwell}}, \citenamefont {{McMorrow}},
  \citenamefont {{Zeiske}},\ and\ \citenamefont {{Godfrey}}}]{Harris1997}%
  \BibitemOpen
  \bibfield  {author} {\bibinfo {author} {\bibfnamefont {M.~J.}\ \bibnamefont
  {{Harris}}}, \bibinfo {author} {\bibfnamefont {S.~T.}\ \bibnamefont
  {{Bramwell}}}, \bibinfo {author} {\bibfnamefont {D.~F.}\ \bibnamefont
  {{McMorrow}}}, \bibinfo {author} {\bibfnamefont {T.}~\bibnamefont
  {{Zeiske}}}, \ and\ \bibinfo {author} {\bibfnamefont {K.~W.}\ \bibnamefont
  {{Godfrey}}},\ }\bibfield  {title} {\enquote {\bibinfo {title} {{Geometrical
  Frustration in the Ferromagnetic Pyrochlore Ho$_{2}$Ti$_{2}$O$_{7}$}},}\
  }\href {\doibase 10.1103/PhysRevLett.79.2554} {\bibfield  {journal} {\bibinfo
   {journal} {Phys. Rev. Lett.}\ }\textbf {\bibinfo {volume} {79}},\ \bibinfo
  {pages} {2554--2557} (\bibinfo {year} {1997})}\BibitemShut {NoStop}%
\bibitem [{\citenamefont {{Ramirez}}\ \emph {et~al.}(1999)\citenamefont
  {{Ramirez}}, \citenamefont {{Hayashi}}, \citenamefont {{Cava}}, \citenamefont
  {{Siddharthan}},\ and\ \citenamefont {{Shastry}}}]{Ramirez1999}%
  \BibitemOpen
  \bibfield  {author} {\bibinfo {author} {\bibfnamefont {A.~P.}\ \bibnamefont
  {{Ramirez}}}, \bibinfo {author} {\bibfnamefont {A.}~\bibnamefont
  {{Hayashi}}}, \bibinfo {author} {\bibfnamefont {R.~J.}\ \bibnamefont
  {{Cava}}}, \bibinfo {author} {\bibfnamefont {R.}~\bibnamefont
  {{Siddharthan}}}, \ and\ \bibinfo {author} {\bibfnamefont {B.~S.}\
  \bibnamefont {{Shastry}}},\ }\bibfield  {title} {\enquote {\bibinfo {title}
  {{Zero-point entropy in `spin ice'}},}\ }\href {\doibase 10.1038/20619}
  {\bibfield  {journal} {\bibinfo  {journal} {Nature}\ }\textbf {\bibinfo
  {volume} {399}},\ \bibinfo {pages} {333--335} (\bibinfo {year}
  {1999})}\BibitemShut {NoStop}%
\bibitem [{\citenamefont {{Bramwell}}\ \emph {et~al.}(2004)\citenamefont
  {{Bramwell}}, \citenamefont {{Gingras}},\ and\ \citenamefont
  {{Holdsworth}}}]{Bramwell_book2004}%
  \BibitemOpen
  \bibfield  {author} {\bibinfo {author} {\bibfnamefont {S.~T.}\ \bibnamefont
  {{Bramwell}}}, \bibinfo {author} {\bibfnamefont {M.~J.~P.}\ \bibnamefont
  {{Gingras}}}, \ and\ \bibinfo {author} {\bibfnamefont {P.~C.~W.}\
  \bibnamefont {{Holdsworth}}},\ }\enquote {\bibinfo {title} {{Spin Ice}},}\
  in\ \href {\doibase 10.1142/9789812567819_0007} {\emph {\bibinfo {booktitle}
  {Frustrated Spin Systems}}},\ \bibinfo {editor} {edited by\ \bibinfo {editor}
  {\bibfnamefont {H.-T.}\ \bibnamefont {{Diep}}}}\ (\bibinfo  {publisher}
  {World Scientific Publishing Co},\ \bibinfo {year} {2004})\ pp.\ \bibinfo
  {pages} {367--456}\BibitemShut {NoStop}%
\bibitem [{\citenamefont {Gingras}(2011)}]{Gingras2011}%
  \BibitemOpen
  \bibfield  {author} {\bibinfo {author} {\bibfnamefont {Michel J.~P.}\
  \bibnamefont {Gingras}},\ }\enquote {\bibinfo {title} {Spin ice},}\ in\ \href
  {\doibase 10.1007/978-3-642-10589-0_12} {\emph {\bibinfo {booktitle}
  {Introduction to Frustrated Magnetism: Materials, Experiments, Theory}}},\
  \bibinfo {editor} {edited by\ \bibinfo {editor} {\bibfnamefont {Claudine}\
  \bibnamefont {Lacroix}}, \bibinfo {editor} {\bibfnamefont {Philippe}\
  \bibnamefont {Mendels}}, \ and\ \bibinfo {editor} {\bibfnamefont
  {Fr{\'e}d{\'e}ric}\ \bibnamefont {Mila}}}\ (\bibinfo  {publisher} {Springer
  Berlin Heidelberg},\ \bibinfo {address} {Berlin, Heidelberg},\ \bibinfo
  {year} {2011})\ pp.\ \bibinfo {pages} {293--329}\BibitemShut {NoStop}%
\bibitem [{\citenamefont {Rau}\ and\ \citenamefont
  {Gingras}(2015)}]{RauPRB2015}%
  \BibitemOpen
  \bibfield  {author} {\bibinfo {author} {\bibfnamefont {Jeffrey~G.}\
  \bibnamefont {Rau}}\ and\ \bibinfo {author} {\bibfnamefont {Michel J.~P.}\
  \bibnamefont {Gingras}},\ }\bibfield  {title} {\enquote {\bibinfo {title}
  {Magnitude of quantum effects in classical spin ices},}\ }\href {\doibase
  10.1103/PhysRevB.92.144417} {\bibfield  {journal} {\bibinfo  {journal} {Phys.
  Rev. B}\ }\textbf {\bibinfo {volume} {92}},\ \bibinfo {pages} {144417}
  (\bibinfo {year} {2015})}\BibitemShut {NoStop}%
\bibitem [{\citenamefont {Henley}(2005)}]{Henley2005}%
  \BibitemOpen
  \bibfield  {author} {\bibinfo {author} {\bibfnamefont {C.~L.}\ \bibnamefont
  {Henley}},\ }\bibfield  {title} {\enquote {\bibinfo {title} {Power-law spin
  correlations in pyrochlore antiferromagnets},}\ }\href {\doibase
  10.1103/PhysRevB.71.014424} {\bibfield  {journal} {\bibinfo  {journal} {Phys.
  Rev. B}\ }\textbf {\bibinfo {volume} {71}},\ \bibinfo {pages} {014424}
  (\bibinfo {year} {2005})}\BibitemShut {NoStop}%
\bibitem [{\citenamefont {Henley}(2010)}]{Henley2010}%
  \BibitemOpen
  \bibfield  {author} {\bibinfo {author} {\bibfnamefont {Christopher~L.}\
  \bibnamefont {Henley}},\ }\bibfield  {title} {\enquote {\bibinfo {title}
  {{The “Coulomb Phase” in Frustrated Systems}},}\ }\href {\doibase
  10.1146/annurev-conmatphys-070909-104138} {\bibfield  {journal} {\bibinfo
  {journal} {Annu. Rev. Condens. Matter Phys.}\ }\textbf {\bibinfo {volume}
  {1}},\ \bibinfo {pages} {179--210} (\bibinfo {year} {2010})}\BibitemShut
  {NoStop}%
\bibitem [{\citenamefont {Castelnovo}\ \emph {et~al.}({2012})\citenamefont
  {Castelnovo}, \citenamefont {Moessner},\ and\ \citenamefont
  {Sondhi}}]{Castelnovo_AnnRevCMP}%
  \BibitemOpen
  \bibfield  {author} {\bibinfo {author} {\bibfnamefont {C.}~\bibnamefont
  {Castelnovo}}, \bibinfo {author} {\bibfnamefont {R.}~\bibnamefont
  {Moessner}}, \ and\ \bibinfo {author} {\bibfnamefont {S.~L.}\ \bibnamefont
  {Sondhi}},\ }\bibfield  {title} {\enquote {\bibinfo {title} {{Spin Ice,
  Fractionalization, and Topological Order}},}\ }\href {\doibase
  10.1146/annurev-conmatphys-020911-125058} {\bibfield  {journal} {\bibinfo
  {journal} {Annu. Rev. Condens. Matter Phys.}\ }\textbf {\bibinfo {volume}
  {{3}}},\ \bibinfo {pages} {33--55} (\bibinfo {year} {{2012}})}\BibitemShut
  {NoStop}%
\bibitem [{\citenamefont {Castelnovo}\ \emph {et~al.}(2008)\citenamefont
  {Castelnovo}, \citenamefont {Moessner},\ and\ \citenamefont
  {Sondhi}}]{Castelnovo2008a}%
  \BibitemOpen
  \bibfield  {author} {\bibinfo {author} {\bibfnamefont {C}~\bibnamefont
  {Castelnovo}}, \bibinfo {author} {\bibfnamefont {R}~\bibnamefont {Moessner}},
  \ and\ \bibinfo {author} {\bibfnamefont {S~L}\ \bibnamefont {Sondhi}},\
  }\bibfield  {title} {\enquote {\bibinfo {title} {{Magnetic monopoles in spin
  ice.}}}\ }\href {\doibase 10.1038/nature06433} {\bibfield  {journal}
  {\bibinfo  {journal} {Nature}\ }\textbf {\bibinfo {volume} {451}},\ \bibinfo
  {pages} {42--45} (\bibinfo {year} {2008})}\BibitemShut {NoStop}%
\bibitem [{\citenamefont {Gingras}\ and\ \citenamefont
  {McClarty}(2014)}]{Gingras2014}%
  \BibitemOpen
  \bibfield  {author} {\bibinfo {author} {\bibfnamefont {M~J~P}\ \bibnamefont
  {Gingras}}\ and\ \bibinfo {author} {\bibfnamefont {P~A}\ \bibnamefont
  {McClarty}},\ }\bibfield  {title} {\enquote {\bibinfo {title} {{Quantum spin
  ice: a search for gapless quantum spin liquids in pyrochlore magnets}},}\
  }\href {\doibase 10.1088/0034-4885/77/5/056501} {\bibfield  {journal}
  {\bibinfo  {journal} {Rep. Prog. Phys.}\ }\textbf {\bibinfo {volume} {77}},\
  \bibinfo {pages} {056501} (\bibinfo {year} {2014})}\BibitemShut {NoStop}%
\bibitem [{\citenamefont {Algara-Siller}\ \emph {et~al.}(2015)\citenamefont
  {Algara-Siller}, \citenamefont {Lehtinen}, \citenamefont {Wang},
  \citenamefont {Nair}, \citenamefont {Kaiser}, \citenamefont {Wu},
  \citenamefont {Geim},\ and\ \citenamefont
  {Grigorieva}}]{algara-siller_square_2015}%
  \BibitemOpen
  \bibfield  {author} {\bibinfo {author} {\bibfnamefont {G.}~\bibnamefont
  {Algara-Siller}}, \bibinfo {author} {\bibfnamefont {O.}~\bibnamefont
  {Lehtinen}}, \bibinfo {author} {\bibfnamefont {F.~C.}\ \bibnamefont {Wang}},
  \bibinfo {author} {\bibfnamefont {R.~R.}\ \bibnamefont {Nair}}, \bibinfo
  {author} {\bibfnamefont {U.}~\bibnamefont {Kaiser}}, \bibinfo {author}
  {\bibfnamefont {H.~A.}\ \bibnamefont {Wu}}, \bibinfo {author} {\bibfnamefont
  {A.~K.}\ \bibnamefont {Geim}}, \ and\ \bibinfo {author} {\bibfnamefont
  {I.~V.}\ \bibnamefont {Grigorieva}},\ }\bibfield  {title} {\enquote {\bibinfo
  {title} {Square ice in graphene nanocapillaries},}\ }\href {\doibase
  10.1038/nature14295} {\bibfield  {journal} {\bibinfo  {journal} {Nature}\
  }\textbf {\bibinfo {volume} {519}},\ \bibinfo {pages} {443--445} (\bibinfo
  {year} {2015})}\BibitemShut {NoStop}%
\bibitem [{\citenamefont {Bovo}\ \emph {et~al.}(2014)\citenamefont {Bovo},
  \citenamefont {Moya}, \citenamefont {Prabhakaran}, \citenamefont {Soh},
  \citenamefont {Boothroyd}, \citenamefont {Mathur}, \citenamefont {Aeppli},\
  and\ \citenamefont {Bramwell}}]{Bovo2014}%
  \BibitemOpen
  \bibfield  {author} {\bibinfo {author} {\bibfnamefont {L}~\bibnamefont
  {Bovo}}, \bibinfo {author} {\bibfnamefont {X}~\bibnamefont {Moya}}, \bibinfo
  {author} {\bibfnamefont {D}~\bibnamefont {Prabhakaran}}, \bibinfo {author}
  {\bibfnamefont {Yeong-Ah}\ \bibnamefont {Soh}}, \bibinfo {author}
  {\bibfnamefont {a~T}\ \bibnamefont {Boothroyd}}, \bibinfo {author}
  {\bibfnamefont {N~D}\ \bibnamefont {Mathur}}, \bibinfo {author}
  {\bibfnamefont {G}~\bibnamefont {Aeppli}}, \ and\ \bibinfo {author}
  {\bibfnamefont {S~T}\ \bibnamefont {Bramwell}},\ }\bibfield  {title}
  {\enquote {\bibinfo {title} {{Restoration of the third law in spin ice thin
  films.}}}\ }\href {\doibase 10.1038/ncomms4439} {\bibfield  {journal}
  {\bibinfo  {journal} {Nat. Commun.}\ }\textbf {\bibinfo {volume} {5}},\
  \bibinfo {pages} {3439} (\bibinfo {year} {2014})}\BibitemShut {NoStop}%
\bibitem [{\citenamefont {Petrenko}(2014)}]{Petrenko2014}%
  \BibitemOpen
  \bibfield  {author} {\bibinfo {author} {\bibfnamefont {O.}~\bibnamefont
  {Petrenko}},\ }\bibfield  {title} {\enquote {\bibinfo {title} {{Oxide
  heterostructures: Thin spin ice under investigation.}}}\ }\href {\doibase
  10.1038/nmat3957} {\bibfield  {journal} {\bibinfo  {journal} {Nat. Mater.}\
  }\textbf {\bibinfo {volume} {13}},\ \bibinfo {pages} {430--1} (\bibinfo
  {year} {2014})}\BibitemShut {NoStop}%
\bibitem [{\citenamefont {Leusink}\ \emph {et~al.}(2014)\citenamefont
  {Leusink}, \citenamefont {Coneri}, \citenamefont {Hoek}, \citenamefont
  {Turner}, \citenamefont {Idrissi}, \citenamefont {{Van Tendeloo}},\ and\
  \citenamefont {Hilgenkamp}}]{Leusink2014}%
  \BibitemOpen
  \bibfield  {author} {\bibinfo {author} {\bibfnamefont {D.~P.}\ \bibnamefont
  {Leusink}}, \bibinfo {author} {\bibfnamefont {F.}~\bibnamefont {Coneri}},
  \bibinfo {author} {\bibfnamefont {M.}~\bibnamefont {Hoek}}, \bibinfo {author}
  {\bibfnamefont {S.}~\bibnamefont {Turner}}, \bibinfo {author} {\bibfnamefont
  {H.}~\bibnamefont {Idrissi}}, \bibinfo {author} {\bibfnamefont
  {G.}~\bibnamefont {{Van Tendeloo}}}, \ and\ \bibinfo {author} {\bibfnamefont
  {H.}~\bibnamefont {Hilgenkamp}},\ }\bibfield  {title} {\enquote {\bibinfo
  {title} {Thin films of the spin ice compound {Ho$_2$Ti$_2$O$_7$}},}\ }\href
  {\doibase 10.1063/1.4867222} {\bibfield  {journal} {\bibinfo  {journal} {APL
  Materials}\ }\textbf {\bibinfo {volume} {2}},\ \bibinfo {pages} {032101}
  (\bibinfo {year} {2014})}\BibitemShut {NoStop}%
\bibitem [{Note1()}]{Note1}%
  \BibitemOpen
  \bibinfo {note} {We note that the results of \protect \citet
  {pomaranski2013absence} report a release of some of the spin ice entropy at
  very low temperatures in bulk {Dy}$_2${Ti}$_2$O$_7$. The origin of this
  release is still a matter of debate~\cite
  {henelius2016,borzi2016intermediate}.}\BibitemShut {Stop}%
\bibitem [{\citenamefont {She}\ \emph {et~al.}(2017)\citenamefont {She},
  \citenamefont {Kim}, \citenamefont {Fennie}, \citenamefont {Lawler},\ and\
  \citenamefont {Kim}}]{She2017}%
  \BibitemOpen
  \bibfield  {author} {\bibinfo {author} {\bibfnamefont {Jian-Huang}\
  \bibnamefont {She}}, \bibinfo {author} {\bibfnamefont {Choong~H}\
  \bibnamefont {Kim}}, \bibinfo {author} {\bibfnamefont {Craig~J}\ \bibnamefont
  {Fennie}}, \bibinfo {author} {\bibfnamefont {Michael~J}\ \bibnamefont
  {Lawler}}, \ and\ \bibinfo {author} {\bibfnamefont {Eun-Ah}\ \bibnamefont
  {Kim}},\ }\bibfield  {title} {\enquote {\bibinfo {title} {{Topological
  superconductivity in metal/quantum-spin-ice heterostructures}},}\ }\href
  {\doibase 10.1038/s41535-017-0063-2} {\bibfield  {journal} {\bibinfo
  {journal} {npj Quantum Materials}\ }\textbf {\bibinfo {volume} {2}},\
  \bibinfo {pages} {64} (\bibinfo {year} {2017})}\BibitemShut {NoStop}%
\bibitem [{\citenamefont {Sasaki}\ \emph {et~al.}(2014)\citenamefont {Sasaki},
  \citenamefont {Imai},\ and\ \citenamefont {Kanazawa}}]{Sasaki2014}%
  \BibitemOpen
  \bibfield  {author} {\bibinfo {author} {\bibfnamefont {T}~\bibnamefont
  {Sasaki}}, \bibinfo {author} {\bibfnamefont {E}~\bibnamefont {Imai}}, \ and\
  \bibinfo {author} {\bibfnamefont {I}~\bibnamefont {Kanazawa}},\ }\bibfield
  {title} {\enquote {\bibinfo {title} {{Witten effect and fractional electric
  charge on the domain wall between topological insulators and spin ice
  compounds}},}\ }\href
  {http://www.scopus.com/inward/record.url?eid=2-s2.0-84919663801{\&}partnerID=40{\&}md5=875cd81d9936a124c100c2abfe94b99d}
  {\bibfield  {journal} {\bibinfo  {journal} {J. Phys. Conf. Ser.}\ }\textbf
  {\bibinfo {volume} {568}} (\bibinfo {year} {2014})}\BibitemShut {NoStop}%
\bibitem [{\citenamefont {Jaubert}\ \emph {et~al.}(2017)\citenamefont
  {Jaubert}, \citenamefont {Lin}, \citenamefont {Opel}, \citenamefont
  {Holdsworth},\ and\ \citenamefont {Gingras}}]{Jaubert2016}%
  \BibitemOpen
  \bibfield  {author} {\bibinfo {author} {\bibfnamefont {L.~D.~C.}\
  \bibnamefont {Jaubert}}, \bibinfo {author} {\bibfnamefont {T.}~\bibnamefont
  {Lin}}, \bibinfo {author} {\bibfnamefont {T.~S.}\ \bibnamefont {Opel}},
  \bibinfo {author} {\bibfnamefont {P.~C.~W.}\ \bibnamefont {Holdsworth}}, \
  and\ \bibinfo {author} {\bibfnamefont {M.~J.~P.}\ \bibnamefont {Gingras}},\
  }\bibfield  {title} {\enquote {\bibinfo {title} {Spin ice thin film: Surface
  ordering, emergent square ice, and strain effects},}\ }\href {\doibase
  10.1103/PhysRevLett.118.207206} {\bibfield  {journal} {\bibinfo  {journal}
  {Phys. Rev. Lett.}\ }\textbf {\bibinfo {volume} {118}},\ \bibinfo {pages}
  {207206} (\bibinfo {year} {2017})}\BibitemShut {NoStop}%
\bibitem [{\citenamefont {Isakov}\ \emph {et~al.}(2004)\citenamefont {Isakov},
  \citenamefont {Gregor}, \citenamefont {Moessner},\ and\ \citenamefont
  {Sondhi}}]{Isakov2004}%
  \BibitemOpen
  \bibfield  {author} {\bibinfo {author} {\bibfnamefont {S.~V.}\ \bibnamefont
  {Isakov}}, \bibinfo {author} {\bibfnamefont {K.}~\bibnamefont {Gregor}},
  \bibinfo {author} {\bibfnamefont {R.}~\bibnamefont {Moessner}}, \ and\
  \bibinfo {author} {\bibfnamefont {S.~L.}\ \bibnamefont {Sondhi}},\ }\bibfield
   {title} {\enquote {\bibinfo {title} {Dipolar spin correlations in classical
  pyrochlore magnets},}\ }\href {\doibase 10.1103/PhysRevLett.93.167204}
  {\bibfield  {journal} {\bibinfo  {journal} {Phys. Rev. Lett.}\ }\textbf
  {\bibinfo {volume} {93}},\ \bibinfo {pages} {167204} (\bibinfo {year}
  {2004})}\BibitemShut {NoStop}%
\bibitem [{\citenamefont {{Dantchev}}\ \emph {et~al.}(2014)\citenamefont
  {{Dantchev}}, \citenamefont {{Bergknoff}},\ and\ \citenamefont
  {{Rudnick}}}]{Dantchev2014}%
  \BibitemOpen
  \bibfield  {author} {\bibinfo {author} {\bibfnamefont {D.}~\bibnamefont
  {{Dantchev}}}, \bibinfo {author} {\bibfnamefont {J.}~\bibnamefont
  {{Bergknoff}}}, \ and\ \bibinfo {author} {\bibfnamefont {J.}~\bibnamefont
  {{Rudnick}}},\ }\bibfield  {title} {\enquote {\bibinfo {title} {{Casimir
  force in the O(n{$\rightarrow \infty$}) model with free boundary
  conditions}},}\ }\href {\doibase 10.1103/PhysRevE.89.042116} {\bibfield
  {journal} {\bibinfo  {journal} {\pre}\ }\textbf {\bibinfo {volume} {89}},\
  \bibinfo {eid} {042116} (\bibinfo {year} {2014})}\BibitemShut {NoStop}%
\bibitem [{\citenamefont {Rehn}\ \emph {et~al.}(2017)\citenamefont {Rehn},
  \citenamefont {Sen},\ and\ \citenamefont {Moessner}}]{Rehn2017}%
  \BibitemOpen
  \bibfield  {author} {\bibinfo {author} {\bibfnamefont {J.}~\bibnamefont
  {Rehn}}, \bibinfo {author} {\bibfnamefont {Arnab}\ \bibnamefont {Sen}}, \
  and\ \bibinfo {author} {\bibfnamefont {R.}~\bibnamefont {Moessner}},\
  }\bibfield  {title} {\enquote {\bibinfo {title} {Fractionalized
  {$\mathbb{Z}_{2}$} classical heisenberg spin liquids},}\ }\href {\doibase
  10.1103/PhysRevLett.118.047201} {\bibfield  {journal} {\bibinfo  {journal}
  {Phys. Rev. Lett.}\ }\textbf {\bibinfo {volume} {118}},\ \bibinfo {pages}
  {047201} (\bibinfo {year} {2017})}\BibitemShut {NoStop}%
\bibitem [{\citenamefont {Anderson}(1956)}]{Anderson1956}%
  \BibitemOpen
  \bibfield  {author} {\bibinfo {author} {\bibfnamefont {P.~W.}\ \bibnamefont
  {Anderson}},\ }\bibfield  {title} {\enquote {\bibinfo {title} {{Ordering and
  antiferromagnetism in ferrites}},}\ }\href {\doibase
  10.1103/PhysRev.102.1008} {\bibfield  {journal} {\bibinfo  {journal} {Phys.
  Rev.}\ }\textbf {\bibinfo {volume} {102}},\ \bibinfo {pages} {1008--1013}
  (\bibinfo {year} {1956})}\BibitemShut {NoStop}%
\bibitem [{\citenamefont {den Hertog}\ and\ \citenamefont
  {Gingras}(2000)}]{DenHertog2000}%
  \BibitemOpen
  \bibfield  {author} {\bibinfo {author} {\bibfnamefont {B~C}\ \bibnamefont
  {den Hertog}}\ and\ \bibinfo {author} {\bibfnamefont {M~J~P}\ \bibnamefont
  {Gingras}},\ }\bibfield  {title} {\enquote {\bibinfo {title} {{Dipolar
  interactions and origin of spin ice in Ising pyrochlore magnets}},}\ }\href
  {\doibase 10.1103/PhysRevLett.84.3430} {\bibfield  {journal} {\bibinfo
  {journal} {Phys. Rev. Lett.}\ }\textbf {\bibinfo {volume} {84}},\ \bibinfo
  {pages} {3430--3433} (\bibinfo {year} {2000})}\BibitemShut {NoStop}%
\bibitem [{\citenamefont {Isakov}\ \emph {et~al.}(2005)\citenamefont {Isakov},
  \citenamefont {Moessner},\ and\ \citenamefont {Sondhi}}]{Isakov2005}%
  \BibitemOpen
  \bibfield  {author} {\bibinfo {author} {\bibfnamefont {S.~V.}\ \bibnamefont
  {Isakov}}, \bibinfo {author} {\bibfnamefont {R.}~\bibnamefont {Moessner}}, \
  and\ \bibinfo {author} {\bibfnamefont {S.~L.}\ \bibnamefont {Sondhi}},\
  }\bibfield  {title} {\enquote {\bibinfo {title} {Why spin ice obeys the ice
  rules},}\ }\href {\doibase 10.1103/PhysRevLett.95.217201} {\bibfield
  {journal} {\bibinfo  {journal} {Phys. Rev. Lett.}\ }\textbf {\bibinfo
  {volume} {95}},\ \bibinfo {pages} {217201} (\bibinfo {year}
  {2005})}\BibitemShut {NoStop}%
\bibitem [{\citenamefont {Gingras}\ and\ \citenamefont {den
  Hertog}(2001)}]{Gingras2001CJP}%
  \BibitemOpen
  \bibfield  {author} {\bibinfo {author} {\bibfnamefont {M~J~P}\ \bibnamefont
  {Gingras}}\ and\ \bibinfo {author} {\bibfnamefont {B~C}\ \bibnamefont {den
  Hertog}},\ }\bibfield  {title} {\enquote {\bibinfo {title} {Origin of
  spin-ice behavior in {I}sing pyrochlore magnets with long-range dipole
  interactions: an insight from mean-field theory},}\ }\href {\doibase
  10.1139/p01-099} {\bibfield  {journal} {\bibinfo  {journal} {Can. J. Phys.}\
  }\textbf {\bibinfo {volume} {79}},\ \bibinfo {pages} {1339--1351} (\bibinfo
  {year} {2001})}\BibitemShut {NoStop}%
\bibitem [{\citenamefont {Zhou}\ \emph {et~al.}(2008)\citenamefont {Zhou},
  \citenamefont {Wiebe}, \citenamefont {Janik}, \citenamefont {Balicas},
  \citenamefont {Yo}, \citenamefont {Qiu}, \citenamefont {Copley},\ and\
  \citenamefont {Gardner}}]{Zhou2008}%
  \BibitemOpen
  \bibfield  {author} {\bibinfo {author} {\bibfnamefont {H.~D.}\ \bibnamefont
  {Zhou}}, \bibinfo {author} {\bibfnamefont {C.~R.}\ \bibnamefont {Wiebe}},
  \bibinfo {author} {\bibfnamefont {J.~A.}\ \bibnamefont {Janik}}, \bibinfo
  {author} {\bibfnamefont {L.}~\bibnamefont {Balicas}}, \bibinfo {author}
  {\bibfnamefont {Y.~J.}\ \bibnamefont {Yo}}, \bibinfo {author} {\bibfnamefont
  {Y.}~\bibnamefont {Qiu}}, \bibinfo {author} {\bibfnamefont {J.~R.~D.}\
  \bibnamefont {Copley}}, \ and\ \bibinfo {author} {\bibfnamefont {J.~S.}\
  \bibnamefont {Gardner}},\ }\bibfield  {title} {\enquote {\bibinfo {title}
  {Dynamic spin ice: {Pr$_{2}$Sn$_{2}$O$_{7}$}},}\ }\href {\doibase
  10.1103/PhysRevLett.101.227204} {\bibfield  {journal} {\bibinfo  {journal}
  {Phys. Rev. Lett.}\ }\textbf {\bibinfo {volume} {101}},\ \bibinfo {pages}
  {227204} (\bibinfo {year} {2008})}\BibitemShut {NoStop}%
\bibitem [{\citenamefont {Kimura}\ \emph {et~al.}(2013)\citenamefont {Kimura},
  \citenamefont {Nakatsuji}, \citenamefont {Wen}, \citenamefont {Broholm},
  \citenamefont {Stone}, \citenamefont {Nishibori},\ and\ \citenamefont
  {Sawa}}]{kimura2013quantum}%
  \BibitemOpen
  \bibfield  {author} {\bibinfo {author} {\bibfnamefont {Kenta}\ \bibnamefont
  {Kimura}}, \bibinfo {author} {\bibfnamefont {S}~\bibnamefont {Nakatsuji}},
  \bibinfo {author} {\bibfnamefont {JJ}~\bibnamefont {Wen}}, \bibinfo {author}
  {\bibfnamefont {C}~\bibnamefont {Broholm}}, \bibinfo {author} {\bibfnamefont
  {MB}~\bibnamefont {Stone}}, \bibinfo {author} {\bibfnamefont {E}~\bibnamefont
  {Nishibori}}, \ and\ \bibinfo {author} {\bibfnamefont {H}~\bibnamefont
  {Sawa}},\ }\bibfield  {title} {\enquote {\bibinfo {title} {Quantum
  fluctuations in spin-ice-like {Pr$_2$Zr$_2$O$_7$}},}\ }\href {\doibase
  10.1038/ncomms2914} {\bibfield  {journal} {\bibinfo  {journal} {Nat.
  Commun.}\ }\textbf {\bibinfo {volume} {4}},\ \bibinfo {pages} {1934}
  (\bibinfo {year} {2013})}\BibitemShut {NoStop}%
\bibitem [{\citenamefont {Sibille}\ \emph {et~al.}(2016)\citenamefont
  {Sibille}, \citenamefont {Lhotel}, \citenamefont {Hatnean}, \citenamefont
  {Balakrishnan}, \citenamefont {F\aa{}k}, \citenamefont {Gauthier},
  \citenamefont {Fennell},\ and\ \citenamefont {Kenzelmann}}]{Sibille2016}%
  \BibitemOpen
  \bibfield  {author} {\bibinfo {author} {\bibfnamefont {Romain}\ \bibnamefont
  {Sibille}}, \bibinfo {author} {\bibfnamefont {Elsa}\ \bibnamefont {Lhotel}},
  \bibinfo {author} {\bibfnamefont {Monica~Ciomaga}\ \bibnamefont {Hatnean}},
  \bibinfo {author} {\bibfnamefont {Geetha}\ \bibnamefont {Balakrishnan}},
  \bibinfo {author} {\bibfnamefont {Bj\"orn}\ \bibnamefont {F\aa{}k}}, \bibinfo
  {author} {\bibfnamefont {Nicolas}\ \bibnamefont {Gauthier}}, \bibinfo
  {author} {\bibfnamefont {Tom}\ \bibnamefont {Fennell}}, \ and\ \bibinfo
  {author} {\bibfnamefont {Michel}\ \bibnamefont {Kenzelmann}},\ }\bibfield
  {title} {\enquote {\bibinfo {title} {Candidate quantum spin ice in the
  pyrochlore {Pr$_{2}$Hf$_{2}$O$_{7}$}},}\ }\href {\doibase
  10.1103/PhysRevB.94.024436} {\bibfield  {journal} {\bibinfo  {journal} {Phys.
  Rev. B}\ }\textbf {\bibinfo {volume} {94}},\ \bibinfo {pages} {024436}
  (\bibinfo {year} {2016})}\BibitemShut {NoStop}%
\bibitem [{\citenamefont {Onoda}\ and\ \citenamefont
  {Tanaka}(2010)}]{onoda2010}%
  \BibitemOpen
  \bibfield  {author} {\bibinfo {author} {\bibfnamefont {Shigeki}\ \bibnamefont
  {Onoda}}\ and\ \bibinfo {author} {\bibfnamefont {Yoichi}\ \bibnamefont
  {Tanaka}},\ }\bibfield  {title} {\enquote {\bibinfo {title} {Quantum melting
  of spin ice: Emergent cooperative quadrupole and chirality},}\ }\href
  {\doibase 10.1103/PhysRevLett.105.047201} {\bibfield  {journal} {\bibinfo
  {journal} {Phys. Rev. Lett.}\ }\textbf {\bibinfo {volume} {105}},\ \bibinfo
  {pages} {047201} (\bibinfo {year} {2010})}\BibitemShut {NoStop}%
\bibitem [{\citenamefont {Onoda}\ and\ \citenamefont
  {Tanaka}(2011)}]{onoda2011}%
  \BibitemOpen
  \bibfield  {author} {\bibinfo {author} {\bibfnamefont {Shigeki}\ \bibnamefont
  {Onoda}}\ and\ \bibinfo {author} {\bibfnamefont {Yoichi}\ \bibnamefont
  {Tanaka}},\ }\bibfield  {title} {\enquote {\bibinfo {title} {Quantum
  fluctuations in the effective pseudospin-$\frac{1}{2}$ model for magnetic
  pyrochlore oxides},}\ }\href {\doibase 10.1103/PhysRevB.83.094411} {\bibfield
   {journal} {\bibinfo  {journal} {Phys. Rev. B}\ }\textbf {\bibinfo {volume}
  {83}},\ \bibinfo {pages} {094411} (\bibinfo {year} {2011})}\BibitemShut
  {NoStop}%
\bibitem [{Note2()}]{Note2}%
  \BibitemOpen
  \bibinfo {note} {Our convention for the orphan-bond exchange $J_{\protect \rm
  O}$ is different than that used in Ref.~[\protect \rev@citealp
  {Jaubert2016}]. In the latter, an orphan bond has an exchange that differs
  from the bulk exchange $J$ by $\delta _O$, with $\delta _O$ positive or
  negative. We thus have $J_{\protect \rm O}=J+\delta _O$ in the notation of
  Ref.~[\protect \rev@citealp {Jaubert2016}].}\BibitemShut {Stop}%
\bibitem [{\citenamefont {Stanley}(1968)}]{Stanley1968}%
  \BibitemOpen
  \bibfield  {author} {\bibinfo {author} {\bibfnamefont {H.~E.}\ \bibnamefont
  {Stanley}},\ }\bibfield  {title} {\enquote {\bibinfo {title} {Spherical model
  as the limit of infinite spin dimensionality},}\ }\href {\doibase
  10.1103/PhysRev.176.718} {\bibfield  {journal} {\bibinfo  {journal} {Phys.
  Rev.}\ }\textbf {\bibinfo {volume} {176}},\ \bibinfo {pages} {718--722}
  (\bibinfo {year} {1968})}\BibitemShut {NoStop}%
\bibitem [{\citenamefont {Bramwell}\ \emph {et~al.}(1994)\citenamefont
  {Bramwell}, \citenamefont {Gingras},\ and\ \citenamefont
  {Reimers}}]{bramwell1994order}%
  \BibitemOpen
  \bibfield  {author} {\bibinfo {author} {\bibfnamefont {S~T}\ \bibnamefont
  {Bramwell}}, \bibinfo {author} {\bibfnamefont {M~J~P}\ \bibnamefont
  {Gingras}}, \ and\ \bibinfo {author} {\bibfnamefont {J~N}\ \bibnamefont
  {Reimers}},\ }\bibfield  {title} {\enquote {\bibinfo {title} {Order by
  disorder in an anisotropic pyrochlore lattice antiferromagnet},}\ }\href
  {\doibase 10.1063/1.355676} {\bibfield  {journal} {\bibinfo  {journal} {J.
  Appl. Phys.}\ }\textbf {\bibinfo {volume} {75}},\ \bibinfo {pages}
  {5523--5525} (\bibinfo {year} {1994})}\BibitemShut {NoStop}%
\bibitem [{\citenamefont {Moessner}\ and\ \citenamefont
  {Chalker}(1998{\natexlab{a}})}]{Moessner1998}%
  \BibitemOpen
  \bibfield  {author} {\bibinfo {author} {\bibfnamefont {R.}~\bibnamefont
  {Moessner}}\ and\ \bibinfo {author} {\bibfnamefont {J.~T.}\ \bibnamefont
  {Chalker}},\ }\bibfield  {title} {\enquote {\bibinfo {title} {{Properties of
  a classical spin liquid: the Heisenberg pyrochlore antiferromagnet}},}\
  }\href {\doibase 10.1103/PhysRevLett.80.2929} {\bibfield  {journal} {\bibinfo
   {journal} {Phys. Rev. Lett.}\ }\textbf {\bibinfo {volume} {80}},\ \bibinfo
  {pages} {2929} (\bibinfo {year} {1998}{\natexlab{a}})}\BibitemShut {NoStop}%
\bibitem [{\citenamefont {Moessner}\ and\ \citenamefont
  {Chalker}(1998{\natexlab{b}})}]{Moessner1998a}%
  \BibitemOpen
  \bibfield  {author} {\bibinfo {author} {\bibfnamefont {R.}~\bibnamefont
  {Moessner}}\ and\ \bibinfo {author} {\bibfnamefont {J.~T.}\ \bibnamefont
  {Chalker}},\ }\bibfield  {title} {\enquote {\bibinfo {title}
  {{Low-temperature properties of classical geometrically frustrated
  antiferromagnets}},}\ }\href {\doibase 10.1103/PhysRevB.58.12049} {\bibfield
  {journal} {\bibinfo  {journal} {Phys. Rev. B}\ }\textbf {\bibinfo {volume}
  {58}},\ \bibinfo {pages} {12049--12062} (\bibinfo {year}
  {1998}{\natexlab{b}})}\BibitemShut {NoStop}%
\bibitem [{\citenamefont {Conlon}\ and\ \citenamefont
  {Chalker}(2010)}]{Conlon2010}%
  \BibitemOpen
  \bibfield  {author} {\bibinfo {author} {\bibfnamefont {P.~H.}\ \bibnamefont
  {Conlon}}\ and\ \bibinfo {author} {\bibfnamefont {J.~T.}\ \bibnamefont
  {Chalker}},\ }\bibfield  {title} {\enquote {\bibinfo {title} {Absent pinch
  points and emergent clusters: Further neighbor interactions in the pyrochlore
  heisenberg antiferromagnet},}\ }\href {\doibase 10.1103/PhysRevB.81.224413}
  {\bibfield  {journal} {\bibinfo  {journal} {Phys. Rev. B}\ }\textbf {\bibinfo
  {volume} {81}},\ \bibinfo {pages} {224413} (\bibinfo {year}
  {2010})}\BibitemShut {NoStop}%
\bibitem [{\citenamefont {Silverstein}\ \emph {et~al.}(2014)\citenamefont
  {Silverstein}, \citenamefont {Fritsch}, \citenamefont {Flicker},
  \citenamefont {Hallas}, \citenamefont {Gardner}, \citenamefont {Qiu},
  \citenamefont {Ehlers}, \citenamefont {Savici}, \citenamefont {Yamani},
  \citenamefont {Ross}, \citenamefont {Gaulin}, \citenamefont {Gingras},
  \citenamefont {Paddison}, \citenamefont {Foyevtsova}, \citenamefont
  {Valenti}, \citenamefont {Hawthorne}, \citenamefont {Wiebe},\ and\
  \citenamefont {Zhou}}]{Silverstein2014}%
  \BibitemOpen
  \bibfield  {author} {\bibinfo {author} {\bibfnamefont {H.~J.}\ \bibnamefont
  {Silverstein}}, \bibinfo {author} {\bibfnamefont {K.}~\bibnamefont
  {Fritsch}}, \bibinfo {author} {\bibfnamefont {F.}~\bibnamefont {Flicker}},
  \bibinfo {author} {\bibfnamefont {A.~M.}\ \bibnamefont {Hallas}}, \bibinfo
  {author} {\bibfnamefont {J.~S.}\ \bibnamefont {Gardner}}, \bibinfo {author}
  {\bibfnamefont {Y.}~\bibnamefont {Qiu}}, \bibinfo {author} {\bibfnamefont
  {G.}~\bibnamefont {Ehlers}}, \bibinfo {author} {\bibfnamefont {A.~T.}\
  \bibnamefont {Savici}}, \bibinfo {author} {\bibfnamefont {Z.}~\bibnamefont
  {Yamani}}, \bibinfo {author} {\bibfnamefont {K.~A.}\ \bibnamefont {Ross}},
  \bibinfo {author} {\bibfnamefont {B.~D.}\ \bibnamefont {Gaulin}}, \bibinfo
  {author} {\bibfnamefont {M.~J.~P.}\ \bibnamefont {Gingras}}, \bibinfo
  {author} {\bibfnamefont {J.~A.~M.}\ \bibnamefont {Paddison}}, \bibinfo
  {author} {\bibfnamefont {K.}~\bibnamefont {Foyevtsova}}, \bibinfo {author}
  {\bibfnamefont {R.}~\bibnamefont {Valenti}}, \bibinfo {author} {\bibfnamefont
  {F.}~\bibnamefont {Hawthorne}}, \bibinfo {author} {\bibfnamefont {C.~R.}\
  \bibnamefont {Wiebe}}, \ and\ \bibinfo {author} {\bibfnamefont {H.~D.}\
  \bibnamefont {Zhou}},\ }\bibfield  {title} {\enquote {\bibinfo {title}
  {Liquidlike correlations in single-crystalline {Y$_{2}$Mo$_{2}$O$_{7}$}: An
  unconventional spin glass},}\ }\href {\doibase 10.1103/PhysRevB.89.054433}
  {\bibfield  {journal} {\bibinfo  {journal} {Phys. Rev. B}\ }\textbf {\bibinfo
  {volume} {89}},\ \bibinfo {pages} {054433} (\bibinfo {year}
  {2014})}\BibitemShut {NoStop}%
\bibitem [{\citenamefont {Sen}\ \emph {et~al.}(2013)\citenamefont {Sen},
  \citenamefont {Moessner},\ and\ \citenamefont {Sondhi}}]{Sen2013}%
  \BibitemOpen
  \bibfield  {author} {\bibinfo {author} {\bibfnamefont {Arnab}\ \bibnamefont
  {Sen}}, \bibinfo {author} {\bibfnamefont {R.}~\bibnamefont {Moessner}}, \
  and\ \bibinfo {author} {\bibfnamefont {S.~L.}\ \bibnamefont {Sondhi}},\
  }\bibfield  {title} {\enquote {\bibinfo {title} {Coulomb phase diagnostics as
  a function of temperature, interaction range, and disorder},}\ }\href
  {\doibase 10.1103/PhysRevLett.110.107202} {\bibfield  {journal} {\bibinfo
  {journal} {Phys. Rev. Lett.}\ }\textbf {\bibinfo {volume} {110}},\ \bibinfo
  {pages} {107202} (\bibinfo {year} {2013})}\BibitemShut {NoStop}%
\bibitem [{Note3()}]{Note3}%
  \BibitemOpen
  \bibinfo {note} {In principle, the $\lambda _{l}$ could also depend on the
  sublattice index $\alpha $. However, we find that for the cases of interest,
  the symmetries of the film enforce uniformity of the constraint fields within
  each layer (independent of $\alpha $)}\BibitemShut {NoStop}%
\bibitem [{\citenamefont {Otsuka}(2014)}]{otsuka-2014-cluster}%
  \BibitemOpen
  \bibfield  {author} {\bibinfo {author} {\bibfnamefont {Hiromi}\ \bibnamefont
  {Otsuka}},\ }\bibfield  {title} {\enquote {\bibinfo {title} {Cluster
  algorithm for monte carlo simulations of spin ice},}\ }\href {\doibase
  10.1103/PhysRevB.90.220406} {\bibfield  {journal} {\bibinfo  {journal} {Phys.
  Rev. B}\ }\textbf {\bibinfo {volume} {90}},\ \bibinfo {pages} {220406}
  (\bibinfo {year} {2014})}\BibitemShut {NoStop}%
\bibitem [{\citenamefont {Melko}\ \emph {et~al.}(2001)\citenamefont {Melko},
  \citenamefont {den Hertog},\ and\ \citenamefont {Gingras}}]{Melko2001}%
  \BibitemOpen
  \bibfield  {author} {\bibinfo {author} {\bibfnamefont {R~G}\ \bibnamefont
  {Melko}}, \bibinfo {author} {\bibfnamefont {B~C}\ \bibnamefont {den Hertog}},
  \ and\ \bibinfo {author} {\bibfnamefont {M~J~P}\ \bibnamefont {Gingras}},\
  }\bibfield  {title} {\enquote {\bibinfo {title} {{Long-range order at low
  temperatures in dipolar spin ice.}}}\ }\href {\doibase
  10.1103/PhysRevLett.87.067203} {\bibfield  {journal} {\bibinfo  {journal}
  {Phys. Rev. Lett.}\ }\textbf {\bibinfo {volume} {87}},\ \bibinfo {pages}
  {067203} (\bibinfo {year} {2001})}\BibitemShut {NoStop}%
\bibitem [{\citenamefont {Melko}\ and\ \citenamefont
  {Gingras}(2004)}]{melko2004monte}%
  \BibitemOpen
  \bibfield  {author} {\bibinfo {author} {\bibfnamefont {Roger~G}\ \bibnamefont
  {Melko}}\ and\ \bibinfo {author} {\bibfnamefont {Michel~JP}\ \bibnamefont
  {Gingras}},\ }\bibfield  {title} {\enquote {\bibinfo {title} {Monte carlo
  studies of the dipolar spin ice model},}\ }\href {\doibase
  10.1088/0953-8984/16/43/R02} {\bibfield  {journal} {\bibinfo  {journal} {J.
  Phys. Condens. Matter}\ }\textbf {\bibinfo {volume} {16}},\ \bibinfo {pages}
  {R1277} (\bibinfo {year} {2004})}\BibitemShut {NoStop}%
\bibitem [{\citenamefont {Swendsen}\ and\ \citenamefont
  {Wang}(1987)}]{swensden-1987-nonuniversal}%
  \BibitemOpen
  \bibfield  {author} {\bibinfo {author} {\bibfnamefont {Robert~H.}\
  \bibnamefont {Swendsen}}\ and\ \bibinfo {author} {\bibfnamefont {Jian-Sheng}\
  \bibnamefont {Wang}},\ }\bibfield  {title} {\enquote {\bibinfo {title}
  {Nonuniversal critical dynamics in monte carlo simulations},}\ }\href
  {\doibase 10.1103/PhysRevLett.58.86} {\bibfield  {journal} {\bibinfo
  {journal} {Phys. Rev. Lett.}\ }\textbf {\bibinfo {volume} {58}},\ \bibinfo
  {pages} {86--88} (\bibinfo {year} {1987})}\BibitemShut {NoStop}%
\bibitem [{\citenamefont {Wolff}(1989)}]{wolff-1989-collective}%
  \BibitemOpen
  \bibfield  {author} {\bibinfo {author} {\bibfnamefont {Ulli}\ \bibnamefont
  {Wolff}},\ }\bibfield  {title} {\enquote {\bibinfo {title} {Collective monte
  carlo updating for spin systems},}\ }\href {\doibase
  10.1103/PhysRevLett.62.361} {\bibfield  {journal} {\bibinfo  {journal} {Phys.
  Rev. Lett.}\ }\textbf {\bibinfo {volume} {62}},\ \bibinfo {pages} {361--364}
  (\bibinfo {year} {1989})}\BibitemShut {NoStop}%
\bibitem [{\citenamefont {Castelnovo}\ \emph {et~al.}(2011)\citenamefont
  {Castelnovo}, \citenamefont {Moessner},\ and\ \citenamefont
  {Sondhi}}]{Castelnovo2011}%
  \BibitemOpen
  \bibfield  {author} {\bibinfo {author} {\bibfnamefont {C.}~\bibnamefont
  {Castelnovo}}, \bibinfo {author} {\bibfnamefont {R.}~\bibnamefont
  {Moessner}}, \ and\ \bibinfo {author} {\bibfnamefont {S.~L.}\ \bibnamefont
  {Sondhi}},\ }\bibfield  {title} {\enquote {\bibinfo {title} {Debye-h\"uckel
  theory for spin ice at low temperature},}\ }\href {\doibase
  10.1103/PhysRevB.84.144435} {\bibfield  {journal} {\bibinfo  {journal} {Phys.
  Rev. B}\ }\textbf {\bibinfo {volume} {84}},\ \bibinfo {pages} {144435}
  (\bibinfo {year} {2011})}\BibitemShut {NoStop}%
\bibitem [{Note4()}]{Note4}%
  \BibitemOpen
  \bibinfo {note} {For the bulk case, there is an (ad-hoc) procedure to cure
  this discrepancy. Specifically, as discussed in Ref.~[\protect \rev@citealp
  {Sen2013}], one can simply replace the temperature dependence of the
  stiffness ($\lambda $) from large-$N$ by the appropriate exponential form.
  This is not straightforward for films since the constraint fields, $\lambda
  _{l}$ now have an explicit layer dependence. While one could define a
  inhomogeneous stiffness for each tetrahedron, it is ambiguous how to do this
  for each layer (since there are two layers per tetrahedron).}\BibitemShut
  {Stop}%
\bibitem [{Note5()}]{Note5}%
  \BibitemOpen
  \bibinfo {note} {In the dumbbell picture, where each spin is represented by a
  pair of (effective) magnetic charges $\pm q$, these surface monopoles carry
  charge $Q = \pm 2q$ ~\cite {Jaubert2016}, as they are the endpoints of
  \protect \emph {two} flux lines (corresponding to the two aligned orphan bond
  spins).}\BibitemShut {Stop}%
\bibitem [{\citenamefont {Brooks-Bartlett}\ \emph {et~al.}(2014)\citenamefont
  {Brooks-Bartlett}, \citenamefont {Banks}, \citenamefont {Jaubert},
  \citenamefont {Harman-Clarke},\ and\ \citenamefont
  {Holdsworth}}]{Brooks-Bartlett2014}%
  \BibitemOpen
  \bibfield  {author} {\bibinfo {author} {\bibfnamefont {M.~E.}\ \bibnamefont
  {Brooks-Bartlett}}, \bibinfo {author} {\bibfnamefont {S.~T.}\ \bibnamefont
  {Banks}}, \bibinfo {author} {\bibfnamefont {L.~D.~C.}\ \bibnamefont
  {Jaubert}}, \bibinfo {author} {\bibfnamefont {A.}~\bibnamefont
  {Harman-Clarke}}, \ and\ \bibinfo {author} {\bibfnamefont {P.~C.~W.}\
  \bibnamefont {Holdsworth}},\ }\bibfield  {title} {\enquote {\bibinfo {title}
  {Magnetic-moment fragmentation and monopole crystallization},}\ }\href
  {\doibase 10.1103/PhysRevX.4.011007} {\bibfield  {journal} {\bibinfo
  {journal} {Phys. Rev. X}\ }\textbf {\bibinfo {volume} {4}},\ \bibinfo {pages}
  {011007} (\bibinfo {year} {2014})}\BibitemShut {NoStop}%
\bibitem [{\citenamefont {Jaubert}\ \emph {et~al.}(2013)\citenamefont
  {Jaubert}, \citenamefont {Harris}, \citenamefont {Fennell}, \citenamefont
  {Melko}, \citenamefont {Bramwell},\ and\ \citenamefont
  {Holdsworth}}]{Jaubert2013}%
  \BibitemOpen
  \bibfield  {author} {\bibinfo {author} {\bibfnamefont {L.~D.~C.}\
  \bibnamefont {Jaubert}}, \bibinfo {author} {\bibfnamefont {M.~J.}\
  \bibnamefont {Harris}}, \bibinfo {author} {\bibfnamefont {T.}~\bibnamefont
  {Fennell}}, \bibinfo {author} {\bibfnamefont {R.~G.}\ \bibnamefont {Melko}},
  \bibinfo {author} {\bibfnamefont {S.~T.}\ \bibnamefont {Bramwell}}, \ and\
  \bibinfo {author} {\bibfnamefont {P.~C.~W.}\ \bibnamefont {Holdsworth}},\
  }\bibfield  {title} {\enquote {\bibinfo {title} {Topological-sector
  fluctuations and curie-law crossover in spin ice},}\ }\href {\doibase
  10.1103/PhysRevX.3.011014} {\bibfield  {journal} {\bibinfo  {journal} {Phys.
  Rev. X}\ }\textbf {\bibinfo {volume} {3}},\ \bibinfo {pages} {011014}
  (\bibinfo {year} {2013})}\BibitemShut {NoStop}%
\bibitem [{\citenamefont {Hermele}\ \emph {et~al.}(2004)\citenamefont
  {Hermele}, \citenamefont {Fisher},\ and\ \citenamefont
  {Balents}}]{Hermele2003}%
  \BibitemOpen
  \bibfield  {author} {\bibinfo {author} {\bibfnamefont {Michael}\ \bibnamefont
  {Hermele}}, \bibinfo {author} {\bibfnamefont {Matthew P.~A.}\ \bibnamefont
  {Fisher}}, \ and\ \bibinfo {author} {\bibfnamefont {Leon}\ \bibnamefont
  {Balents}},\ }\bibfield  {title} {\enquote {\bibinfo {title} {Pyrochlore
  photons: The {$U(1)$} spin liquid in a {$S=\frac{1}{2}$} three-dimensional
  frustrated magnet},}\ }\href {\doibase 10.1103/PhysRevB.69.064404} {\bibfield
   {journal} {\bibinfo  {journal} {Phys. Rev. B}\ }\textbf {\bibinfo {volume}
  {69}},\ \bibinfo {pages} {064404} (\bibinfo {year} {2004})}\BibitemShut
  {NoStop}%
\bibitem [{Note6()}]{Note6}%
  \BibitemOpen
  \bibinfo {note} {Equivalently, these winding numbers can be directly related
  to the three components of the total magnetization, as discussed in
  Ref.~[\protect \rev@citealp {Jaubert2013}]}\BibitemShut {NoStop}%
\bibitem [{Note7()}]{Note7}%
  \BibitemOpen
  \bibinfo {note} {The criterion proposed in Ref.~[\protect \rev@citealp
  {Rehn2017}] for identifying a classical $Z_2$ spin liquid in the large-$N$
  theory also holds here. For $[001]$ films the spectrum of $V(\protect \bm
  {q})$ has a set of zero-energy flat bands. For $J_{\protect \rm O}/J>0$ there
  is no gap between these low-lying bands and the higher bands; for
  $J_{\protect \rm O}/J<0$ there is such a gap. Furthermore, the emergence of
  this classical $Z_2$ spin liquid is related to structure of the dual
  lattice~\cite {Rehn2017}; when $J_{\protect \rm O}/J<0$ the orphan bond spins
  are aligned at low temperature, rendering the dual lattice non-bipartite upon
  identification of the two orphan bond sites}\BibitemShut {NoStop}%
\bibitem [{\citenamefont {Motrunich}(2003)}]{Motrunich2003}%
  \BibitemOpen
  \bibfield  {author} {\bibinfo {author} {\bibfnamefont {O.~I.}\ \bibnamefont
  {Motrunich}},\ }\bibfield  {title} {\enquote {\bibinfo {title} {Bosonic model
  with ${Z}_{3}$ fractionalization},}\ }\href {\doibase
  10.1103/PhysRevB.67.115108} {\bibfield  {journal} {\bibinfo  {journal} {Phys.
  Rev. B}\ }\textbf {\bibinfo {volume} {67}},\ \bibinfo {pages} {115108}
  (\bibinfo {year} {2003})}\BibitemShut {NoStop}%
\bibitem [{\citenamefont {Rehn}\ \emph {et~al.}(2016)\citenamefont {Rehn},
  \citenamefont {Sen}, \citenamefont {Damle},\ and\ \citenamefont
  {Moessner}}]{Rehn2016}%
  \BibitemOpen
  \bibfield  {author} {\bibinfo {author} {\bibfnamefont {J.}~\bibnamefont
  {Rehn}}, \bibinfo {author} {\bibfnamefont {Arnab}\ \bibnamefont {Sen}},
  \bibinfo {author} {\bibfnamefont {Kedar}\ \bibnamefont {Damle}}, \ and\
  \bibinfo {author} {\bibfnamefont {R.}~\bibnamefont {Moessner}},\ }\bibfield
  {title} {\enquote {\bibinfo {title} {Classical spin liquid on the maximally
  frustrated honeycomb lattice},}\ }\href {\doibase
  10.1103/PhysRevLett.117.167201} {\bibfield  {journal} {\bibinfo  {journal}
  {Phys. Rev. Lett.}\ }\textbf {\bibinfo {volume} {117}},\ \bibinfo {pages}
  {167201} (\bibinfo {year} {2016})}\BibitemShut {NoStop}%
\bibitem [{Note8()}]{Note8}%
  \BibitemOpen
  \bibinfo {note} {There is another possibility for $[110]$ and $[111]$ films:
  surface terminations where only one spin per surface tetrahedron remains. In
  that case, the appearance of surface charges is expected for any number $N$
  of spin components.}\BibitemShut {Stop}%
\bibitem [{\citenamefont {Wen}\ \emph {et~al.}(2017)\citenamefont {Wen},
  \citenamefont {Koohpayeh}, \citenamefont {Ross}, \citenamefont {Trump},
  \citenamefont {McQueen}, \citenamefont {Kimura}, \citenamefont {Nakatsuji},
  \citenamefont {Qiu}, \citenamefont {Pajerowski}, \citenamefont {Copley},\
  and\ \citenamefont {Broholm}}]{Wen2017}%
  \BibitemOpen
  \bibfield  {author} {\bibinfo {author} {\bibfnamefont {J.-J.}\ \bibnamefont
  {Wen}}, \bibinfo {author} {\bibfnamefont {S.~M.}\ \bibnamefont {Koohpayeh}},
  \bibinfo {author} {\bibfnamefont {K.~A.}\ \bibnamefont {Ross}}, \bibinfo
  {author} {\bibfnamefont {B.~A.}\ \bibnamefont {Trump}}, \bibinfo {author}
  {\bibfnamefont {T.~M.}\ \bibnamefont {McQueen}}, \bibinfo {author}
  {\bibfnamefont {K.}~\bibnamefont {Kimura}}, \bibinfo {author} {\bibfnamefont
  {S.}~\bibnamefont {Nakatsuji}}, \bibinfo {author} {\bibfnamefont
  {Y.}~\bibnamefont {Qiu}}, \bibinfo {author} {\bibfnamefont {D.~M.}\
  \bibnamefont {Pajerowski}}, \bibinfo {author} {\bibfnamefont {J.~R.~D.}\
  \bibnamefont {Copley}}, \ and\ \bibinfo {author} {\bibfnamefont {C.~L.}\
  \bibnamefont {Broholm}},\ }\bibfield  {title} {\enquote {\bibinfo {title}
  {Disordered route to the coulomb quantum spin liquid: Random transverse
  fields on spin ice in {Pr$_{2}$Zr$_{2}$O$_{7}$}},}\ }\href {\doibase
  10.1103/PhysRevLett.118.107206} {\bibfield  {journal} {\bibinfo  {journal}
  {Phys. Rev. Lett.}\ }\textbf {\bibinfo {volume} {118}},\ \bibinfo {pages}
  {107206} (\bibinfo {year} {2017})}\BibitemShut {NoStop}%
\bibitem [{\citenamefont {Jaubert}\ \emph {et~al.}(2010)\citenamefont
  {Jaubert}, \citenamefont {Chalker}, \citenamefont {Holdsworth},\ and\
  \citenamefont {Moessner}}]{Jaubert2010}%
  \BibitemOpen
  \bibfield  {author} {\bibinfo {author} {\bibfnamefont {L.~D.~C.}\
  \bibnamefont {Jaubert}}, \bibinfo {author} {\bibfnamefont {J.~T.}\
  \bibnamefont {Chalker}}, \bibinfo {author} {\bibfnamefont {P.~C.~W.}\
  \bibnamefont {Holdsworth}}, \ and\ \bibinfo {author} {\bibfnamefont
  {R.}~\bibnamefont {Moessner}},\ }\bibfield  {title} {\enquote {\bibinfo
  {title} {Spin ice under pressure: Symmetry enhancement and infinite order
  multicriticality},}\ }\href {\doibase 10.1103/PhysRevLett.105.087201}
  {\bibfield  {journal} {\bibinfo  {journal} {Phys. Rev. Lett.}\ }\textbf
  {\bibinfo {volume} {105}},\ \bibinfo {pages} {087201} (\bibinfo {year}
  {2010})}\BibitemShut {NoStop}%
\bibitem [{\citenamefont {Wannier}(1950)}]{Wannier1950}%
  \BibitemOpen
  \bibfield  {author} {\bibinfo {author} {\bibfnamefont {G.~H.}\ \bibnamefont
  {Wannier}},\ }\bibfield  {title} {\enquote {\bibinfo {title}
  {Antiferromagnetism. the triangular ising net},}\ }\href {\doibase
  10.1103/PhysRev.79.357} {\bibfield  {journal} {\bibinfo  {journal} {Phys.
  Rev.}\ }\textbf {\bibinfo {volume} {79}},\ \bibinfo {pages} {357--364}
  (\bibinfo {year} {1950})}\BibitemShut {NoStop}%
\bibitem [{\citenamefont {Lee}\ \emph {et~al.}(2010)\citenamefont {Lee},
  \citenamefont {Takagi}, \citenamefont {Louca}, \citenamefont {Matsuda},
  \citenamefont {Ji}, \citenamefont {Ueda}, \citenamefont {Ueda}, \citenamefont
  {Katsufuji}, \citenamefont {Chung}, \citenamefont {Park}, \citenamefont
  {Cheong},\ and\ \citenamefont {Broholm}}]{Lee2010}%
  \BibitemOpen
  \bibfield  {author} {\bibinfo {author} {\bibfnamefont {Seung~Hun}\
  \bibnamefont {Lee}}, \bibinfo {author} {\bibfnamefont {Hidenori}\
  \bibnamefont {Takagi}}, \bibinfo {author} {\bibfnamefont {Despina}\
  \bibnamefont {Louca}}, \bibinfo {author} {\bibfnamefont {Masaaki}\
  \bibnamefont {Matsuda}}, \bibinfo {author} {\bibfnamefont {Sungdae}\
  \bibnamefont {Ji}}, \bibinfo {author} {\bibfnamefont {Hiroaki}\ \bibnamefont
  {Ueda}}, \bibinfo {author} {\bibfnamefont {Yutaka}\ \bibnamefont {Ueda}},
  \bibinfo {author} {\bibfnamefont {Takuro}\ \bibnamefont {Katsufuji}},
  \bibinfo {author} {\bibfnamefont {Jae~Ho}\ \bibnamefont {Chung}}, \bibinfo
  {author} {\bibfnamefont {Sungil}\ \bibnamefont {Park}}, \bibinfo {author}
  {\bibfnamefont {Sang~Wook}\ \bibnamefont {Cheong}}, \ and\ \bibinfo {author}
  {\bibfnamefont {Collin}\ \bibnamefont {Broholm}},\ }\bibfield  {title}
  {\enquote {\bibinfo {title} {{Frustrated magnetism and cooperative phase
  transitions in spinels}},}\ }\href {\doibase 10.1143/JPSJ.79.011004}
  {\bibfield  {journal} {\bibinfo  {journal} {J. Phys. Soc. Jpn.}\ }\textbf
  {\bibinfo {volume} {79}},\ \bibinfo {pages} {1--14} (\bibinfo {year}
  {2010})}\BibitemShut {NoStop}%
\bibitem [{\citenamefont {Krizan}\ and\ \citenamefont {Cava}(2014)}]{krizan1}%
  \BibitemOpen
  \bibfield  {author} {\bibinfo {author} {\bibfnamefont {J.~W.}\ \bibnamefont
  {Krizan}}\ and\ \bibinfo {author} {\bibfnamefont {R.~J.}\ \bibnamefont
  {Cava}},\ }\bibfield  {title} {\enquote {\bibinfo {title}
  {{NaCaCo$_{2}$F$_{7}$}: A single-crystal high-temperature pyrochlore
  antiferromagnet},}\ }\href {\doibase 10.1103/PhysRevB.89.214401} {\bibfield
  {journal} {\bibinfo  {journal} {Phys. Rev. B}\ }\textbf {\bibinfo {volume}
  {89}},\ \bibinfo {pages} {214401} (\bibinfo {year} {2014})}\BibitemShut
  {NoStop}%
\bibitem [{\citenamefont {Krizan}\ and\ \citenamefont {Cava}(2015)}]{krizan2}%
  \BibitemOpen
  \bibfield  {author} {\bibinfo {author} {\bibfnamefont {J.~W.}\ \bibnamefont
  {Krizan}}\ and\ \bibinfo {author} {\bibfnamefont {R.~J.}\ \bibnamefont
  {Cava}},\ }\bibfield  {title} {\enquote {\bibinfo {title}
  {{NaCaNi$_{2}$F$_{7}$}: A frustrated high-temperature pyrochlore
  antiferromagnet with {$S=1Ni^{2+}$}},}\ }\href {\doibase
  10.1103/PhysRevB.92.014406} {\bibfield  {journal} {\bibinfo  {journal} {Phys.
  Rev. B}\ }\textbf {\bibinfo {volume} {92}},\ \bibinfo {pages} {014406}
  (\bibinfo {year} {2015})}\BibitemShut {NoStop}%
\bibitem [{\citenamefont {Champion}\ \emph {et~al.}(2003)\citenamefont
  {Champion}, \citenamefont {Harris}, \citenamefont {Holdsworth}, \citenamefont
  {Wills}, \citenamefont {Balakrishnan}, \citenamefont {Bramwell},
  \citenamefont {\ifmmode \check{C}\else \v{C}\fi{}i\ifmmode~\check{z}\else
  \v{z}\fi{}m\'ar}, \citenamefont {Fennell}, \citenamefont {Gardner},
  \citenamefont {Lago}, \citenamefont {McMorrow}, \citenamefont
  {Orend\'a\ifmmode~\check{c}\else \v{c}\fi{}}, \citenamefont
  {Orend\'a\ifmmode~\check{c}\else \v{c}\fi{}ov\'a}, \citenamefont {Paul},
  \citenamefont {Smith}, \citenamefont {Telling},\ and\ \citenamefont
  {Wildes}}]{Champion}%
  \BibitemOpen
  \bibfield  {author} {\bibinfo {author} {\bibfnamefont {J.~D.~M.}\
  \bibnamefont {Champion}}, \bibinfo {author} {\bibfnamefont {M.~J.}\
  \bibnamefont {Harris}}, \bibinfo {author} {\bibfnamefont {P.~C.~W.}\
  \bibnamefont {Holdsworth}}, \bibinfo {author} {\bibfnamefont {A.~S.}\
  \bibnamefont {Wills}}, \bibinfo {author} {\bibfnamefont {G.}~\bibnamefont
  {Balakrishnan}}, \bibinfo {author} {\bibfnamefont {S.~T.}\ \bibnamefont
  {Bramwell}}, \bibinfo {author} {\bibfnamefont {E.}~\bibnamefont {\ifmmode
  \check{C}\else \v{C}\fi{}i\ifmmode~\check{z}\else \v{z}\fi{}m\'ar}}, \bibinfo
  {author} {\bibfnamefont {T.}~\bibnamefont {Fennell}}, \bibinfo {author}
  {\bibfnamefont {J.~S.}\ \bibnamefont {Gardner}}, \bibinfo {author}
  {\bibfnamefont {J.}~\bibnamefont {Lago}}, \bibinfo {author} {\bibfnamefont
  {D.~F.}\ \bibnamefont {McMorrow}}, \bibinfo {author} {\bibfnamefont
  {M.}~\bibnamefont {Orend\'a\ifmmode~\check{c}\else \v{c}\fi{}}}, \bibinfo
  {author} {\bibfnamefont {A.}~\bibnamefont {Orend\'a\ifmmode~\check{c}\else
  \v{c}\fi{}ov\'a}}, \bibinfo {author} {\bibfnamefont {D.~McK.}\ \bibnamefont
  {Paul}}, \bibinfo {author} {\bibfnamefont {R.~I.}\ \bibnamefont {Smith}},
  \bibinfo {author} {\bibfnamefont {M.~T.~F.}\ \bibnamefont {Telling}}, \ and\
  \bibinfo {author} {\bibfnamefont {A.}~\bibnamefont {Wildes}},\ }\bibfield
  {title} {\enquote {\bibinfo {title} {{Er$_{2}$Ti$_{2}$O$_{7}$}: Evidence of
  quantum order by disorder in a frustrated antiferromagnet},}\ }\href
  {\doibase 10.1103/PhysRevB.68.020401} {\bibfield  {journal} {\bibinfo
  {journal} {Phys. Rev. B}\ }\textbf {\bibinfo {volume} {68}},\ \bibinfo
  {pages} {020401} (\bibinfo {year} {2003})}\BibitemShut {NoStop}%
\bibitem [{\citenamefont {Zhitomirsky}\ \emph {et~al.}(2012)\citenamefont
  {Zhitomirsky}, \citenamefont {Gvozdikova}, \citenamefont {Holdsworth},\ and\
  \citenamefont {Moessner}}]{Zhito}%
  \BibitemOpen
  \bibfield  {author} {\bibinfo {author} {\bibfnamefont {M.~E.}\ \bibnamefont
  {Zhitomirsky}}, \bibinfo {author} {\bibfnamefont {M.~V.}\ \bibnamefont
  {Gvozdikova}}, \bibinfo {author} {\bibfnamefont {P.~C.~W.}\ \bibnamefont
  {Holdsworth}}, \ and\ \bibinfo {author} {\bibfnamefont {R.}~\bibnamefont
  {Moessner}},\ }\bibfield  {title} {\enquote {\bibinfo {title} {Quantum order
  by disorder and accidental soft mode in {Er$_{2}$Ti$_{2}$O$_{7}$}},}\ }\href
  {\doibase 10.1103/PhysRevLett.109.077204} {\bibfield  {journal} {\bibinfo
  {journal} {Phys. Rev. Lett.}\ }\textbf {\bibinfo {volume} {109}},\ \bibinfo
  {pages} {077204} (\bibinfo {year} {2012})}\BibitemShut {NoStop}%
\bibitem [{\citenamefont {Savary}\ \emph {et~al.}(2012)\citenamefont {Savary},
  \citenamefont {Ross}, \citenamefont {Gaulin}, \citenamefont {Ruff},\ and\
  \citenamefont {Balents}}]{Savary}%
  \BibitemOpen
  \bibfield  {author} {\bibinfo {author} {\bibfnamefont {Lucile}\ \bibnamefont
  {Savary}}, \bibinfo {author} {\bibfnamefont {Kate~A.}\ \bibnamefont {Ross}},
  \bibinfo {author} {\bibfnamefont {Bruce~D.}\ \bibnamefont {Gaulin}}, \bibinfo
  {author} {\bibfnamefont {Jacob P.~C.}\ \bibnamefont {Ruff}}, \ and\ \bibinfo
  {author} {\bibfnamefont {Leon}\ \bibnamefont {Balents}},\ }\bibfield  {title}
  {\enquote {\bibinfo {title} {Order by quantum disorder in
  {Er$_{2}$Ti$_{2}$O$_{7}$}},}\ }\href {\doibase
  10.1103/PhysRevLett.109.167201} {\bibfield  {journal} {\bibinfo  {journal}
  {Phys. Rev. Lett.}\ }\textbf {\bibinfo {volume} {109}},\ \bibinfo {pages}
  {167201} (\bibinfo {year} {2012})}\BibitemShut {NoStop}%
\bibitem [{\citenamefont {Dun}\ \emph {et~al.}(2014)\citenamefont {Dun},
  \citenamefont {Lee}, \citenamefont {Choi}, \citenamefont {Hallas},
  \citenamefont {Wiebe}, \citenamefont {Gardner}, \citenamefont {Arrighi},
  \citenamefont {Freitas}, \citenamefont {Arevalo-Lopez}, \citenamefont
  {Attfield}, \citenamefont {Zhou},\ and\ \citenamefont {Cheng}}]{Dun2014}%
  \BibitemOpen
  \bibfield  {author} {\bibinfo {author} {\bibfnamefont {Z.~L.}\ \bibnamefont
  {Dun}}, \bibinfo {author} {\bibfnamefont {M.}~\bibnamefont {Lee}}, \bibinfo
  {author} {\bibfnamefont {E.~S.}\ \bibnamefont {Choi}}, \bibinfo {author}
  {\bibfnamefont {A.~M.}\ \bibnamefont {Hallas}}, \bibinfo {author}
  {\bibfnamefont {C.~R.}\ \bibnamefont {Wiebe}}, \bibinfo {author}
  {\bibfnamefont {J.~S.}\ \bibnamefont {Gardner}}, \bibinfo {author}
  {\bibfnamefont {E.}~\bibnamefont {Arrighi}}, \bibinfo {author} {\bibfnamefont
  {R.~S.}\ \bibnamefont {Freitas}}, \bibinfo {author} {\bibfnamefont {A.~M.}\
  \bibnamefont {Arevalo-Lopez}}, \bibinfo {author} {\bibfnamefont {J.~P.}\
  \bibnamefont {Attfield}}, \bibinfo {author} {\bibfnamefont {H.~D.}\
  \bibnamefont {Zhou}}, \ and\ \bibinfo {author} {\bibfnamefont {J.~G.}\
  \bibnamefont {Cheng}},\ }\bibfield  {title} {\enquote {\bibinfo {title}
  {Chemical pressure effects on magnetism in the quantum spin liquid candidates
  {Yb$_{2}X_{2}$O$_{7}$ ($X=$Sn, Ti, Ge)}},}\ }\href {\doibase
  10.1103/PhysRevB.89.064401} {\bibfield  {journal} {\bibinfo  {journal} {Phys.
  Rev. B}\ }\textbf {\bibinfo {volume} {89}},\ \bibinfo {pages} {064401}
  (\bibinfo {year} {2014})}\BibitemShut {NoStop}%
\bibitem [{\citenamefont {Dun}\ \emph {et~al.}(2015)\citenamefont {Dun},
  \citenamefont {Li}, \citenamefont {Freitas}, \citenamefont {Arrighi},
  \citenamefont {Dela~Cruz}, \citenamefont {Lee}, \citenamefont {Choi},
  \citenamefont {Cao}, \citenamefont {Silverstein}, \citenamefont {Wiebe},
  \citenamefont {Cheng},\ and\ \citenamefont {Zhou}}]{Dun2015}%
  \BibitemOpen
  \bibfield  {author} {\bibinfo {author} {\bibfnamefont {Z.~L.}\ \bibnamefont
  {Dun}}, \bibinfo {author} {\bibfnamefont {X.}~\bibnamefont {Li}}, \bibinfo
  {author} {\bibfnamefont {R.~S.}\ \bibnamefont {Freitas}}, \bibinfo {author}
  {\bibfnamefont {E.}~\bibnamefont {Arrighi}}, \bibinfo {author} {\bibfnamefont
  {C.~R.}\ \bibnamefont {Dela~Cruz}}, \bibinfo {author} {\bibfnamefont
  {M.}~\bibnamefont {Lee}}, \bibinfo {author} {\bibfnamefont {E.~S.}\
  \bibnamefont {Choi}}, \bibinfo {author} {\bibfnamefont {H.~B.}\ \bibnamefont
  {Cao}}, \bibinfo {author} {\bibfnamefont {H.~J.}\ \bibnamefont
  {Silverstein}}, \bibinfo {author} {\bibfnamefont {C.~R.}\ \bibnamefont
  {Wiebe}}, \bibinfo {author} {\bibfnamefont {J.~G.}\ \bibnamefont {Cheng}}, \
  and\ \bibinfo {author} {\bibfnamefont {H.~D.}\ \bibnamefont {Zhou}},\
  }\bibfield  {title} {\enquote {\bibinfo {title} {Antiferromagnetic order in
  the pyrochlores {$R_{2}$Ge$_{2}$O$_{7}$ ($R=$ Er, Yb)}},}\ }\href {\doibase
  10.1103/PhysRevB.92.140407} {\bibfield  {journal} {\bibinfo  {journal} {Phys.
  Rev. B}\ }\textbf {\bibinfo {volume} {92}},\ \bibinfo {pages} {140407}
  (\bibinfo {year} {2015})}\BibitemShut {NoStop}%
\bibitem [{\citenamefont {Hallas}\ \emph {et~al.}(2016)\citenamefont {Hallas},
  \citenamefont {Gaudet}, \citenamefont {Butch}, \citenamefont {Tachibana},
  \citenamefont {Freitas}, \citenamefont {Luke}, \citenamefont {Wiebe},\ and\
  \citenamefont {Gaulin}}]{Hallas}%
  \BibitemOpen
  \bibfield  {author} {\bibinfo {author} {\bibfnamefont {A.~M.}\ \bibnamefont
  {Hallas}}, \bibinfo {author} {\bibfnamefont {J.}~\bibnamefont {Gaudet}},
  \bibinfo {author} {\bibfnamefont {N.~P.}\ \bibnamefont {Butch}}, \bibinfo
  {author} {\bibfnamefont {M.}~\bibnamefont {Tachibana}}, \bibinfo {author}
  {\bibfnamefont {R.~S.}\ \bibnamefont {Freitas}}, \bibinfo {author}
  {\bibfnamefont {G.~M.}\ \bibnamefont {Luke}}, \bibinfo {author}
  {\bibfnamefont {C.~R.}\ \bibnamefont {Wiebe}}, \ and\ \bibinfo {author}
  {\bibfnamefont {B.~D.}\ \bibnamefont {Gaulin}},\ }\bibfield  {title}
  {\enquote {\bibinfo {title} {Universal dynamic magnetism in {Y}b pyrochlores
  with disparate ground states},}\ }\href {\doibase 10.1103/PhysRevB.93.100403}
  {\bibfield  {journal} {\bibinfo  {journal} {Phys. Rev. B}\ }\textbf {\bibinfo
  {volume} {93}},\ \bibinfo {pages} {100403} (\bibinfo {year}
  {2016})}\BibitemShut {NoStop}%
\bibitem [{\citenamefont {Hallas}\ \emph {et~al.}(2018)\citenamefont {Hallas},
  \citenamefont {Gaudet},\ and\ \citenamefont {Gaulin}}]{hallas2017review}%
  \BibitemOpen
  \bibfield  {author} {\bibinfo {author} {\bibfnamefont {Alannah~M.}\
  \bibnamefont {Hallas}}, \bibinfo {author} {\bibfnamefont {Jonathan}\
  \bibnamefont {Gaudet}}, \ and\ \bibinfo {author} {\bibfnamefont {Bruce~D.}\
  \bibnamefont {Gaulin}},\ }\bibfield  {title} {\enquote {\bibinfo {title}
  {Experimental insights into ground-state selection of quantum {XY}
  pyrochlores},}\ }\href {\doibase 10.1146/annurev-conmatphys-031016-025218}
  {\bibfield  {journal} {\bibinfo  {journal} {Annu. Rev. Condens. Matter
  Phys.}\ }\textbf {\bibinfo {volume} {9}},\ \bibinfo {pages} {105--124}
  (\bibinfo {year} {2018})}\BibitemShut {NoStop}%
\bibitem [{\citenamefont {Taillefumier}\ \emph {et~al.}(2017)\citenamefont
  {Taillefumier}, \citenamefont {Benton}, \citenamefont {Yan}, \citenamefont
  {Jaubert},\ and\ \citenamefont {Shannon}}]{taillefumier2017frustrating}%
  \BibitemOpen
  \bibfield  {author} {\bibinfo {author} {\bibfnamefont {Mathieu}\ \bibnamefont
  {Taillefumier}}, \bibinfo {author} {\bibfnamefont {Owen}\ \bibnamefont
  {Benton}}, \bibinfo {author} {\bibfnamefont {Han}\ \bibnamefont {Yan}},
  \bibinfo {author} {\bibfnamefont {L.~D.~C.}\ \bibnamefont {Jaubert}}, \ and\
  \bibinfo {author} {\bibfnamefont {Nic}\ \bibnamefont {Shannon}},\ }\bibfield
  {title} {\enquote {\bibinfo {title} {Competing spin liquids and hidden
  spin-nematic order in spin ice with frustrated transverse exchange},}\ }\href
  {\doibase 10.1103/PhysRevX.7.041057} {\bibfield  {journal} {\bibinfo
  {journal} {Phys. Rev. X}\ }\textbf {\bibinfo {volume} {7}},\ \bibinfo {pages}
  {041057} (\bibinfo {year} {2017})}\BibitemShut {NoStop}%
\bibitem [{\citenamefont {Moessner}\ and\ \citenamefont
  {Sondhi}(2003)}]{moessner2003}%
  \BibitemOpen
  \bibfield  {author} {\bibinfo {author} {\bibfnamefont {R.}~\bibnamefont
  {Moessner}}\ and\ \bibinfo {author} {\bibfnamefont {S.~L.}\ \bibnamefont
  {Sondhi}},\ }\bibfield  {title} {\enquote {\bibinfo {title} {Theory of the
  [111] magnetization plateau in spin ice},}\ }\href {\doibase
  10.1103/PhysRevB.68.064411} {\bibfield  {journal} {\bibinfo  {journal} {Phys.
  Rev. B}\ }\textbf {\bibinfo {volume} {68}},\ \bibinfo {pages} {064411}
  (\bibinfo {year} {2003})}\BibitemShut {NoStop}%
\bibitem [{\citenamefont {Jaubert}\ \emph {et~al.}(2008)\citenamefont
  {Jaubert}, \citenamefont {Chalker}, \citenamefont {Holdsworth},\ and\
  \citenamefont {Moessner}}]{Jaubert2008}%
  \BibitemOpen
  \bibfield  {author} {\bibinfo {author} {\bibfnamefont {L.~D.~C.}\
  \bibnamefont {Jaubert}}, \bibinfo {author} {\bibfnamefont {J.~T.}\
  \bibnamefont {Chalker}}, \bibinfo {author} {\bibfnamefont {P.~C.~W.}\
  \bibnamefont {Holdsworth}}, \ and\ \bibinfo {author} {\bibfnamefont
  {R.}~\bibnamefont {Moessner}},\ }\bibfield  {title} {\enquote {\bibinfo
  {title} {Three-dimensional {K}asteleyn transition: Spin ice in a [100]
  field},}\ }\href {\doibase 10.1103/PhysRevLett.100.067207} {\bibfield
  {journal} {\bibinfo  {journal} {Phys. Rev. Lett.}\ }\textbf {\bibinfo
  {volume} {100}},\ \bibinfo {pages} {067207} (\bibinfo {year}
  {2008})}\BibitemShut {NoStop}%
\bibitem [{\citenamefont {Ruff}\ \emph {et~al.}(2005)\citenamefont {Ruff},
  \citenamefont {Melko},\ and\ \citenamefont {Gingras}}]{Ruff2005}%
  \BibitemOpen
  \bibfield  {author} {\bibinfo {author} {\bibfnamefont {Jacob P.~C.}\
  \bibnamefont {Ruff}}, \bibinfo {author} {\bibfnamefont {Roger~G.}\
  \bibnamefont {Melko}}, \ and\ \bibinfo {author} {\bibfnamefont {Michel
  J.~P.}\ \bibnamefont {Gingras}},\ }\bibfield  {title} {\enquote {\bibinfo
  {title} {Finite-temperature transitions in dipolar spin ice in a large
  magnetic field},}\ }\href {\doibase 10.1103/PhysRevLett.95.097202} {\bibfield
   {journal} {\bibinfo  {journal} {Phys. Rev. Lett.}\ }\textbf {\bibinfo
  {volume} {95}},\ \bibinfo {pages} {097202} (\bibinfo {year}
  {2005})}\BibitemShut {NoStop}%
\bibitem [{\citenamefont {Polyakov}(1977)}]{polyakov1977quark}%
  \BibitemOpen
  \bibfield  {author} {\bibinfo {author} {\bibfnamefont {Alexander~M}\
  \bibnamefont {Polyakov}},\ }\bibfield  {title} {\enquote {\bibinfo {title}
  {Quark confinement and topology of gauge theories},}\ }\href {\doibase
  10.1016/0550-3213(77)90086-4} {\bibfield  {journal} {\bibinfo  {journal}
  {Nucl. Phys. B}\ }\textbf {\bibinfo {volume} {120}},\ \bibinfo {pages}
  {429--458} (\bibinfo {year} {1977})}\BibitemShut {NoStop}%
\bibitem [{\citenamefont {Henry}\ and\ \citenamefont
  {Roscilde}(2014)}]{Henry2014}%
  \BibitemOpen
  \bibfield  {author} {\bibinfo {author} {\bibfnamefont {Louis-Paul}\
  \bibnamefont {Henry}}\ and\ \bibinfo {author} {\bibfnamefont {Tommaso}\
  \bibnamefont {Roscilde}},\ }\bibfield  {title} {\enquote {\bibinfo {title}
  {Order-by-disorder and quantum coulomb phase in quantum square ice},}\ }\href
  {\doibase 10.1103/PhysRevLett.113.027204} {\bibfield  {journal} {\bibinfo
  {journal} {Phys. Rev. Lett.}\ }\textbf {\bibinfo {volume} {113}},\ \bibinfo
  {pages} {027204} (\bibinfo {year} {2014})}\BibitemShut {NoStop}%
\bibitem [{\citenamefont {Carrasquilla}\ \emph {et~al.}(2015)\citenamefont
  {Carrasquilla}, \citenamefont {Hao},\ and\ \citenamefont
  {Melko}}]{carrasquilla2015two}%
  \BibitemOpen
  \bibfield  {author} {\bibinfo {author} {\bibfnamefont {Juan}\ \bibnamefont
  {Carrasquilla}}, \bibinfo {author} {\bibfnamefont {Zhihao}\ \bibnamefont
  {Hao}}, \ and\ \bibinfo {author} {\bibfnamefont {Roger~G}\ \bibnamefont
  {Melko}},\ }\bibfield  {title} {\enquote {\bibinfo {title} {A two-dimensional
  spin liquid in quantum kagome ice},}\ }\href {\doibase 10.1038/ncomms8421}
  {\bibfield  {journal} {\bibinfo  {journal} {Nat. Commun.}\ }\textbf {\bibinfo
  {volume} {6}},\ \bibinfo {pages} {7421} (\bibinfo {year} {2015})}\BibitemShut
  {NoStop}%
\bibitem [{\citenamefont {Kandel}\ and\ \citenamefont
  {Domany}(1991)}]{kandel1991}%
  \BibitemOpen
  \bibfield  {author} {\bibinfo {author} {\bibfnamefont {Daniel}\ \bibnamefont
  {Kandel}}\ and\ \bibinfo {author} {\bibfnamefont {Eytan}\ \bibnamefont
  {Domany}},\ }\bibfield  {title} {\enquote {\bibinfo {title} {General cluster
  monte carlo dynamics},}\ }\href {\doibase 10.1103/PhysRevB.43.8539}
  {\bibfield  {journal} {\bibinfo  {journal} {Phys. Rev. B}\ }\textbf {\bibinfo
  {volume} {43}},\ \bibinfo {pages} {8539--8548} (\bibinfo {year}
  {1991})}\BibitemShut {NoStop}%
\bibitem [{\citenamefont {Evertz}(2003)}]{evertz2003loop}%
  \BibitemOpen
  \bibfield  {author} {\bibinfo {author} {\bibfnamefont {Hans~Gerd}\
  \bibnamefont {Evertz}},\ }\bibfield  {title} {\enquote {\bibinfo {title} {The
  loop algorithm},}\ }\href {\doibase 10.1080/0001873021000049195} {\bibfield
  {journal} {\bibinfo  {journal} {Adv. Phys.}\ }\textbf {\bibinfo {volume}
  {52}},\ \bibinfo {pages} {1--66} (\bibinfo {year} {2003})}\BibitemShut
  {NoStop}%
\bibitem [{\citenamefont {Pomaranski}\ \emph {et~al.}(2013)\citenamefont
  {Pomaranski}, \citenamefont {Yaraskavitch}, \citenamefont {Meng},
  \citenamefont {Ross}, \citenamefont {Noad}, \citenamefont {Dabkowska},
  \citenamefont {Gaulin},\ and\ \citenamefont {Kycia}}]{pomaranski2013absence}%
  \BibitemOpen
  \bibfield  {author} {\bibinfo {author} {\bibfnamefont {D}~\bibnamefont
  {Pomaranski}}, \bibinfo {author} {\bibfnamefont {LR}~\bibnamefont
  {Yaraskavitch}}, \bibinfo {author} {\bibfnamefont {S}~\bibnamefont {Meng}},
  \bibinfo {author} {\bibfnamefont {KA}~\bibnamefont {Ross}}, \bibinfo {author}
  {\bibfnamefont {HML}\ \bibnamefont {Noad}}, \bibinfo {author} {\bibfnamefont
  {HA}~\bibnamefont {Dabkowska}}, \bibinfo {author} {\bibfnamefont
  {BD}~\bibnamefont {Gaulin}}, \ and\ \bibinfo {author} {\bibfnamefont
  {JB}~\bibnamefont {Kycia}},\ }\bibfield  {title} {\enquote {\bibinfo {title}
  {Absence of {P}auling's residual entropy in thermally equilibrated
  {Dy$_2$Ti$_2$O$_7$}},}\ }\href {\doibase 10.1038/nphys2591} {\bibfield
  {journal} {\bibinfo  {journal} {Nat. Phys.}\ }\textbf {\bibinfo {volume}
  {9}},\ \bibinfo {pages} {353} (\bibinfo {year} {2013})}\BibitemShut {NoStop}%
\bibitem [{\citenamefont {Henelius}\ \emph {et~al.}(2016)\citenamefont
  {Henelius}, \citenamefont {Lin}, \citenamefont {Enjalran}, \citenamefont
  {Hao}, \citenamefont {Rau}, \citenamefont {Altosaar}, \citenamefont
  {Flicker}, \citenamefont {Yavors'kii},\ and\ \citenamefont
  {Gingras}}]{henelius2016}%
  \BibitemOpen
  \bibfield  {author} {\bibinfo {author} {\bibfnamefont {P.}~\bibnamefont
  {Henelius}}, \bibinfo {author} {\bibfnamefont {T.}~\bibnamefont {Lin}},
  \bibinfo {author} {\bibfnamefont {M.}~\bibnamefont {Enjalran}}, \bibinfo
  {author} {\bibfnamefont {Z.}~\bibnamefont {Hao}}, \bibinfo {author}
  {\bibfnamefont {J.~G.}\ \bibnamefont {Rau}}, \bibinfo {author} {\bibfnamefont
  {J.}~\bibnamefont {Altosaar}}, \bibinfo {author} {\bibfnamefont
  {F.}~\bibnamefont {Flicker}}, \bibinfo {author} {\bibfnamefont
  {T.}~\bibnamefont {Yavors'kii}}, \ and\ \bibinfo {author} {\bibfnamefont
  {M.~J.~P.}\ \bibnamefont {Gingras}},\ }\bibfield  {title} {\enquote {\bibinfo
  {title} {Refrustration and competing orders in the prototypical {
  Dy$_{2}$Ti$_{2}$O$_{7}$} spin ice material},}\ }\href {\doibase
  10.1103/PhysRevB.93.024402} {\bibfield  {journal} {\bibinfo  {journal} {Phys.
  Rev. B}\ }\textbf {\bibinfo {volume} {93}},\ \bibinfo {pages} {024402}
  (\bibinfo {year} {2016})}\BibitemShut {NoStop}%
\bibitem [{\citenamefont {Borzi}\ \emph {et~al.}(2016)\citenamefont {Borzi},
  \citenamefont {Albarrac{\'\i}n}, \citenamefont {Rosales}, \citenamefont
  {Rossini}, \citenamefont {Steppke}, \citenamefont {Prabhakaran},
  \citenamefont {Mackenzie}, \citenamefont {Cabra},\ and\ \citenamefont
  {Grigera}}]{borzi2016intermediate}%
  \BibitemOpen
  \bibfield  {author} {\bibinfo {author} {\bibfnamefont {RA}~\bibnamefont
  {Borzi}}, \bibinfo {author} {\bibfnamefont {FA~G{\'o}mez}\ \bibnamefont
  {Albarrac{\'\i}n}}, \bibinfo {author} {\bibfnamefont {HD}~\bibnamefont
  {Rosales}}, \bibinfo {author} {\bibfnamefont {GL}~\bibnamefont {Rossini}},
  \bibinfo {author} {\bibfnamefont {Alexander}\ \bibnamefont {Steppke}},
  \bibinfo {author} {\bibfnamefont {D}~\bibnamefont {Prabhakaran}}, \bibinfo
  {author} {\bibfnamefont {AP}~\bibnamefont {Mackenzie}}, \bibinfo {author}
  {\bibfnamefont {DC}~\bibnamefont {Cabra}}, \ and\ \bibinfo {author}
  {\bibfnamefont {SA}~\bibnamefont {Grigera}},\ }\bibfield  {title} {\enquote
  {\bibinfo {title} {Intermediate magnetization state and competing orders in {
  Dy$_{2}$Ti$_{2}$O$_{7}$} and {Ho$_{2}$Ti$_{2}$O$_{7}$}},}\ }\href {\doibase
  10.1038/ncomms12592} {\bibfield  {journal} {\bibinfo  {journal} {Nat.
  Commun.}\ }\textbf {\bibinfo {volume} {7}},\ \bibinfo {pages} {12592}
  (\bibinfo {year} {2016})}\BibitemShut {NoStop}%
\end{thebibliography}%

\end{document}